\RequirePackage{fix-cm}
\documentclass[twocolumn]{svjour3}          %
\smartqed  %
\usepackage{graphicx}
\usepackage{subcaption}
\usepackage{booktabs}
\usepackage{algorithm}
\usepackage{algorithmic}
\usepackage{gensymb}
\usepackage{amssymb}
\usepackage{amsmath}
\usepackage[switch]{lineno}
\usepackage{dblfloatfix}
\newcommand{\expnumber}[2]{{#1}\mathrm{e}{#2}}
\journalname{Autonomous Agents and Multi-Agent Systems}
\begin{document}

\title{Quantifying the effects of environment and population diversity in multi-agent reinforcement learning%
}

\author{Kevin R. McKee*\thanks{*equal contribution} \and
        Joel Z. Leibo          \and
        Charlie Beattie        \and
        Richard Everett*
}

\authorrunning{Kevin R. McKee et al.} %

\institute{ K. McKee \at
            \email{kevinrmckee@deepmind.com}
            \and
            R. Everett \at
            \email{reverett@deepmind.com}
}

\date{Received: 30 Jul 2021 / Accepted: 05 Feb 2022}

\maketitle

\begin{abstract}
Generalization is a major challenge for multi-agent reinforcement learning. How well does an agent perform when placed in novel environments and in interactions with new co-players? In this paper, we investigate and quantify the relationship between generalization and \textit{diversity} in the multi-agent domain. %
Across the range of multi-agent environments considered here, procedurally generating training levels significantly improves agent performance on held-out levels. %
However, agent performance on the specific levels used in training sometimes declines as a result.
To better understand the effects of co-player variation, our experiments introduce a new environment-agnostic measure of behavioral diversity. Results demonstrate that population size and intrinsic motivation are both effective methods of generating greater population diversity. In turn, training with a diverse set of co-players strengthens agent performance in some (but not all) cases.
\end{abstract}
\section{Introduction}
\label{sec:introduction}

An emerging theme in single-agent reinforcement learning research is the effect of environment diversity on learning and generalization \cite{juliani2019obstacle,justesen2018illuminating,wang2019paired}. Reinforcement learning agents are typically trained and tested on a single level, which produces high performance and brittle generalization. Such overfitting stems from agents' capacity to memorize a mapping from environmental states observed in training to specific actions \cite{zhang2018study}. Single-agent research has counteracted and alleviated overfitting by incorporating environment diversity into training. For example, procedural generation can be used to produce larger sets of training levels and thereby encourage policy generality \cite{cobbe2019leveraging,cobbe2019quantifying}.

In multi-agent settings, the tendency of agents to overfit to their co-players is another large challenge to generalization \cite{lanctot2017unified}. Generalization performance tends to be more robust when agents train with a heterogeneous set of co-players. Prior studies have induced policy generality through population-based training \cite{carroll2019utility,jaderberg2019human}, policy ensembles \cite{lowe2017multi}, the application of diverse leagues of game opponents \cite{vinyals2019grandmaster}, and the diversification of architectures or hyperparameters for the agents within the population \cite{hu2020other,mckee2020social}.

Of course, the environment is still a major component of multi-agent reinforcement learning. In multi-agent games, an agent’s learning is shaped by both the other co-players and the environment \cite{littman1994markov}. Despite this structure, only a handful of studies have explicitly assessed the effects of environmental variation on multi-agent learning. Jaderberg et al. \cite{jaderberg2019human} developed agents for Capture the Flag that were capable of responding to a variety of opponents and match conditions. They argued that this generalizability was produced in part by the use of procedurally generated levels during training. Other multi-agent experiments using procedurally generated levels (e.g., \cite{everett2019optimising,leibo2019malthusian}) stop short of rigorously measuring generalization. It thus remains an open question whether procedural generation of training levels benefits generalization in multi-agent learning.

\begin{figure*}[t!]
    \begin{subfigure}{0.23\textwidth}
        \centering
        \includegraphics[height=16em]{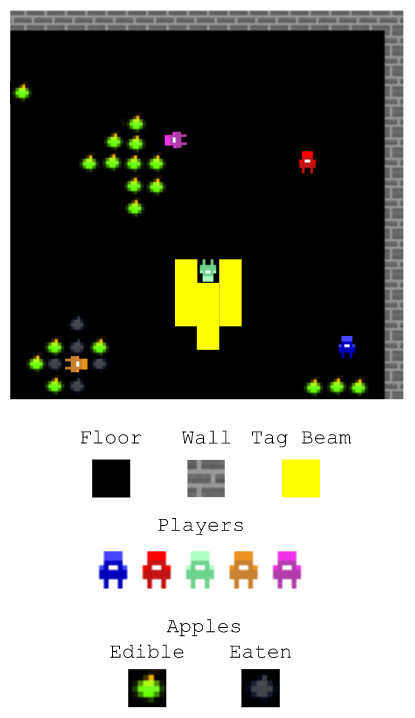}
        \caption{HarvestPatch}
        \label{fig:2/environments/hp}        
    \end{subfigure}
    \hfill    
    \begin{subfigure}{0.23\textwidth}
        \centering
        \includegraphics[height=16em]{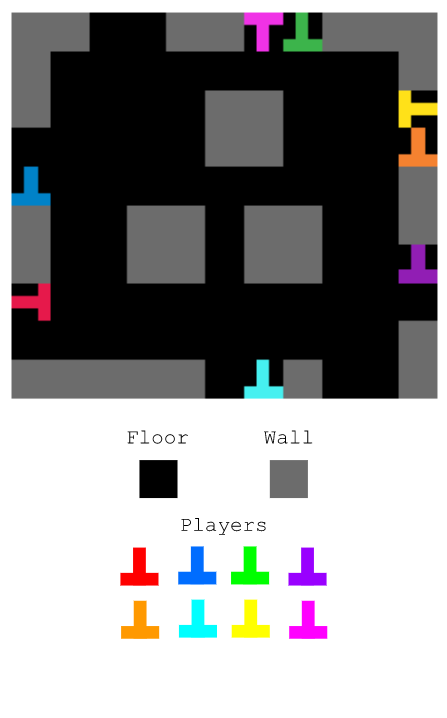}
        \caption{Traffic Navigation}
        \label{fig:2/environments/tn}        
    \end{subfigure}   
    \hfill
    \begin{subfigure}{0.23\textwidth}
        \centering
        \includegraphics[height=16em]{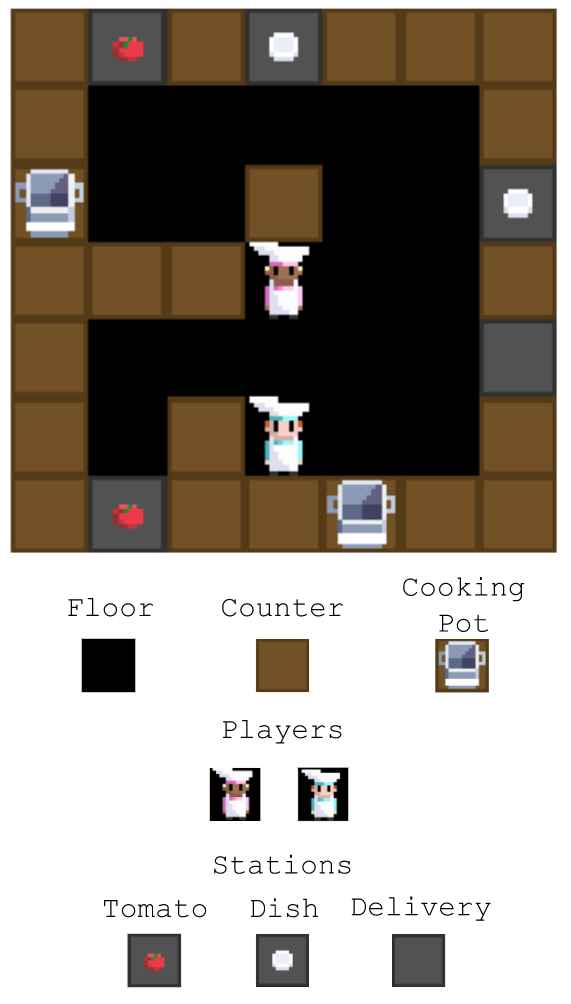}
        \caption{Overcooked}
        \label{fig:2/environments/oc}
    \end{subfigure}    
    \hfill
    \begin{subfigure}{0.23\textwidth}
        \centering
        \includegraphics[height=16em]{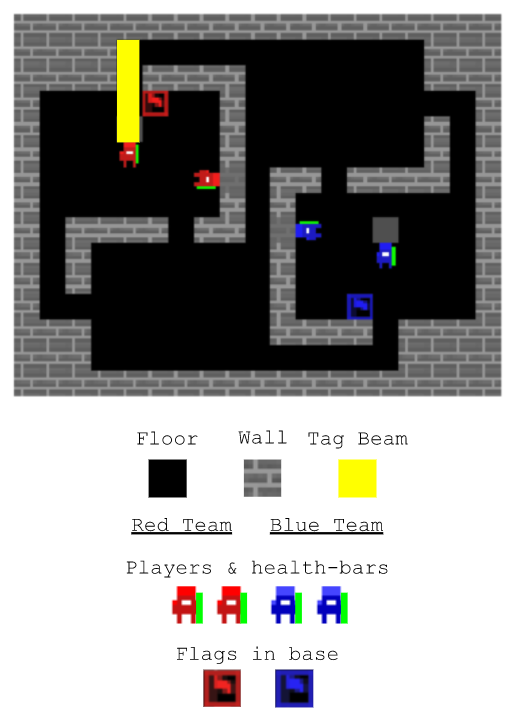}
        \caption{Capture the Flag}
        \label{fig:2/environments/ctf}
    \end{subfigure}
    \caption{We investigate the influence of environment and population diversity on agent performance across four distinct $n$-player Markov games: (a) HarvestPatch (a six-player mixed-motive game), (b) Traffic Navigation (an eight-player coordination game), (c) Overcooked (a two-player common-payoff game), and (d) Capture the Flag (a four-player team-competition game).}    
    \label{fig:2/environments}
\end{figure*}

Here we build from prior research and rigorously characterize the effects of environment and population diversity on multi-agent reinforcement learning. Specifically, we use procedural generation and population play to investigate performance and generalization in four distinct multi-agent environments drawn from prior studies: HarvestPatch, Traffic Navigation, Overcooked, and Capture the Flag.
These experiments make three contributions to multi-agent reinforcement learning research:
\begin{enumerate}
    \item 
    Agents trained with greater environment diversity exhibit stronger generalization to new levels. However, in some environments and with certain co-players, these improvements come at the expense of performance on an agent's training set.
    \item Expected action variation---a new, domain-agnostic metric introduced here---can be used to assess behavioral diversity in a population.
    \item Behavioral diversity tends to increase with population size, and in some (but not all) environments is associated with increases in performance and generalization.
\end{enumerate}

\section{Environments}

\subsection{Markov games and multi-agent reinforcement learning}
\label{app:markov_games}

This paper aims to explore the influence of diversity on agent behavior and generalization in $n$-player Markov games \cite{littman1994markov}. A partially observable Markov game $\mathcal{M}$ is played by $n$ players within a finite set of states $\mathcal{S}$. The game is parameterized by an observation function $O: \mathcal{S} \times \{1,\dots,n\} \rightarrow \mathbb{R}^d$, sets of available actions for each player $\mathcal{A}_1,\dots,\mathcal{A}_n$, and a stochastic transition function $\mathcal{T}: \mathcal{S} \times \mathcal{A}_1 \times \cdots \times \mathcal{A}_n \rightarrow \Delta(\mathcal{S})$, mapping from joint actions at each state to the set of discrete probability distributions over states. 

Each player $i$ independently experiences the game and receives its own observation $o_i = O(s,i)$. The observations of the $n$ players in the game can be represented jointly as $\vec{o} = (o_1,\dots, o_n)$. Following this notation, we can also refer to the vector of player actions $\vec{a} = (a_1,\dots,a_n) \in \mathcal{A}_1,\dots,\mathcal{A}_n$ for convenience. Each agent $i$ independently learns a behavior policy $\pi(a_i|o_i)$ based on its observation $o_i$ and its extrinsic reward $r_i(s,\vec{a})$. Agent $i$ learns a policy which maximizes a long-term $\gamma$-discounted payoff defined as:

\begin{equation}
    V_{\vec{\pi}_i}(s_0) = \mathbb{E} \left[ \sum \limits_{t=0}^{\infty} \gamma^t U_i(s_t, \vec{o}_t, \vec{a}_t) | \vec{a}_t \sim \vec{\pi}_t, s_{t+1} \sim \mathcal{T}(s_t, \vec{a}_t) \right]
    \label{eq:max_v}
\end{equation}

\noindent where $U_i(s_t, \vec{o}_t, \vec{a}_t)$ is the utility function for agent $i$. In the absence of reward sharing \cite{ibrahim2020reward} or instrinsic motivation \cite{hughes2018inequity,singh2005intrinsically}, the utility function maps directly to the extrinsic reward provided by the environment.

A key source of diversity in Markov games is the environment itself. To this end, we train agents on distributions of environment levels produced by procedural generators. Our investigation explores four distinct environments drawn from prior studies: HarvestPatch (a mixed-motive game), Traffic Navigation (a coordination game), Overcooked (a common-payoff game), and Capture the Flag (a competitive game). The following subsections provide an overview of the game rules for each of these games. All environments were implemented with the open-source engine DeepMind Lab2D \cite{beattie2020deepmind}. Full details on the environments and the procedural generation methods are available in the Appendix~\ref{app:environments}.

\subsection{HarvestPatch}
HarvestPatch \cite{mckee2020social} (Figure~\ref{fig:2/environments/hp}) is a mixed-motive game, played by $n = 6$ players in the experiments here (see also \cite{hughes2018inequity,jaques2019intrinsic,kramar2020should,perolat2017multi}).

Players inhabit a gridworld environment containing harvestable apples. Players can harvest apples by moving over them, receiving a small reward for each apple collected ($+1$ reward). Apples regrow after being harvested at a rate determined by the number of unharvested apples within the regrowth radius $r$. An apple cannot regrow if there are no apples within its radius. This property induces a social dilemma for the players. The group as a whole will perform better if its members are abstentious in their apple consumption, but in the short term individuals can always do better by harvesting greedily.

Levels are arranged with patches of apples scattered throughout the environment in varying densities. Every step, players can either stand still, move around the level, or fire a short tag-out beam. If another player is hit by the beam, they are removed from play for a number of steps. They also observe a partial egocentric window of the environment.

\subsection{Traffic Navigation}
Traffic Navigation \cite{lerer2019learning} (Figure~\ref{fig:2/environments/tn}) is a coordination game, played by $n = 8$ players. 

Players are placed at the edges of a gridworld environment and tasked with reaching specific goal locations within the environment. When a player reaches their goal, they receive a reward and a new goal location. If they collide with another player, they receive a negative reward. Consequently, each player's objective is to reach their goal locations as fast as possible while avoiding collisions with other players.

To make coordinated navigation more challenging, blocking walls are scattered throughout the environment, creating narrow paths which limit the number of players that can pass at a time. On each step of the game, players can either stand still or move around the level. Players observe both a small egocentric window of the environment and their relative offset to their current goal location.

\subsection{Overcooked}
Overcooked \cite{carroll2019utility} (Figure~\ref{fig:2/environments/oc}) is a common-payoff game, played in the experiments here by $n = 2$ players (see also \cite{charakorn2020investigating,knott2021evaluating,wang2020too}).

Players are placed in a kitchen-inspired gridworld environment and tasked with cooking as many dishes of tomato soup as possible. Cooking a dish is a sequential task: players must deposit three tomatoes into a cooking pot, let the tomatoes cook, remove the cooked soup with a dish, and then deliver the dish. Both players receive a reward upon the delivery of a plated dish.

Environment levels contain multiple cooking pots and stations. Players can coordinate their actions to maximize their shared reward. On each step, players can stand still, move around the level, or interact with the entity the object are facing (e.g., pick up tomato, place tomato onto counter, or deliver soup). Players observe a partial egocentric window of the level.

\subsection{Capture the Flag}
Capture the Flag (Figure~\ref{fig:2/environments/ctf}) is a competitive game. Jaderberg et al. \cite{jaderberg2019human} studied Capture the Flag using the Quake engine. Here, we implement a gridworld version of Capture the Flag played by $n = 4$ players.

Players are split into red and blue teams and compete to capture the opposing team's flag by strategically navigating, evading, and tagging members of the opposing team. The team that captures the greater number of flags by the end of the episode wins. %

Walls partition environment levels into rooms and corridors, generating strategic spaces for players to navigate and exploit to gain an advantage over the other team. On each step, players can stand still, move around the level, or fire a tag-out beam. If another player is hit by the tag-out beam three times, they are removed from play for a set number of steps. Each player observes a partial egocentric window oriented in the direction they are facing, as well as whether each of the two flags is held by its opposing team.

\section{Agents}

We use a distributed, asynchronous framework for training, deploying a set of ``arenas'' to train each population of $N$ reinforcement learning agents. Arenas run in parallel; each arena instantiates a copy of the environment, running one episode at a time. To begin an episode, an arena selects a population $i$ of size $N_i$ with an associated set of $L_i$ training levels. The arena samples one level $l$ from the population's training set and $n$ agents from the population (with replacement). The episode lasts $T$ steps, with the resulting trajectories used by the sampled agents to update their weights. Agents are trained until episodic rewards converge. After training ends, we run various evaluation experiments with agents sampled after the convergence point.

For the learning algorithm of our agents, we use V-MPO \cite{song2019v}, an on-policy variant of Maximum a Posteriori Policy Optimization (MPO). In later experiments, we additionally endow these agents with the Social Value Orientation (SVO) component, encoding an intrinsic motivation to maintain certain group distributions of reward \cite{mckee2020social}. These augmented agents act as a baseline for our behavioral diversity analysis, following suggestions from prior research that imposing variation in important hyperparameters can lead to greater population diversity.
More details on the algorithm (including hyperparameters) are available in Appendix~\ref{app:agents}.

\section{Methods} \label{sec:methods}

\subsection{Investigating Environment Diversity} \label{sec:4/env_diversity_methods}
To assess how environment diversity (i.e., the number of unique levels encountered during training) affects an agent's ability to generalize, we follow single-agent work on quantifying agent generalization in procedurally generated environments \cite{cobbe2019leveraging,cobbe2019quantifying,zhang2018study}.

\begin{figure*}[!t]
    \centering
    \begin{subfigure}{0.45\linewidth}
        \centering
        \includegraphics[height=10em]{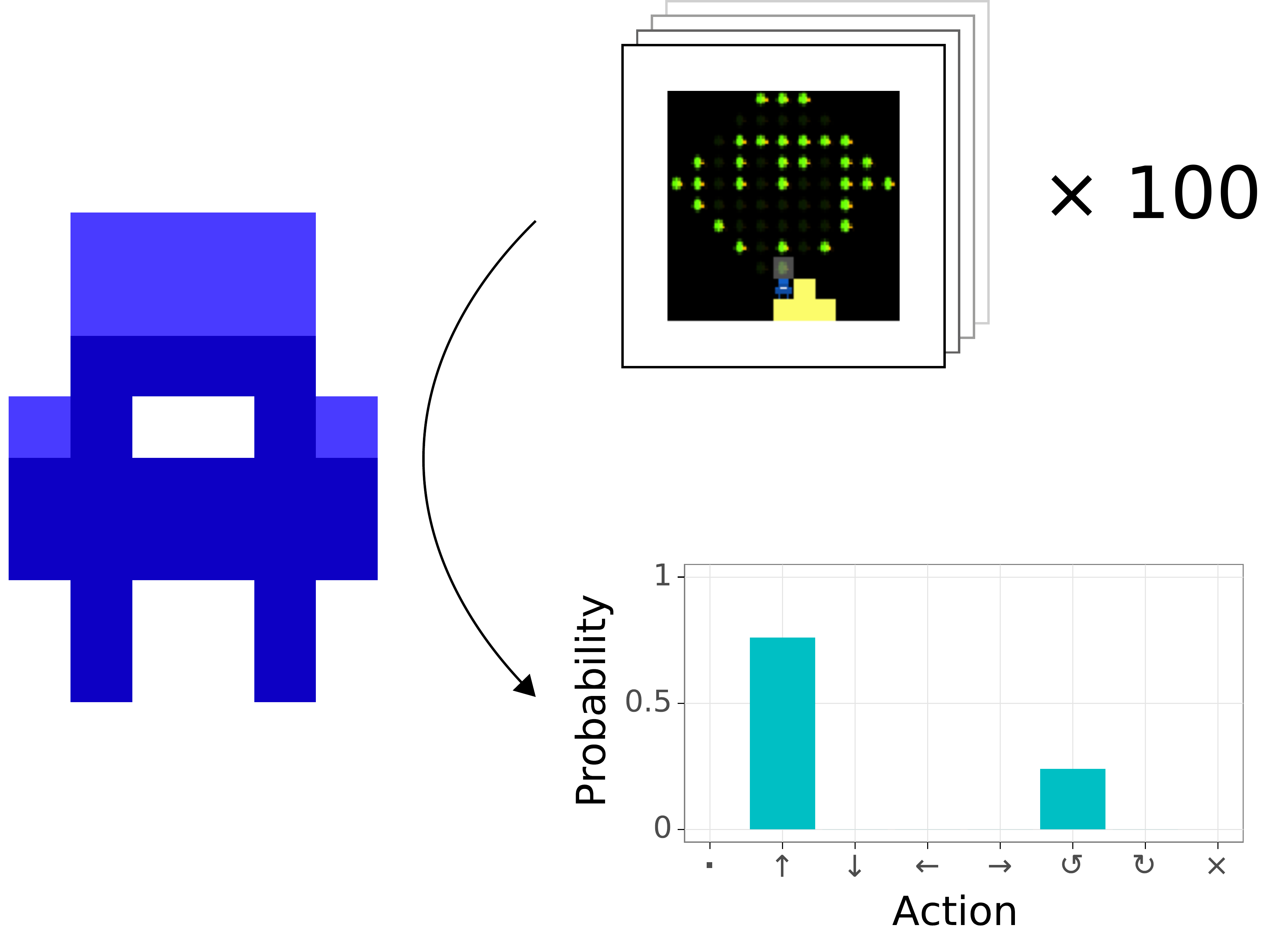}
        \caption{To assess expected action variation, each agent in a population is prompted multiple times with a number of agent states. The probabilistic action outputs for each state are recorded. Here, an agent (Agent 1) is prompted 100 times with a state (State 1) from HarvestPatch. The process will be repeated for both other states and other agents in the population.}
        \label{fig:eav_abstract}
    \end{subfigure}
    \hfill
    \begin{subfigure}{0.45\linewidth}
        \centering
        \includegraphics[height=10em]{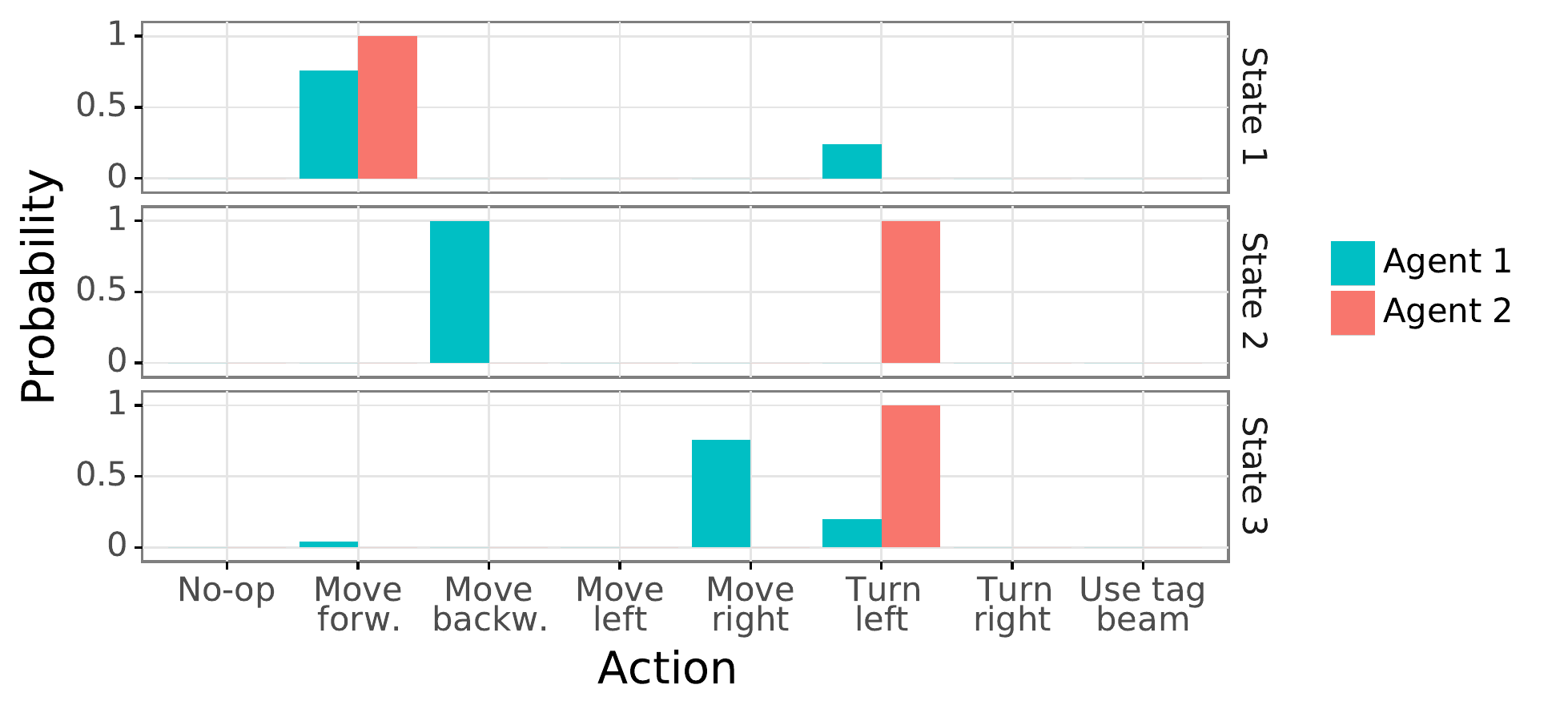}
        \caption{The action outputs are then compared for each pair of agents in the population. Here we see an example set of action outputs from Agent 1 and another agent over three states. For a population comprising these two agents, the computed expected action variation is $0.68$.\newline\newline}
        \label{fig:eav_example}
    \end{subfigure}
    \caption{For each population, we calculate \textit{expected action variation} (EAV), a new measure of behavioral diversity. The exact procedure for calculating this measure is detailed in the Appendix~\ref{app:eav}.}
    \label{fig:eav}
\end{figure*}

Specifically, we train multiple populations of $N = 1$ agents with different sets of training levels. We procedurally generate training levels in sets of size $L \in \{1, \expnumber{1}{1}, \expnumber{1}{2}, \expnumber{1}{3}, \expnumber{1}{4}\}$, where each training set is a subset of any larger sets. We also procedurally generate a separate test set containing 100 held-out levels. These held-out levels are not played by agents during training. For each training set of size $L$, we launch ten independent training runs and train each population until their rewards converge.

\paragraph{Generalization Gap} 
Following prior work, we compare the performance of populations on the levels from their training set with their performance on the 100 held-out test levels. We focus on the size of the \textit{generalization gap}, defined as the absolute difference between the population's performance on the test-set levels and training-set levels.

\paragraph{Cross-Play Evaluation}
We also assess population performance through \textit{cross-play evaluation}---that is, by evaluating agents in groups formed from two different training populations. We evaluate populations in cross-play both on the level(s) in their training set and on the held-out test levels. Specifically, for every pair of populations A and B, the training-level evaluation places agents sampled from population A (e.g., trained on $L = 1$ level) with agents sampled from population B (e.g., trained on $L = \expnumber{1}{1}$ levels) in a level from the intersection of the populations' training sets. The held-out evaluation similarly samples agents from populations A and B, but uses a level not found in either of the populations' training sets.

As each environment requires a different number of players, we group agents from populations A and B as shown in Table~\ref{tab:411/cross_play_groupings}. For HarvestPatch, Traffic Navigation, and Overcooked, we report the individual rewards achieved by the agents sampled from population A. For Capture the Flag, we analogously report the win rate for the agents from population A.

\begin{table}[h]
    \centering
    \begin{tabular}[h]{c c c}
        \toprule
        & \multicolumn{2}{c}{Population:} \\
        Environment & \makebox[15pt]{A} & \makebox[15pt]{B} \\
        \midrule
        HarvestPatch & 1 & 5 \\
        Traffic Navigation & 1 & 7 \\
        Overcooked & 1 & 1 \\
        Capture the Flag & 2 & 2 \\
        \bottomrule
    \end{tabular}
    \caption{Number of agents sampled from populations A and B for cross-play evaluation in each environment.}
    \label{tab:411/cross_play_groupings}    
\end{table}

\subsection{Investigating Population Diversity} \label{sec:pop_div_methods}
We run a second set of experiments to investigate how population diversity affects generalization. Measuring and optimizing for agent diversity are established challenges in reinforcement learning research. In single-agent domains, diversity is often estimated over behavioral trajectories or state-action distributions \cite{dai2022diversity,haarnoja2017reinforcement}. Prior multi-agent projects have largely focused on two-player zero-sum games \cite{balduzzi2019open,nieves2021modelling}. In these environments, the diversity of a set of agent strategies can be directly estimated from the empirical payoff matrix, rather than behavioral trajectories.

Given the varied environments used in our experiments (and particularly the cooperative and competitive natures of their payoff structures), we draw inspiration from the former approach, focusing on heterogeneity in agent behavior. This paper uses the term ``population diversity'' to refer to variation in the set of potential co-player policies that a new individual joining a population might face \cite{mckee2020social}. High policy diversity maximizes coverage over the set of all possible behaviors in an environment (including potentially suboptimal or useless behaviors) \cite{eysenbach2018diversity}, while low policy diversity consistently maps a given state to the same behavior, regardless of the agents involved.

In multi-agent research, the increasing prevalence of population play and population-based training make population size a commonly tuned feature of experimental design. Prior studies suggest that larger population sizes can increase policy diversity \cite{czarnecki2020real,sanjaya2021measuring}. We directly examine how varying population size affects diversity by training agents in populations of size $N \in \{1, 2, 4, 8\}$ for each environment. For Capture the Flag experiments, we additionally train populations with $N = 16$.

\paragraph{Expected Action Variation}
Researchers should ideally be able to estimate population diversity in a task-agnostic manner, but in practice diversity is often evaluated using specific knowledge about the environment (e.g., unit composition in StarCraft II \cite{vinyals2019grandmaster}). %

To address this challenge, we introduce a new method of measuring behavioral diversity in a population that we call \textit{expected action variation} (EAV; Algorithm \ref{alg:behavioral_diversity} in Appendix~\ref{app:eav}). Intuitively, this metric captures the probability that two agents sampled at random from the population will select different actions when provided the same state (Figure~\ref{fig:eav}). At a high level, we compute expected action variation by simulating a number of rollouts for each policy in the population and calculating the total variational distance between the resulting action distributions. One of the key advantages of this metric is that it can be na\"{i}vely applied to a set of stochastic policies generated in any environment.

Expected action variation ranges in value from $0$ to $1$: a value of $0$ indicates that the population is behaviorally homogeneous (all agents select the same action for any given agent state), whereas a value of $1$ indicates that the population is maximally behaviorally diverse (all agents select different actions for any given agent state). An expected action variation of $0.5$ indicates that if two agents are sampled at random from the population and provided a representative state, they are just as likely to select the same action as they are to select different actions.

This procedure is designed to help compare diversity across populations and to reason about the way a focal agent's experience of the game might change as a function of which co-players are encountered. Expected action variation is affected by stochasticity in policies, since such stochasticity can affect the state transitions that a focal agent experiences. Expected action variation is not intended to test whether the behavioral diversity of a population is significantly different from zero (or from $1$), since such a difference could emerge for trivial reasons.

We leverage expected action variation to assess the effect of population size on behavioral diversity. We also include additional baselines to help explore the dynamics of co-player diversity. Specifically, we train several $N = 4$ populations parameterized with an intrinsic motivation module \cite{singh2005intrinsically} on $L = \expnumber{1}{3}$ levels. In particular, we use the SVO component to motivate agents in these populations to maintain target distributions of reward \cite{mckee2020social}. Each population is parameterized with either a homogeneous or heterogeneous distribution of SVO targets (see Appendix~\ref{app:svo}).

\begin{figure*}[t]
    \begin{subfigure}{0.24\textwidth}
        \centering
        \includegraphics[width=\linewidth]{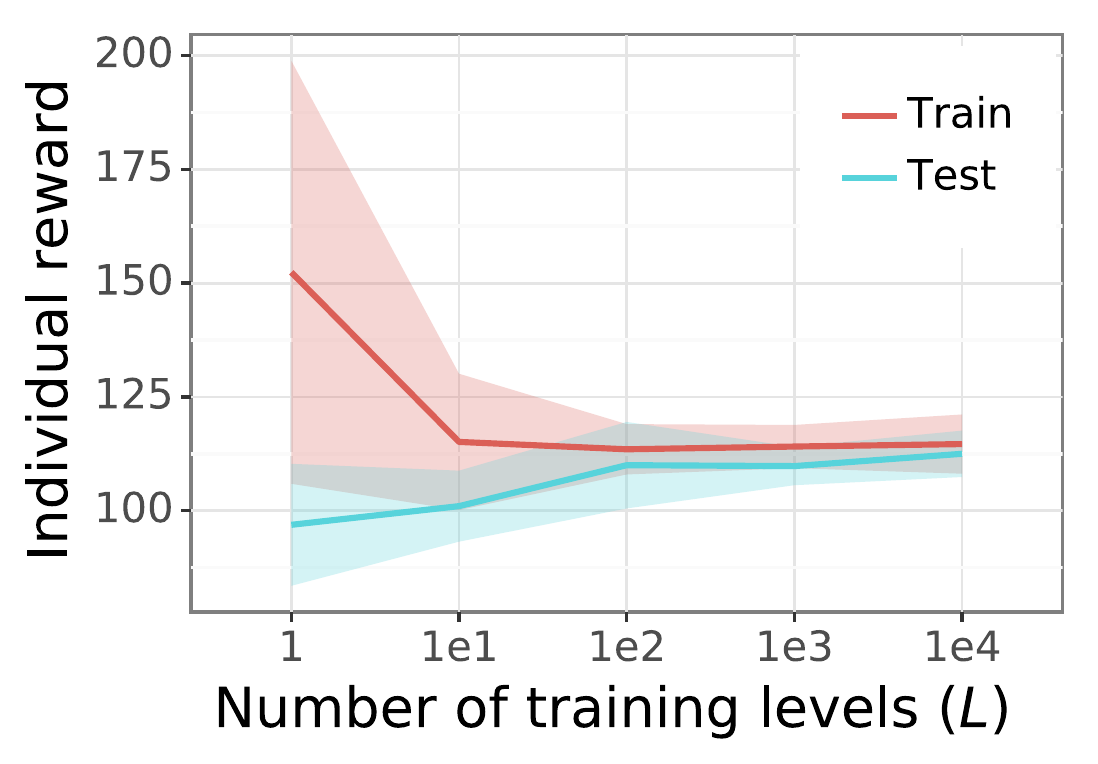}
    \end{subfigure}
    \hfill
    \begin{subfigure}{0.24\textwidth}
        \centering
        \includegraphics[width=\linewidth]{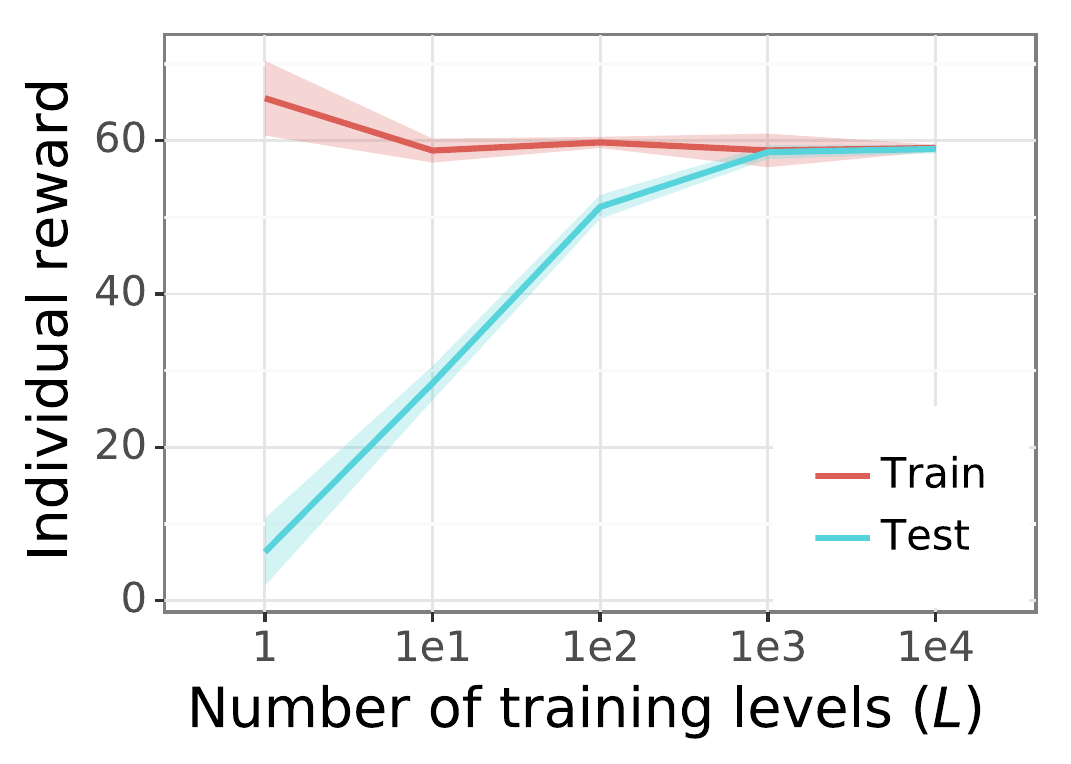}
    \end{subfigure}
    \hfill
    \begin{subfigure}{0.24\textwidth}
        \centering
        \includegraphics[width=\linewidth]{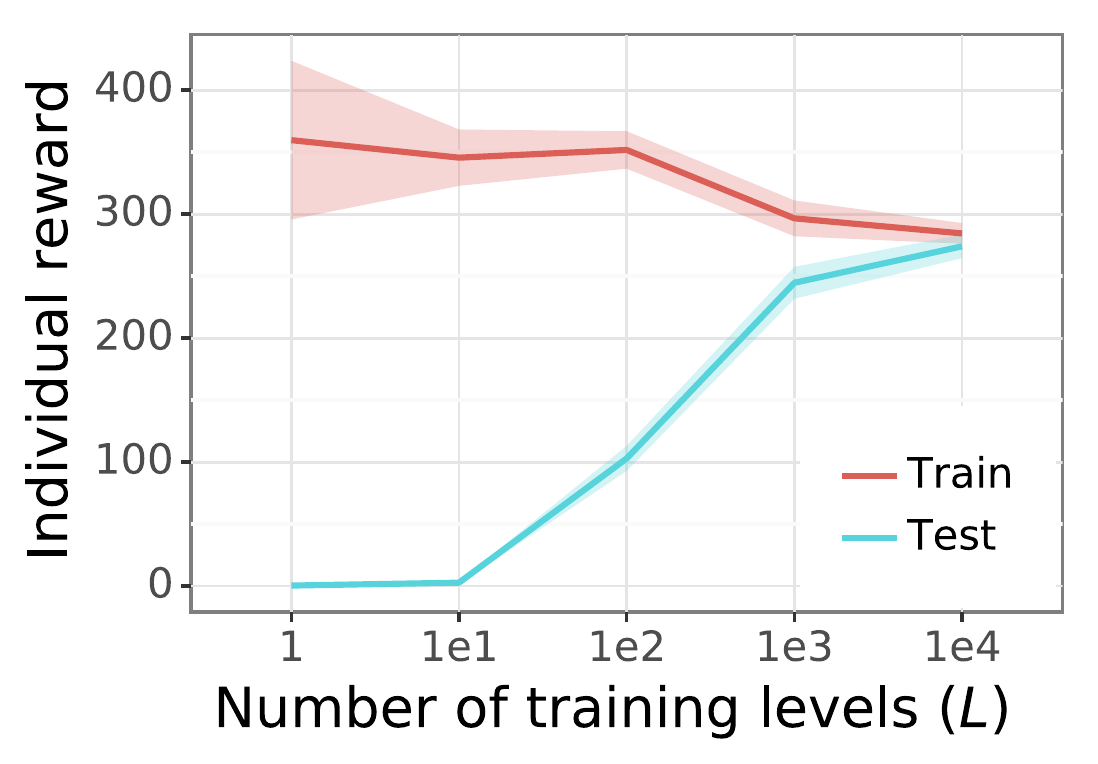}
    \end{subfigure}
    \hfill
    \begin{subfigure}{0.24\textwidth}
        \centering
        \includegraphics[width=\linewidth]{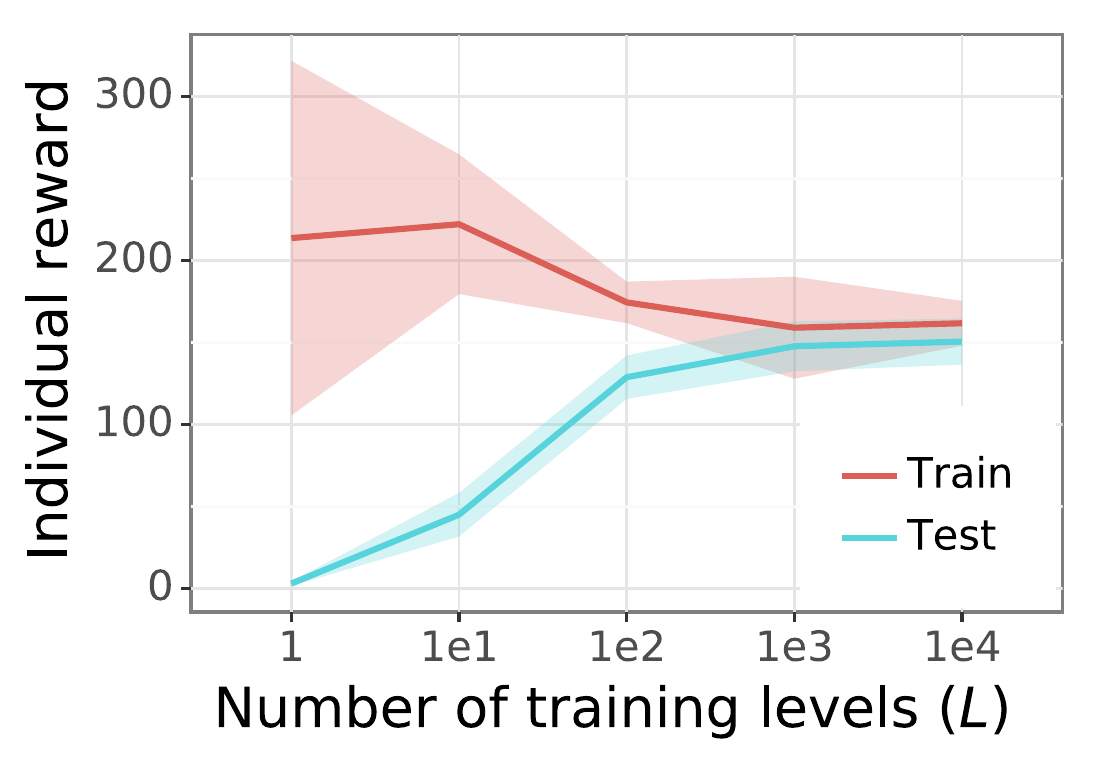}
    \end{subfigure}
    \vspace{1em}
    \newline
    \begin{subfigure}{0.24\textwidth}
        \centering
        \includegraphics[width=\linewidth]{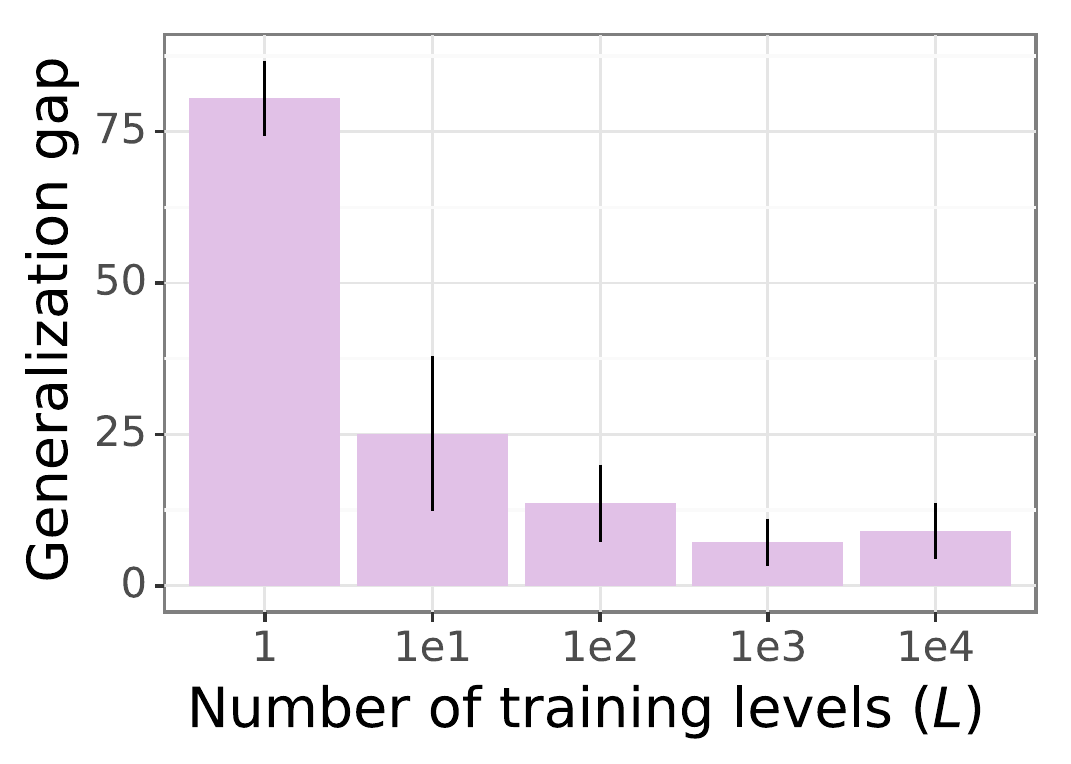}
        \caption{HarvestPatch.}
    \end{subfigure}
    \hfill
    \begin{subfigure}{0.24\textwidth}
        \centering
        \includegraphics[width=\linewidth]{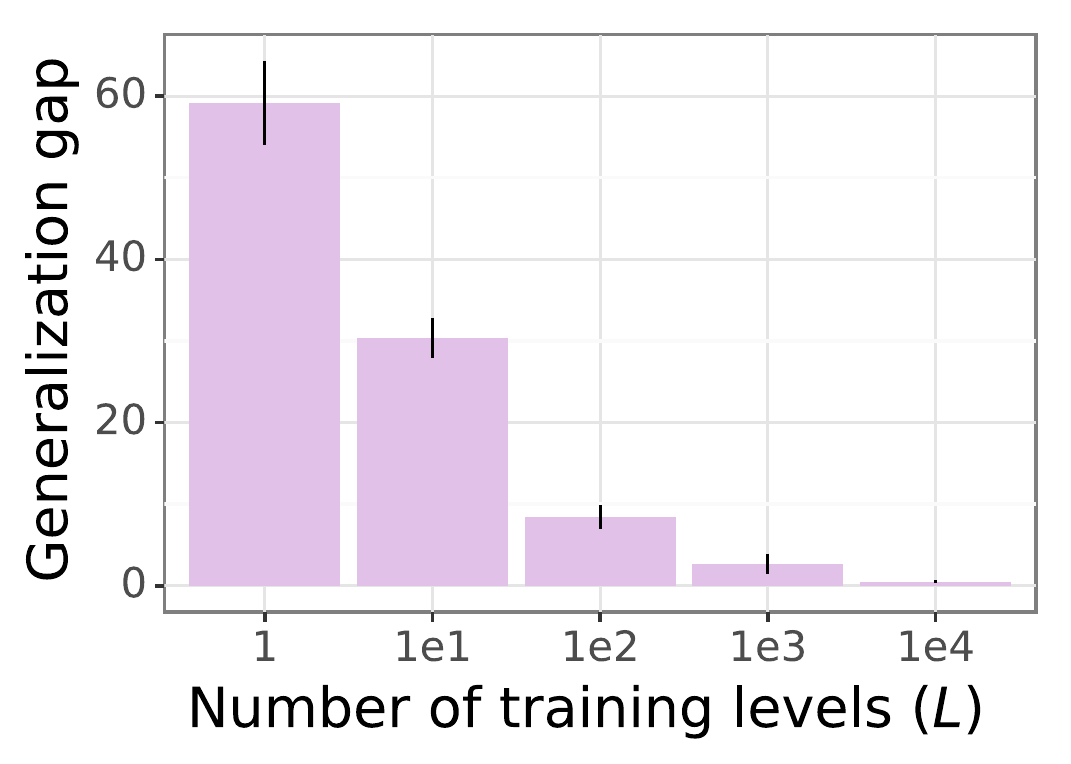}
        \caption{Traffic Navigation.}
    \end{subfigure}
    \hfill
    \begin{subfigure}{0.24\textwidth}
        \centering
        \includegraphics[width=\linewidth]{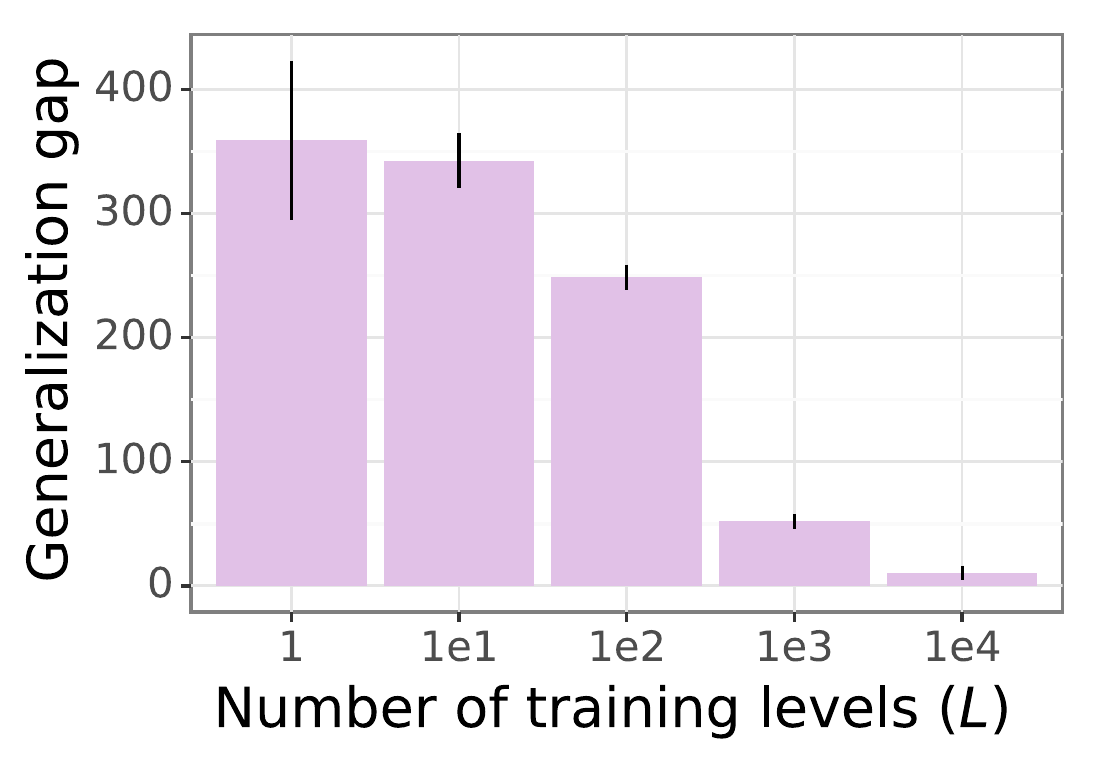}
        \caption{Overcooked.}
    \end{subfigure}    
    \hfill
    \begin{subfigure}{0.24\textwidth}
        \centering
        \includegraphics[width=\linewidth]{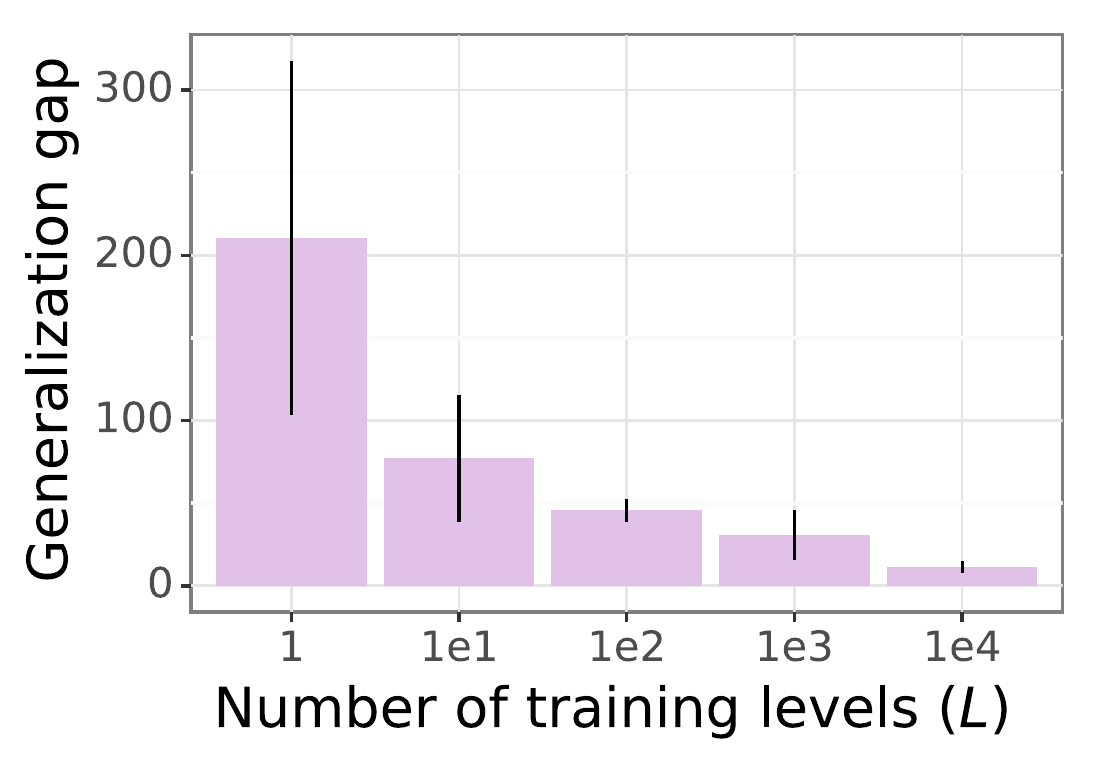}
        \caption{Capture the Flag.}
    \end{subfigure}
    \caption{\textbf{Top row:} Effect of training set size $L$ on group performance on train vs. test levels for each environment. Error bands reflect 95\% confidence intervals calculated over 10 independent runs (nine for Capture the Flag). \textbf{Bottom row:} Effect of training set size $L$ on the generalization gap between training and test levels for each environment. Error bars correspond to 95\% confidence intervals calculated over 10 independent runs (nine for Capture the Flag). \textbf{Result:} As environment diversity increases, test performance tends to improve. Training performance and the generalization gap experience concomitant decreases.}
    \label{fig:511/env/generalization_gap}
\end{figure*}

\paragraph{Cross-Play Evaluation}
We employ a cross-play evaluation procedure to measure the performance resulting from varying population sizes, following Section~\ref{sec:4/env_diversity_methods}. Specifically, we group agents sampled from populations A and B and then evaluate group performance on a level from the intersection of the populations' training sets. We use the same grouping and reporting procedure as before.

\subsection{Quantifying Performance}
For the majority of our environments, we quantify and analyze the individual rewards earned by the agents. In Capture the Flag, we evaluate agents in team competition. Consequently, we record the result of each match from which we calculate win rates and skill ratings. To estimate each population's skill, we use the Elo rating system \cite{elo1978rating}, an evaluation metric commonly used in games such as chess (see Appendix~\ref{app:elo} for details). 

\subsection{Statistical Analysis}
In our experiments, we launch multiple independent training runs for each value of $L$ and each value of $N$ being investigated. Critically, we match the training sets of these independent runs across values of $L$ and $N$. For example, the first run of the $N = 1$ HarvestPatch experiment trains on the exact same training set as the first runs of the $N \in \{2, 4, 8\}$ experiments. Similarly, the second runs for each of the $N = 1$ to $N = 8$ experiments use the same training set, and so on. This allows us to avoid confounding the effects of $N$ with those of $L$ and vice versa.

For our statistical analyses, we primarily leverage the Analysis of Variance (ANOVA) method \cite{fisher1928statistical}. The ANOVA allows us to test whether changing the value of an independent variable (e.g., environment diversity) significantly affects the value of a specified dependent variable (e.g., individual reward). Each ANOVA is summarized with an $F$-statistic and a $p$-value. In cases where we repeat ANOVAs for each environment, we apply a Holm--Bonferroni correction to control the probability of false positives \cite{holm1979simple}. %

\section{Results}

\subsection{Environment Diversity}
To begin, we assess how environment diversity (i.e., the number of unique levels used for training) affects generalization. Agents are trained on $L \in \{1, \expnumber{1}{1}, \expnumber{1}{2}, \expnumber{1}{3}, \expnumber{1}{4}\}$ levels in populations of size $N = 1$. %

\begin{figure*}[t]
    \begin{subfigure}{0.23\textwidth}
        \centering
        \includegraphics[width=\linewidth]{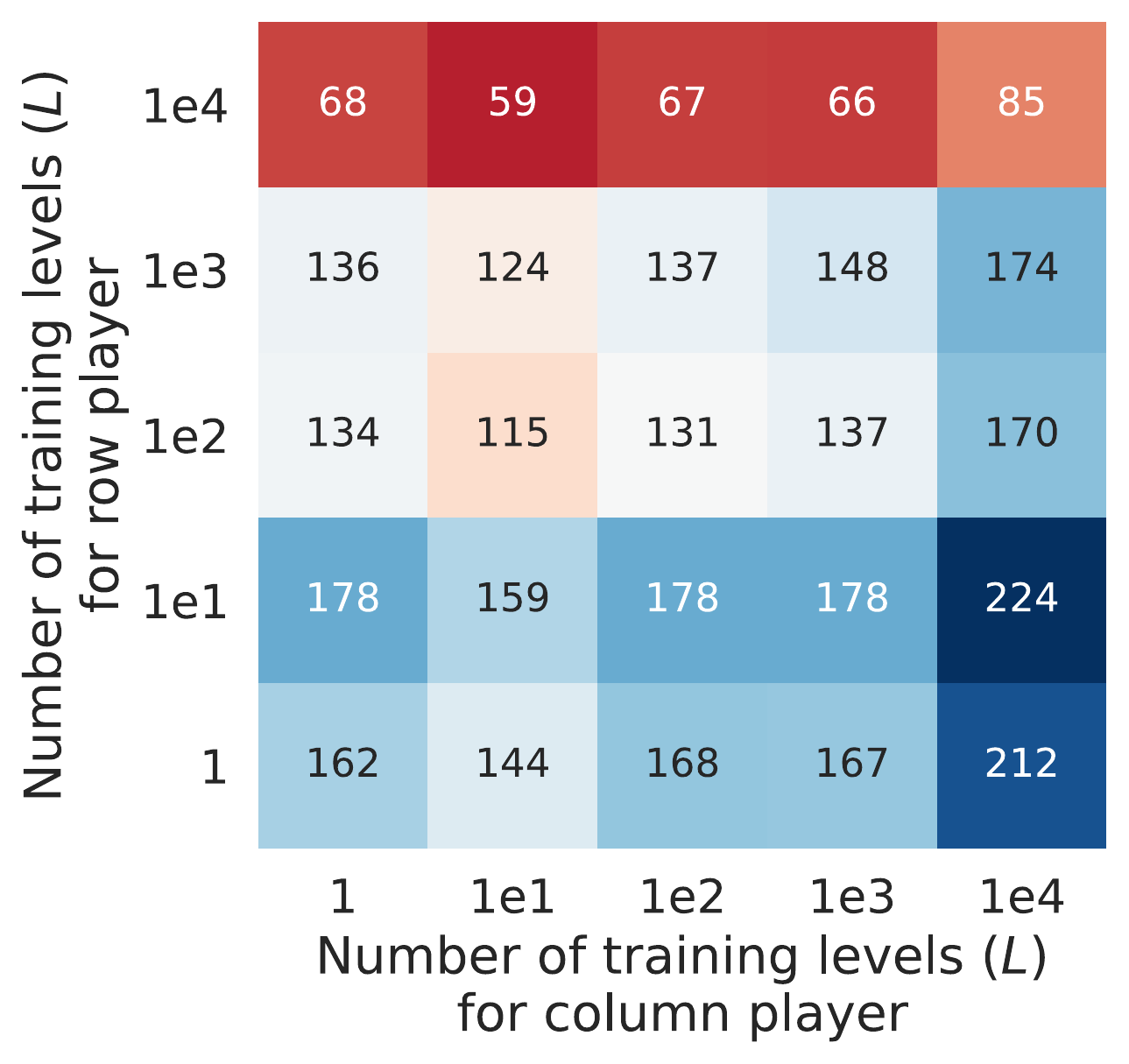}
    \end{subfigure}
    \hfill
    \begin{subfigure}{0.23\textwidth}
        \centering
        \includegraphics[width=\linewidth]{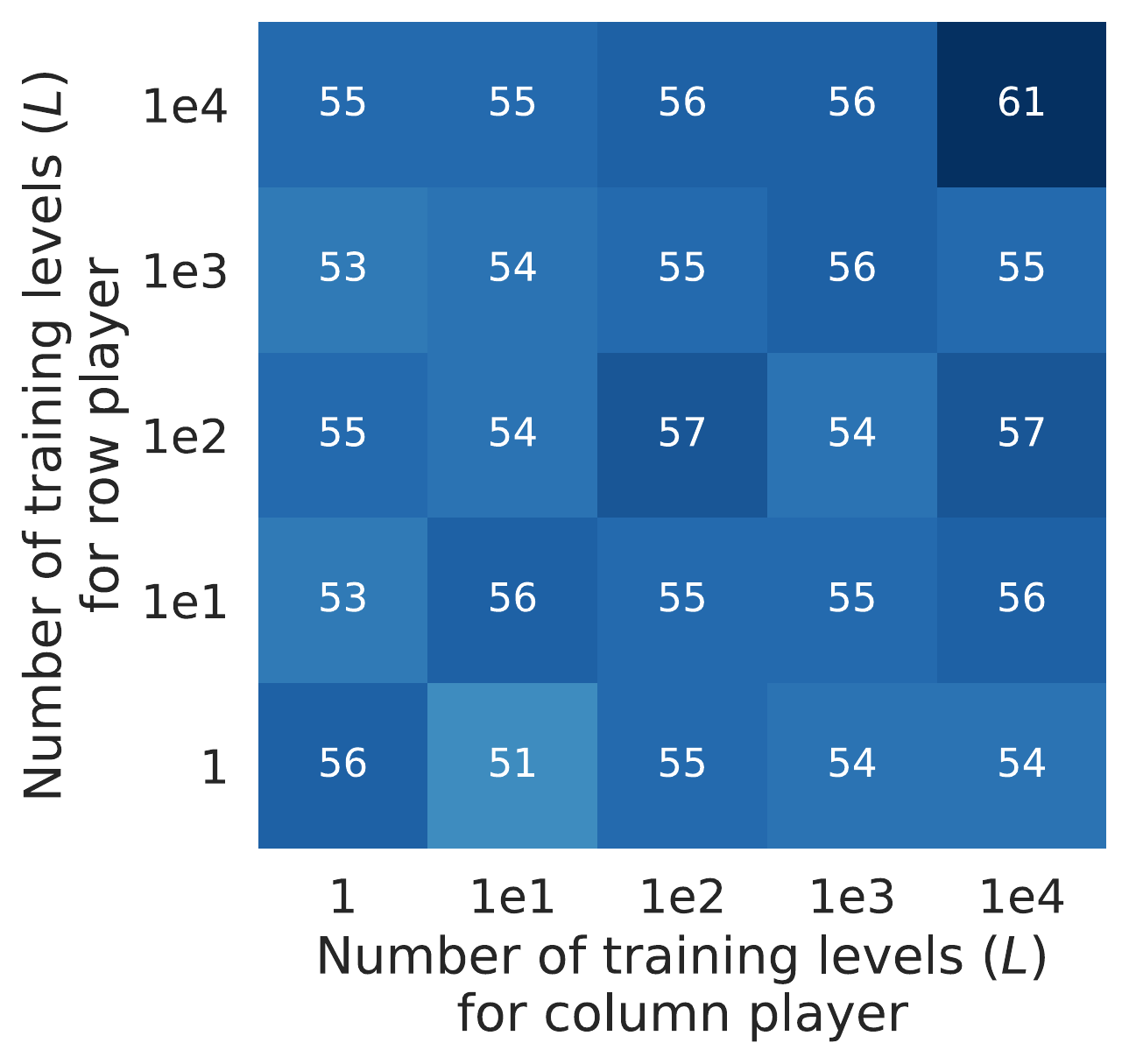}
    \end{subfigure}
    \hfill
    \begin{subfigure}{0.23\textwidth}
        \centering
        \includegraphics[width=\linewidth]{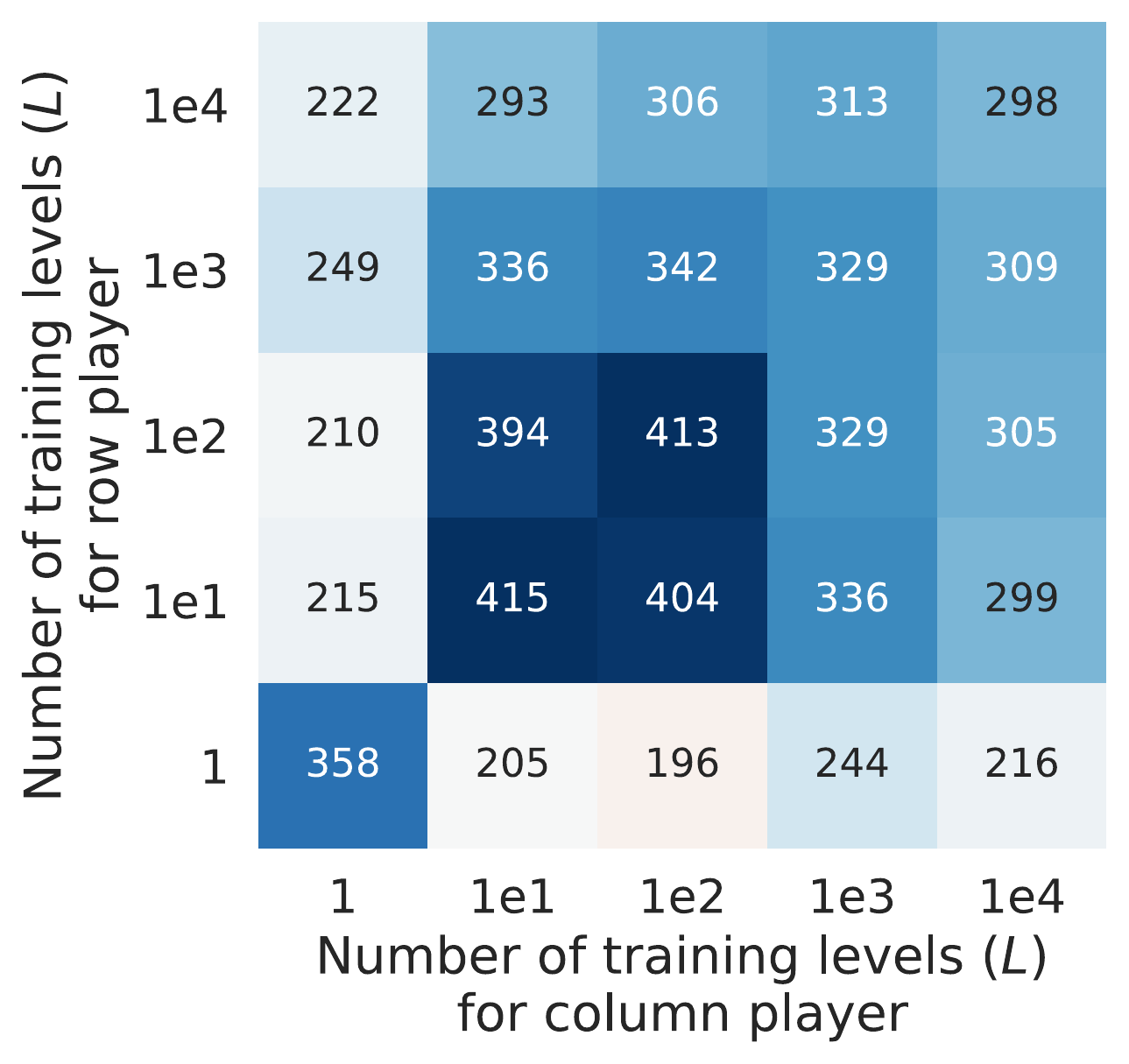}
    \end{subfigure}    
    \hfill
    \begin{subfigure}{0.23\textwidth}
        \centering
        \includegraphics[width=\linewidth]{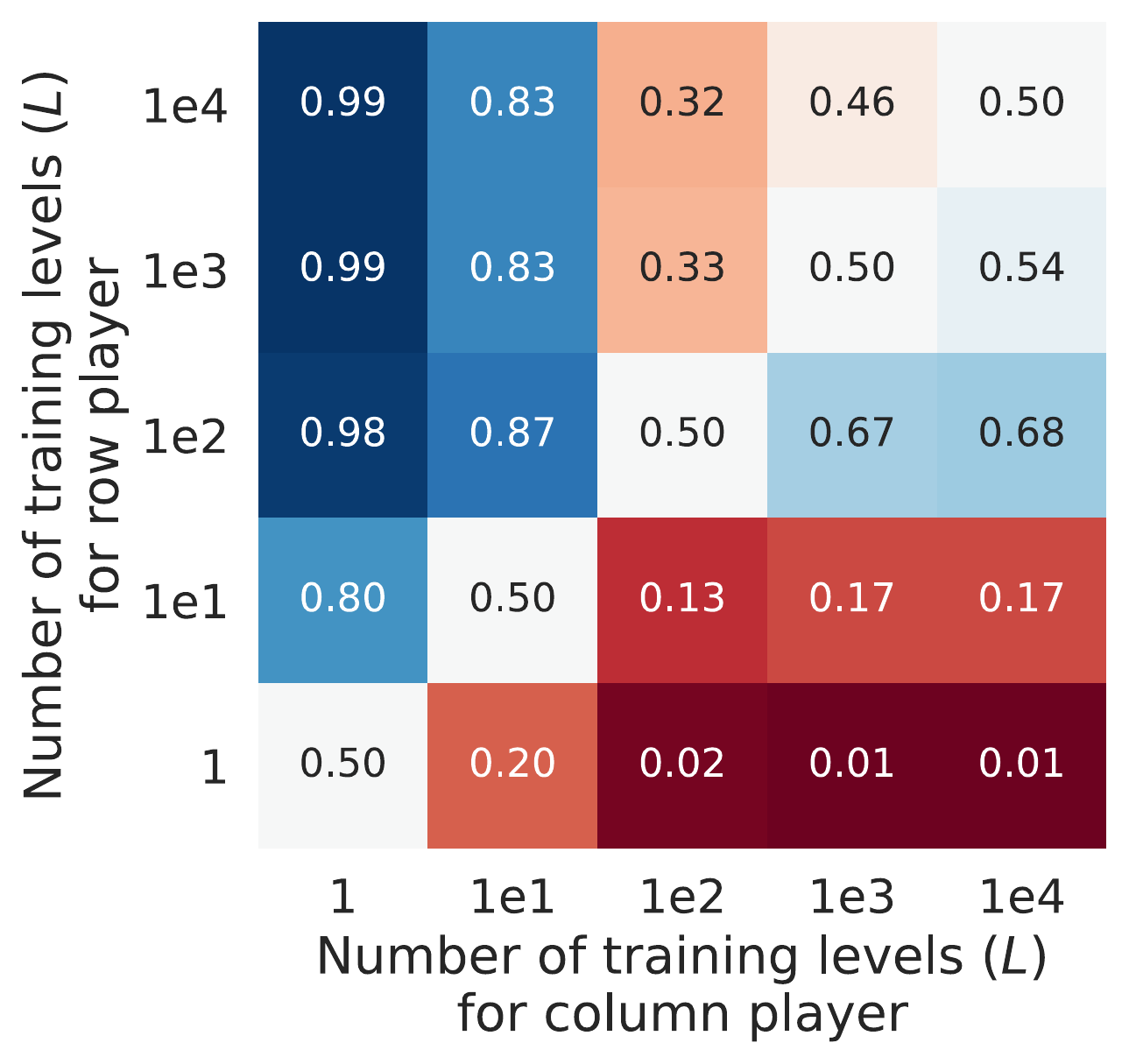}
    \end{subfigure}
    \vspace{1em}
    \newline
    \begin{subfigure}{0.23\textwidth}
        \centering
        \includegraphics[width=\linewidth]{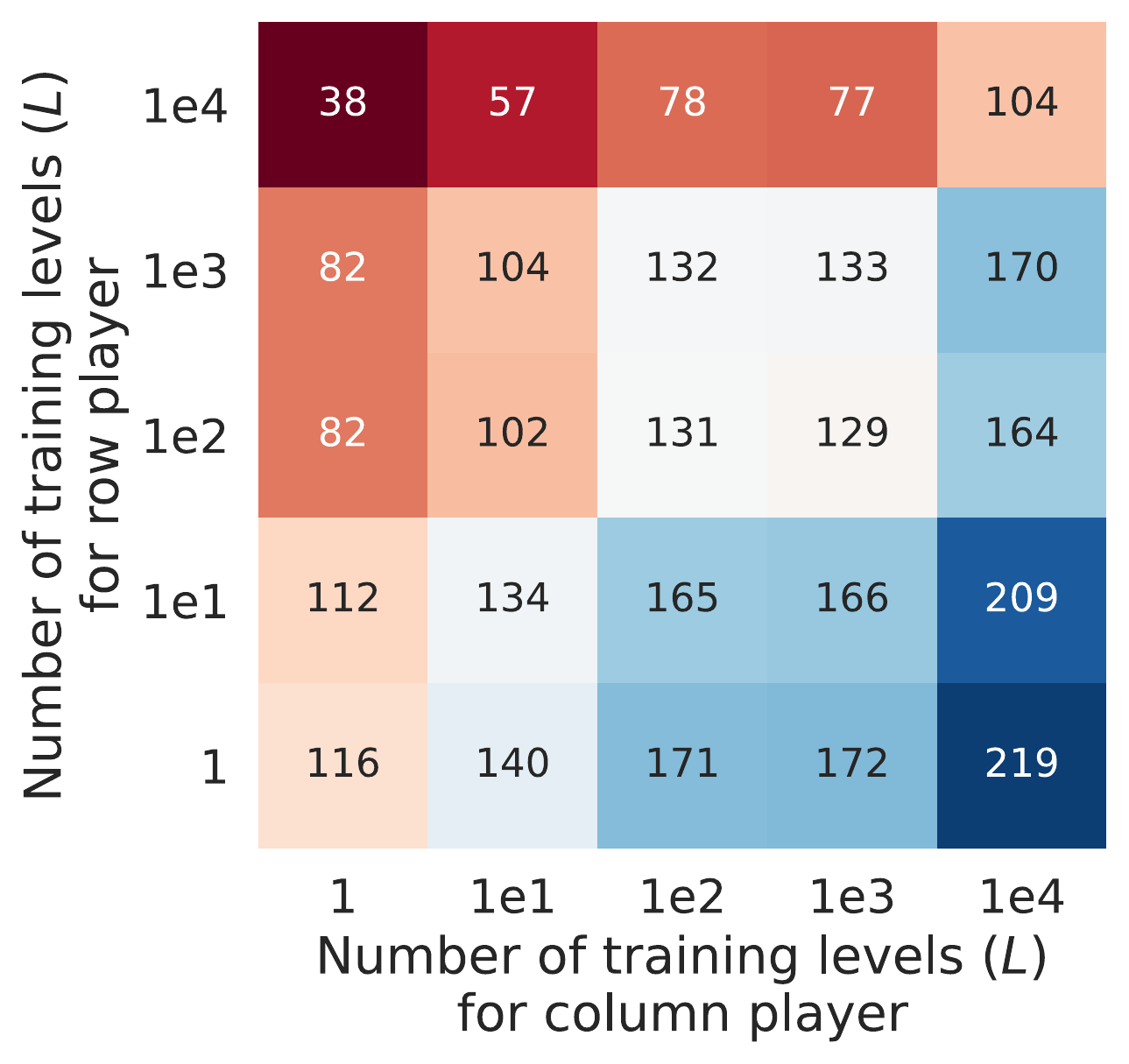}
        \caption{HarvestPatch: Reward of one row player when grouped with five column players.}
    \end{subfigure}
    \hfill
    \begin{subfigure}{0.23\textwidth}
        \centering
        \includegraphics[width=\linewidth]{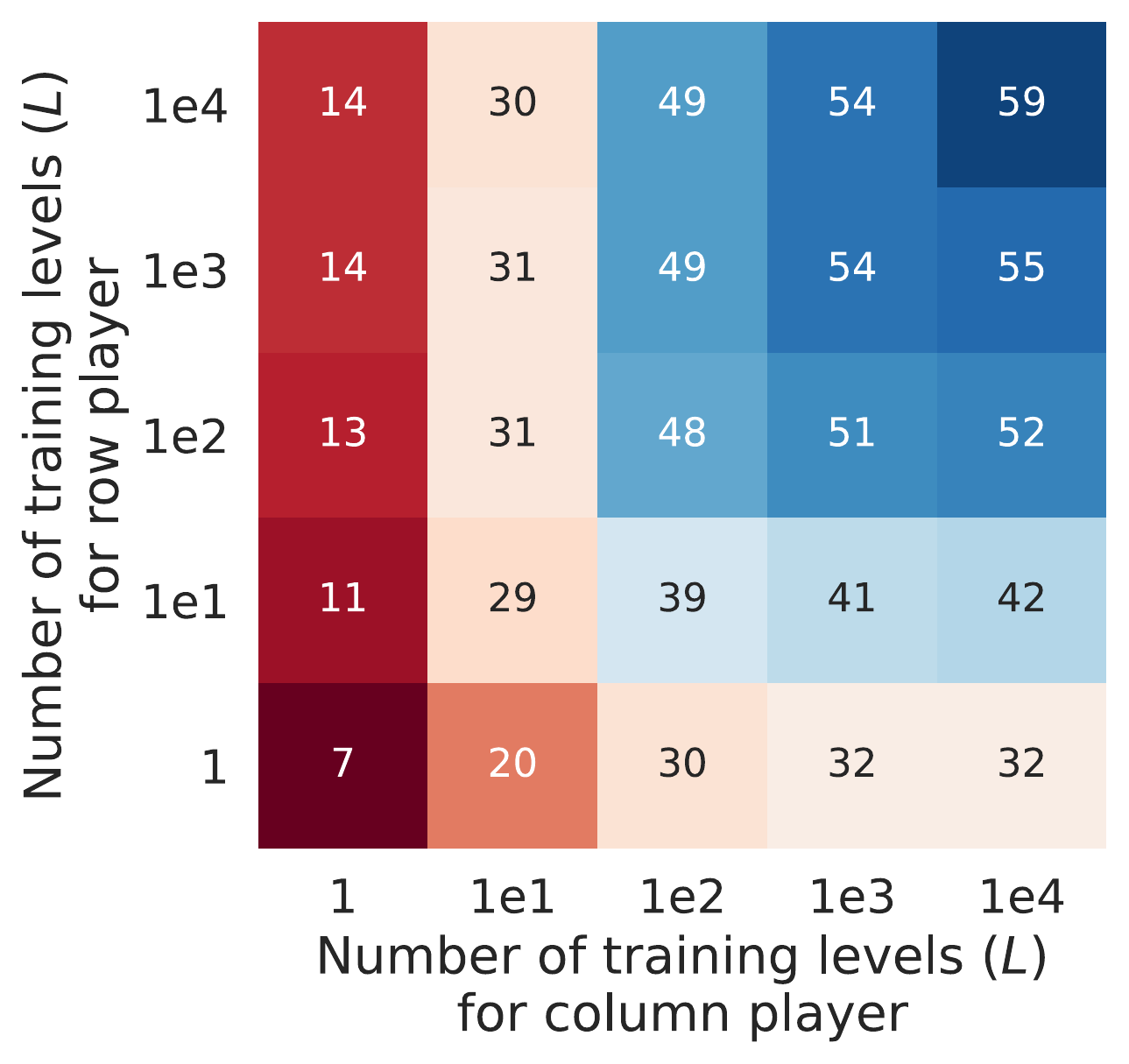}
        \caption{Traffic Navigation: Reward of one row player when grouped with seven column players.}
    \end{subfigure}
    \hfill
    \begin{subfigure}{0.23\textwidth}
        \centering
        \includegraphics[width=\linewidth]{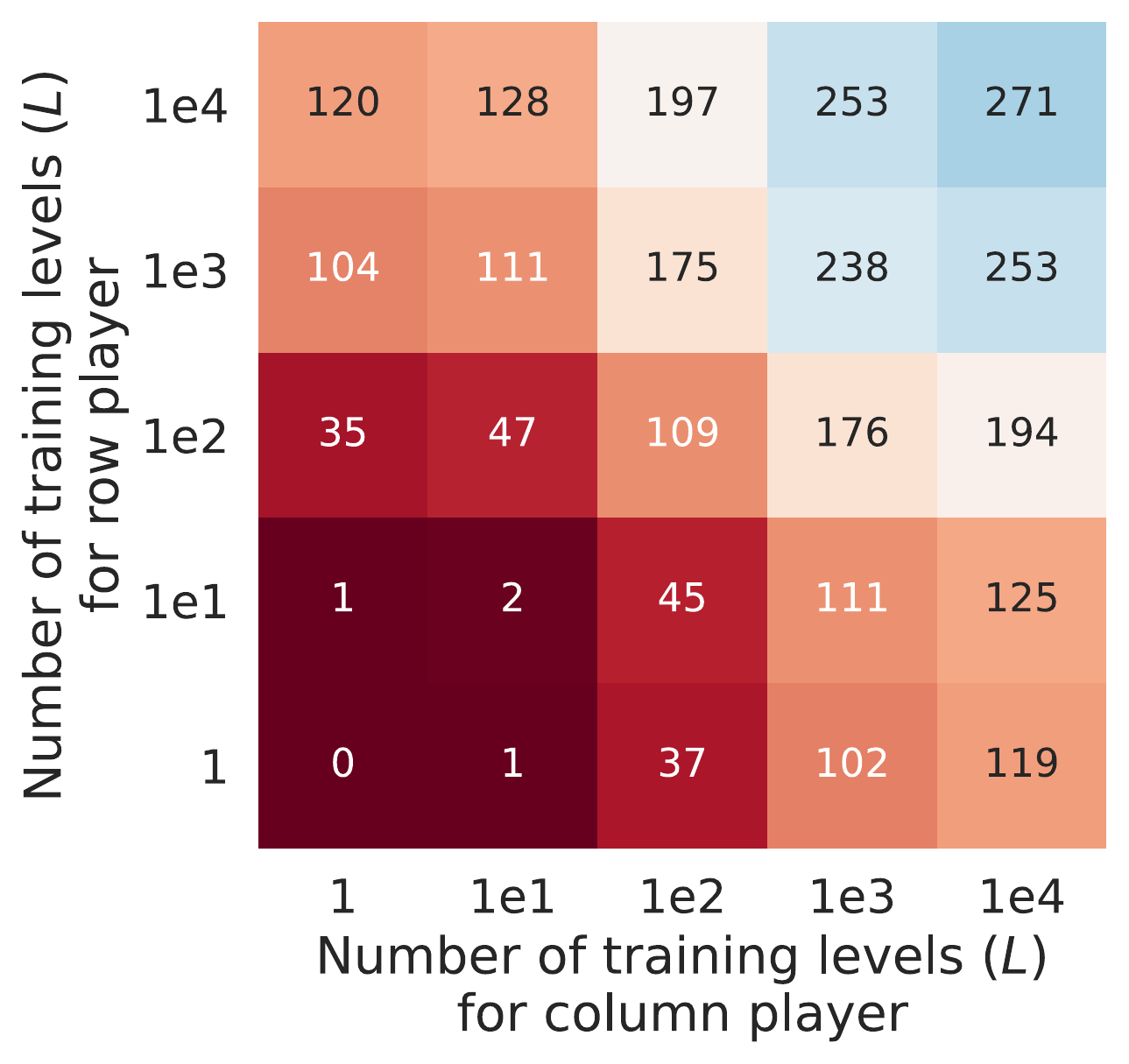}
        \caption{Overcooked: Reward of one row player when paired with one column player.}
    \end{subfigure}    
    \hfill
    \begin{subfigure}{0.23\textwidth}
        \centering
        \includegraphics[width=\linewidth]{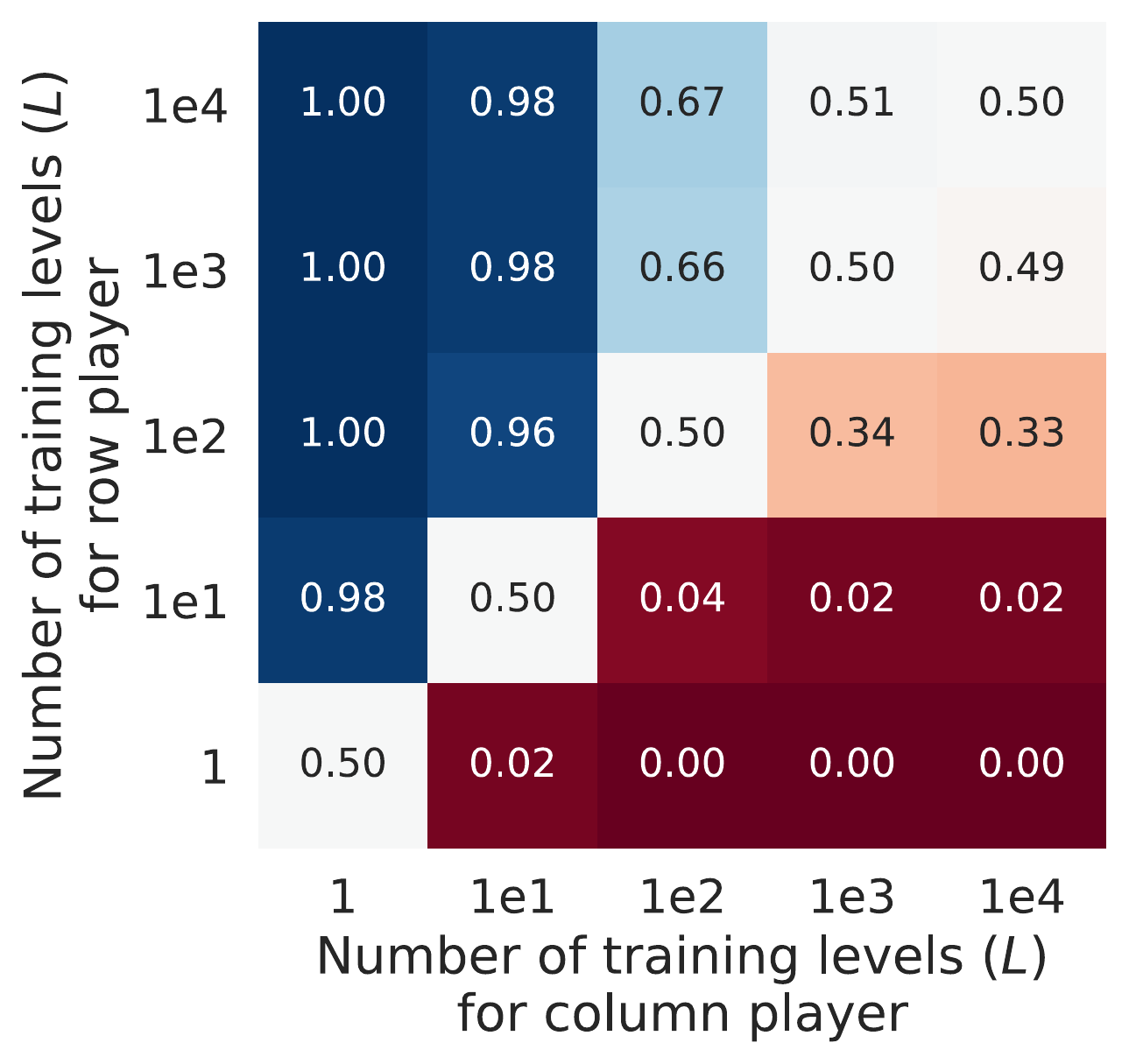}
        \caption{Capture the Flag: Win rate of two row players versus two column players.}
    \end{subfigure}
    \caption{\textbf{Top row:} Cross-play evaluation of agent performance for each environment, using levels drawn from their training set. \textbf{Bottom row:} Cross-play evaluation of agent performance for each environment, using held-out test levels. \textbf{Result:} Environment diversity exerts strong effects on agent performance, though the exact pattern varies substantially across environments.}
    \label{fig:512/env/cross_play}
\end{figure*}

\subsubsection{Generalization Gap Analysis}

As shown in Figure~\ref{fig:511/env/generalization_gap}, for all environments generalization improves as the number of levels used for training increases. Performance on the test set increases as $L$ increases, while performance on the training set tends to decrease with greater values of $L$ (Figure~\ref{fig:511/env/generalization_gap}, top row).

Performance on the training set experiences a minor decrease from low to high values of $L$. In contrast, the variance in training-set performance declines considerably as environment diversity increases. The variance in training-set performance is notably large for HarvestPatch and Capture the Flag when $L = 1$. This variability likely results from the wide distribution of possible rewards in the generated levels (e.g., due to varying apple density in HarvestPatch or map size in Capture the Flag). For Capture the Flag, the observed variance may also stem from the inherent difficulty of learning navigation behaviors on a singular large level where the rewards are sparse (i.e., without the availability of a natural curriculum).

To avoid ecological fallacy \cite{freedman1999ecological}, we directly quantify and analyze the generalization gap with the procedure outlined in Section~\ref{sec:4/env_diversity_methods}. As environment diversity increases, the generalization gap between the training and test sets decreases substantially (Figure~\ref{fig:511/env/generalization_gap}, bottom row). The trend is sizeable, materializing even in the shift from $L = 1$ to $L = \expnumber{1}{1}$. Across the environments considered here, the generalization gap approaches zero around values of $L=\expnumber{1}{3}$. A set of ANOVAs confirm that $L$ has a statistically significant effect on generalization in HarvestPatch, $F(4,45) = 4.8$, $p = 2.5 \times 10^{-3}$, Traffic Navigation, $F(4,45) = 314.9$, $p = 1.1 \times 10^{-31}$, Overcooked, $F(4,45) = 106.8$, $p = 6.9 \times 10^{-22}$, and Capture the Flag, $F(4,40) = 12.8$, $p = 1.6 \times 10^{-6}$ ($p$-values adjusted for multiple comparisons with a Holm--Bonferroni correction).

\begin{figure*}[b]
    \begin{subfigure}{0.24\textwidth}
        \centering
        \includegraphics[width=\linewidth]{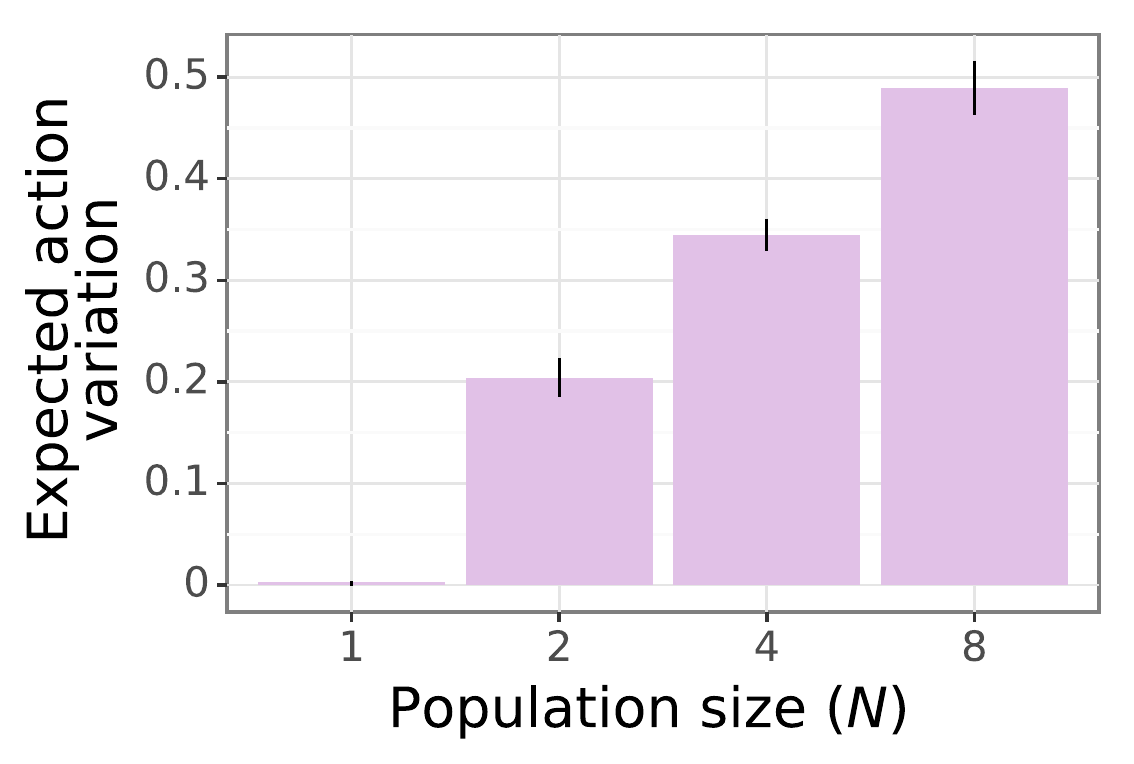}
        \caption{HarvestPatch.}        
    \end{subfigure}
    \hfill
    \begin{subfigure}{0.24\textwidth}
        \centering
        \includegraphics[width=\linewidth]{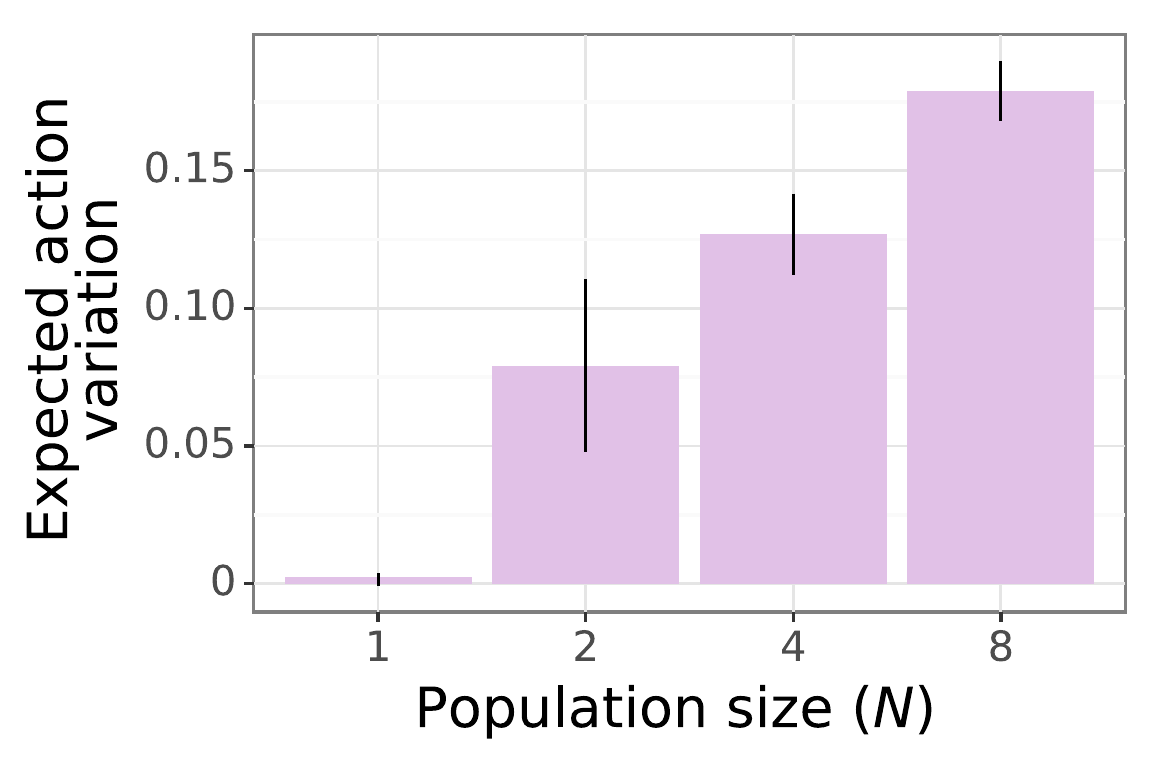}
        \caption{Traffic.}        
    \end{subfigure}
    \hfill
    \begin{subfigure}{0.24\textwidth}
        \centering
        \includegraphics[width=\linewidth]{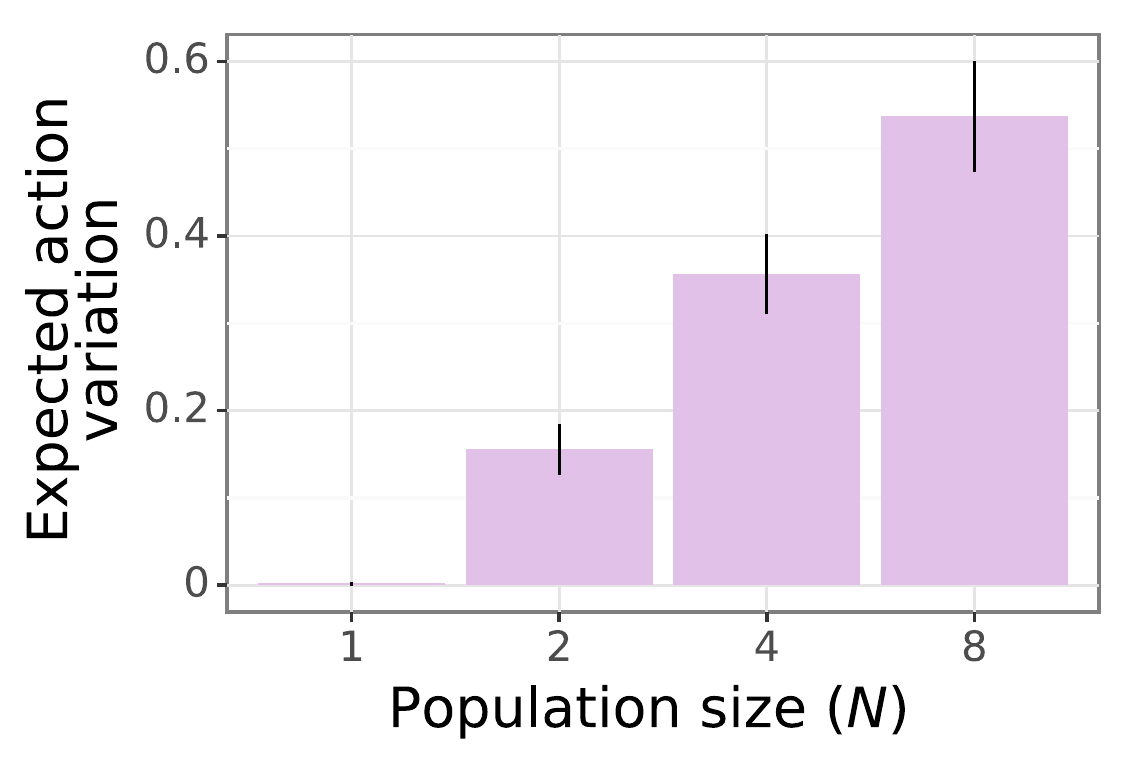}
        \caption{Overcooked.}        
    \end{subfigure}
    \hfill
    \begin{subfigure}{0.24\textwidth}
        \centering
        \includegraphics[width=\linewidth]{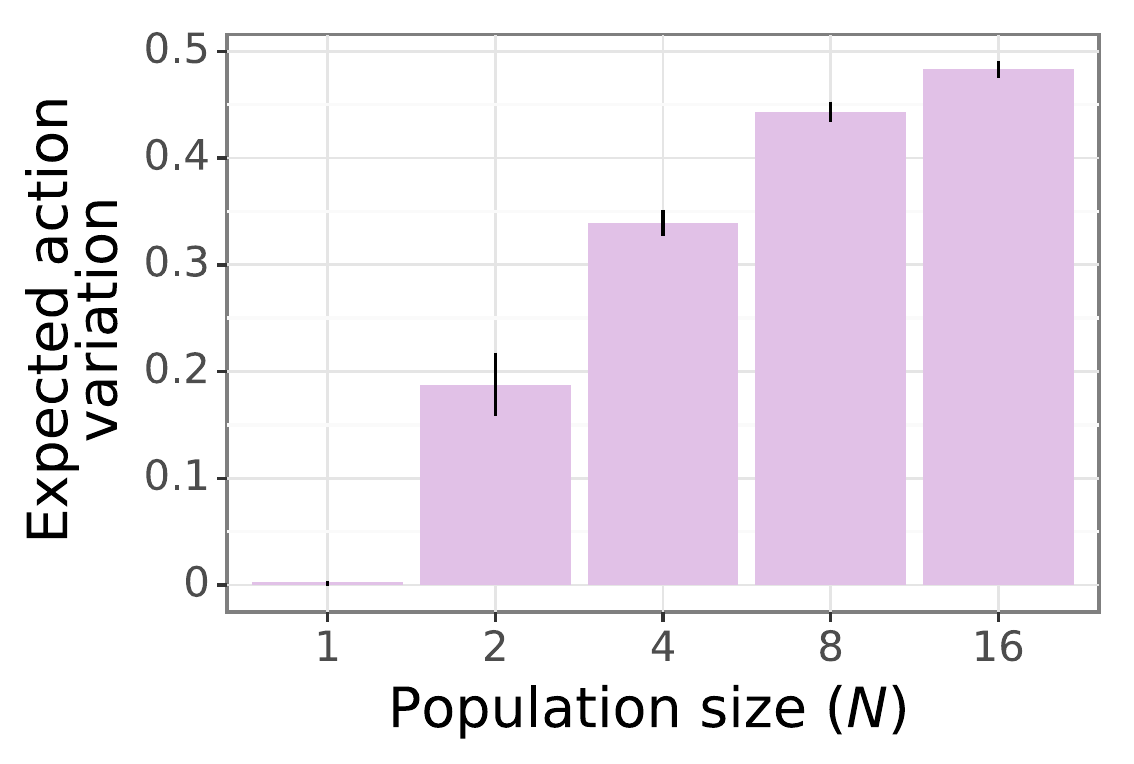}
        \caption{Capture the Flag.}        
    \end{subfigure}
    \caption{Effect of population size $N$ on behavioral diversity, as measured by expected action variation. Error bars represent 95\% confidence intervals calculated over five independent runs. \textbf{Result:} Increasing population size induces greater behavioral diversity.}
    \label{fig:522/pop/eav}
\end{figure*}

\subsubsection{Cross-Play Evaluation}
\label{sec:env/cross_play}
Next, we conduct cross-play evaluations of all populations following the procedure outlined in Section~\ref{sec:4/env_diversity_methods}. We separately evaluate populations on the single level included in the training set for all populations (Figure~\ref{fig:512/env/cross_play}, top row) and the held-out test levels (Figure~\ref{fig:512/env/cross_play}, bottom row).

Overall, the effects of environment diversity vary substantially across environments.

\paragraph{HarvestPatch} We observe the highest level of performance for the agents playing with a group trained on a large number of levels (column $L = \expnumber{1}{4}$) after themselves training on a small number of levels (row $L = 1$). In contrast, the worst-performing agents play with a group trained on a small number of levels (column $L = 1$) after themselves training on a large number of levels (row $L = \expnumber{1}{4}$). These patterns emerge in both the training-level and test-level evaluations.

\paragraph{Traffic Navigation} Agents perform equally well on their training level across all values for their training set size and for the training set size of the group's other members. In contrast, when evaluating agents on held-out test levels, an agent's performance strongly depends on how many levels the agent and its group were trained on. Average rewards increase monotonically from column $L = 1$ to column $L = \expnumber{1}{4}$, and increase near-monotonically from row $L = 1$ to row $L = \expnumber{1}{4}$. Navigation appears more successful with increasing experience of various level layouts and with increasingly experienced groupmates. 

\paragraph{Overcooked} In the held-out evaluations, we observe a consistent improvement in rewards earned from the bottom left ($L=1$ grouped with $L=1$) to the top right ($L=\expnumber{1}{4}$ grouped with $L=\expnumber{1}{4}$). An agent benefits both from playing with a partner with diverse training and from itself training with environment diversity.

A different pattern emerges in the training-level evaluation. Team performance generally decreases when one of the two agents trains on just $L = 1$ levels. However, when both agents train on $L = 1$, they collaborate fairly effectively. The highest scores occur at the intermediate values $L = \expnumber{1}{2}$ and $L = \expnumber{1}{3}$, rather than at $L = \expnumber{1}{4}$. Population skill on training levels declines with increasing environment diversity.

\paragraph{Capture the Flag} Team performance is closely tied to environment diversity. A team's odds of winning are quite low when they train on a smaller level set than the opposing team, and the win rate tends to jump considerably as soon as a team's training set is larger than their opponents'. However, echoing the results in Overcooked, agents trained on an intermediate level-set size achieve the highest performance on training levels. Population skill actually \textit{decreases} above $L = \expnumber{1}{2}$ on these levels (Table~\ref{tab:512/env/cross_play/ctf/elos}, middle column). In contrast, in held-out evaluation, environment diversity consistently strengthens performance; Elo ratings monotonically increase as $L$ increases (Table~\ref{tab:512/env/cross_play/ctf/elos}, right column).

\begin{table}[h]
    \centering
    \begin{tabular}[h]{ccc}
        \toprule
        & \multicolumn{2}{c}{Elo rating on:} \\ 
        $L$ & Training level & Test set \\
        \midrule
        $1$ & 604 & 258 \\
        $\expnumber{1}{1}$ & 882 & 773 \\
        $\expnumber{1}{2}$ & \textbf{1245} & 1248 \\
        $\expnumber{1}{3}$ & 1142 & 1349 \\
        $\expnumber{1}{4}$ & 1124 & \textbf{1369} \\
        \bottomrule
    \end{tabular}
    \caption{Elo ratings across number of training levels ($L$): Results from evaluating all populations in direct competition with one another. Training with greater environment diversity (i.e., number of training levels $L$) yields stronger populations with diminishing returns as $L$ increases.}
    \label{tab:512/env/cross_play/ctf/elos}
\end{table}

\subsection{Population Diversity}
We next delve into the effects of population diversity on agent policies and performance. Agents are trained in populations of size $N \in \{1, 2, 4, 8\}$ on $L = 1$ levels. In Capture the Flag, a set of additional populations are trained with size $N = 16$.

\subsubsection{Expected Action Variation Analysis}
We investigate the behavioral diversity of each population by calculating their expected action variation (see Section~\ref{sec:pop_div_methods}). 

\setcounter{figure}{6}
\begin{figure*}[b]
    \centering
    \begin{subfigure}{0.24\textwidth}
        \centering
        \includegraphics[width=\linewidth]{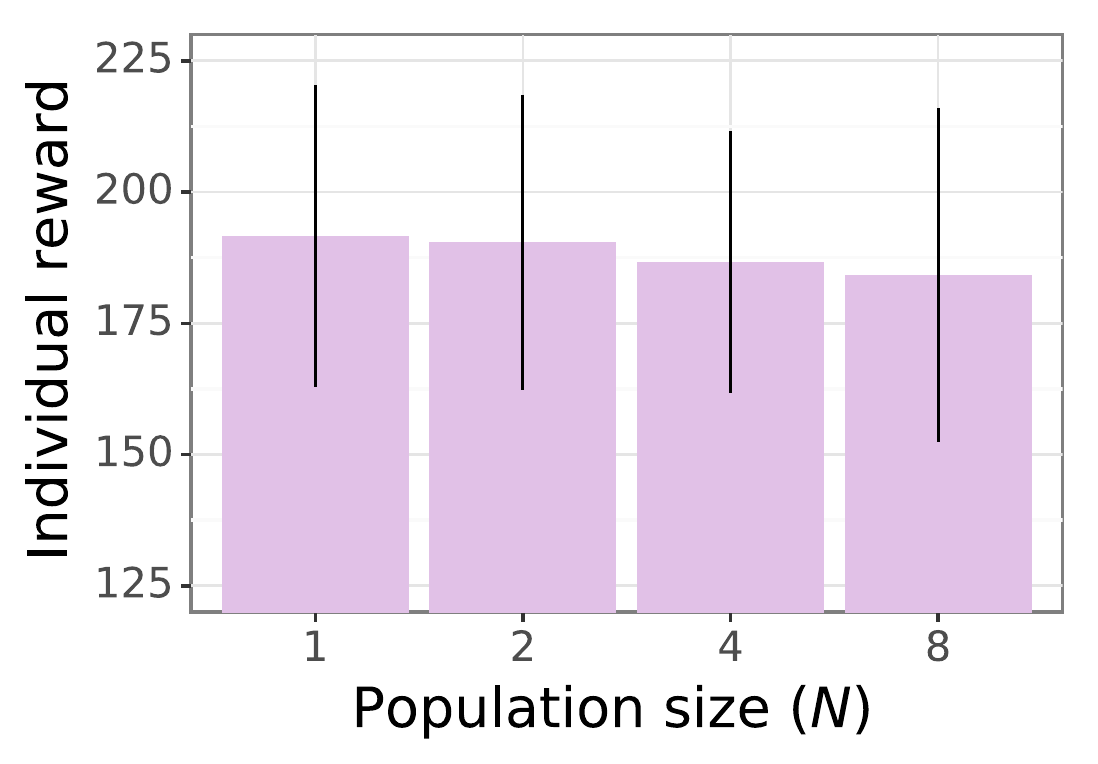}
        \caption{HarvestPatch}
        \label{fig:521/pop/cross_play/hp}
    \end{subfigure}
    \hfill
    \begin{subfigure}{0.24\textwidth}
        \centering
        \includegraphics[width=\linewidth]{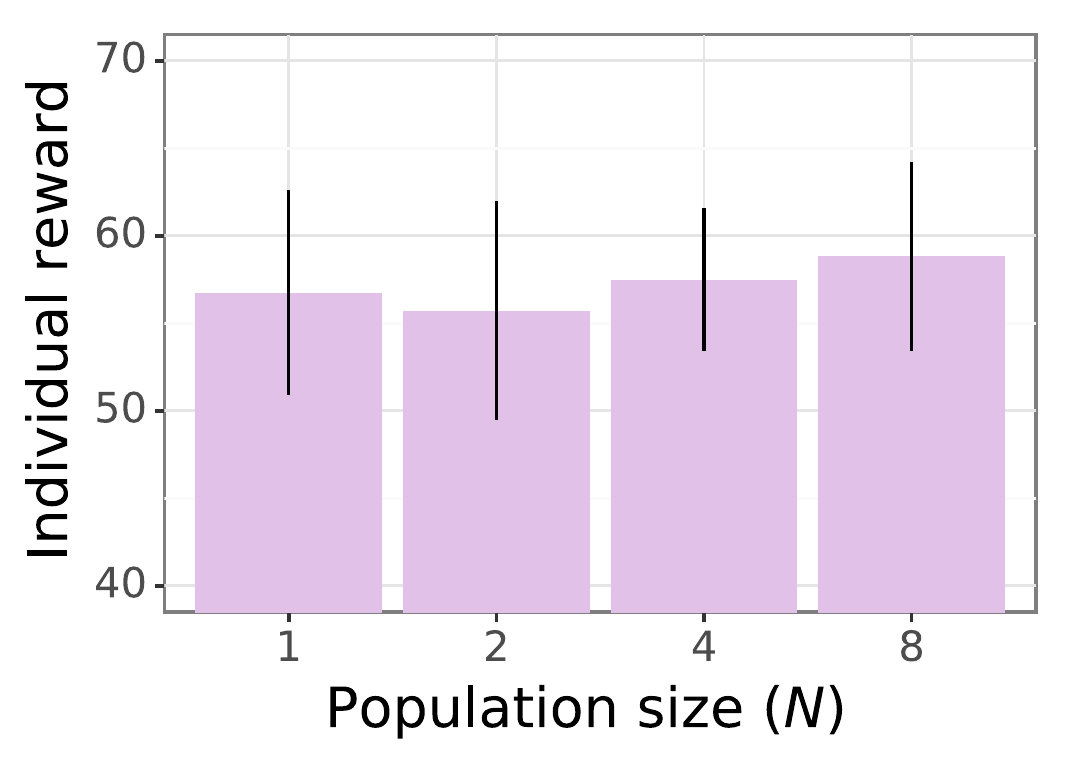}
        \caption{Traffic Navigation}
        \label{fig:521/pop/cross_play/tn}
    \end{subfigure}
    \hfill
    \begin{subfigure}{0.24\textwidth}
        \centering
        \includegraphics[width=\linewidth]{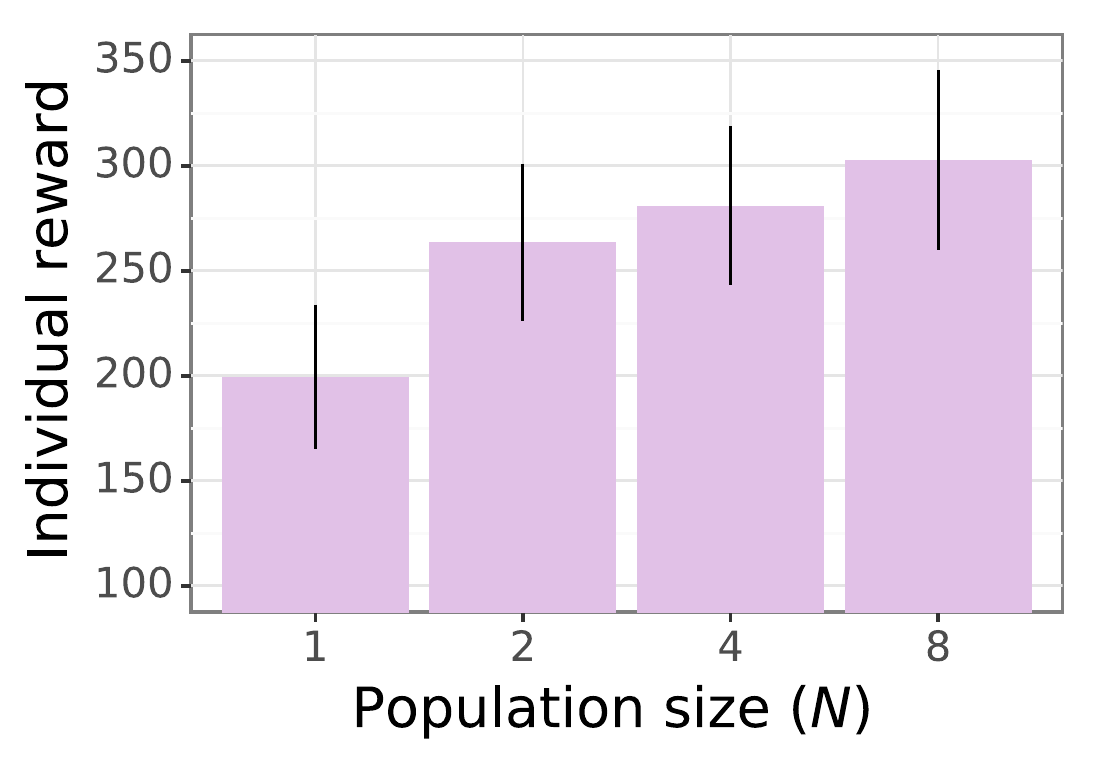}
        \caption{Overcooked}
        \label{fig:521/pop/cross_play/oc}
    \end{subfigure}    
    \hfill
    \begin{subfigure}{0.24\textwidth}
        \centering
        \begin{tabular}[h]{cc}
            \toprule
            $N$ & Elo rating \\
            \midrule
            $1$ & 751  \\
            $2$ & 936  \\
            $4$ & 1042 \\
            $8$ & 1123  \\
            $16$ & \textbf{1145}  \\
            \bottomrule
        \end{tabular}
        \label{fig:521/pop/cross_play/ctf}
        \caption{Capture the Flag}
    \end{subfigure} 
    \caption{Effect of population size $N$ on agent performance. Error bars indicate 95\% confidence intervals calculated over five independent runs (20 for Overcooked). \textbf{Result:} Training population size has no influence on the rewards of agents for HarvestPatch and Traffic Navigation. For Overcooked and Capture the Flag, larger populations produced stronger agents. The increase in performance is especially salient moving from $N = 1$ to $N = 2$, with diminishing returns as $N$ increases further.}
    \label{fig:521/pop/cross_play}
\end{figure*}

As shown in Figure~\ref{fig:522/pop/eav}, population size positively associates with behavioral diversity among agents trained in each environment. A set of ANOVAs confirm that $N$ has a statistically significant effect on expected action variation in HarvestPatch, $F(3,16) = 367.7$, $p = 1.7 \times 10^{-14}$, Traffic Navigation, $F(3,16) = 70.5$, $p = 1.9 \times 10^{-9}$, Overcooked, $F(3,16) = 126.7$, $p = 4.6 \times 10^{-11}$, and Capture the Flag, $F(4,20) = 634.6$, $p = 3.7 \times 10^{-20}$ ($p$-values adjusted for multiple comparisons with a Holm--Bonferroni correction).
Increasing population size amplifies co-player diversity, without requiring any explicit optimization for variation.
Multiple sources likely contribute to this diversity, including random initialization for agents and independent exploration and learning. 

\setcounter{figure}{5}
\begin{figure}[t]
    \centering
    \includegraphics[width=0.60\linewidth]{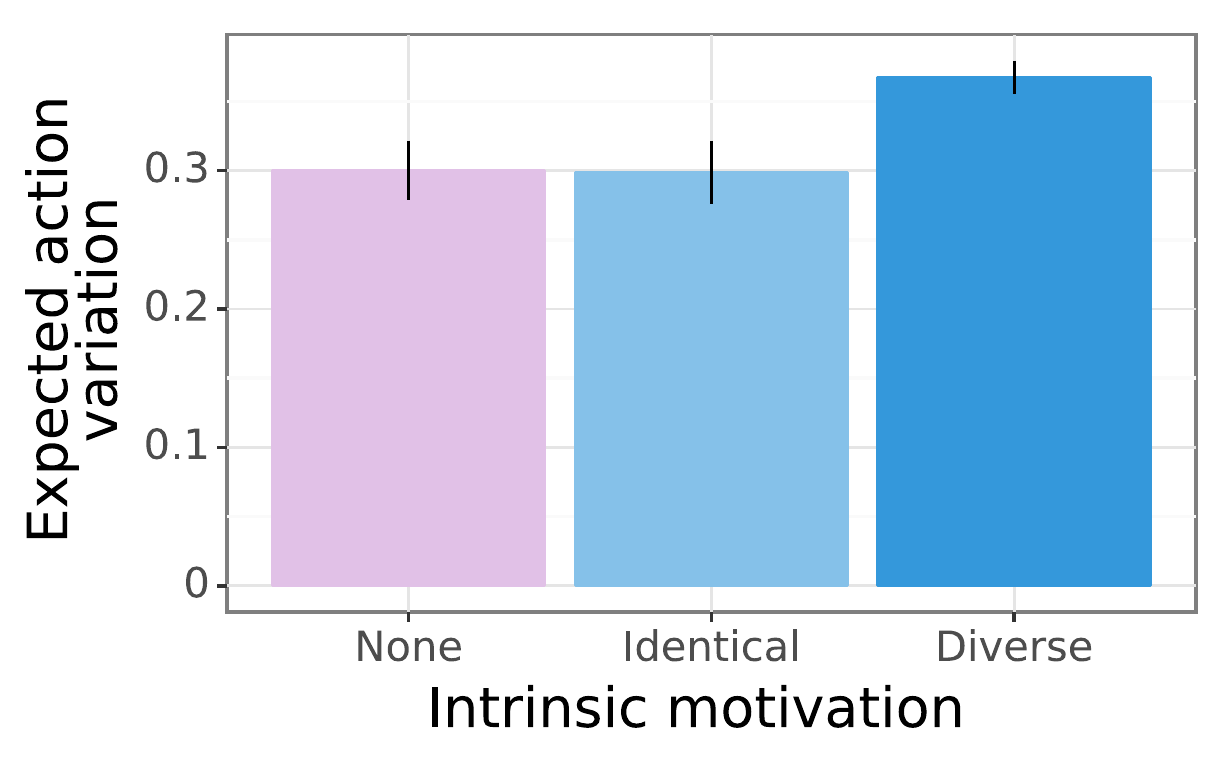}
    \caption{Effect of variation in intrinsic motivation on behavioral diversity on HarvestPatch. Error bands reflect 95\% confidence intervals calculated over five independent runs. \textbf{Result:} Populations with a heterogeneous distribution of intrinsic motivation exhibit significantly greater behavioral diversity than populations with no intrinsic motivation or with a homogeneous distribution.}
    \label{fig:522/pop/eav/svo}
\end{figure}
\setcounter{figure}{7}

\paragraph{Intrinsic Motivation and Behavioral Diversity}
Prior studies demonstrate that parameterizing an agent population with heterogeneous levels of intrinsic motivation can induce behavioral diversity, as measured through task-specific, hard-coded metrics \cite{mckee2020social}. These agent populations benefited from the resulting diversity in social dilemma tasks, including HarvestPatch. We run an experiment to confirm that this behavioral diversity can be detected through the measurement of expected action variation. Following prior work on HarvestPatch, we endow several $N = 4$ populations with SVO, an intrinsic motivation for maintaining a target distribution of reward among group members, and then train them on $L = \expnumber{1}{3}$ levels. We parameterize these populations with either a homogeneous or heterogeneous distribution of SVO (see Appendix~\ref{app:svo} for details).

As seen in Figure \ref{fig:522/pop/eav/svo}, populations with heterogeneous intrinsic motivation exhibit significantly greater behavioral diversity than populations without intrinsic motivation, $p = 4.9 \times 10^{-4}$. In contrast, behavioral diversity does not differ significantly between populations of agents lacking intrinsic motivation and those parameterized with homogeneous intrinsic motivation, $p = 0.99$. These results help baseline the diversity induced by increasing population size and demonstrate that expected action variation can be used to assess established sources of behavioral heterogeneity.

\subsubsection{Cross-Play Evaluation}
We next conduct a cross-play evaluation of all populations following the procedure outlined in Section~\ref{sec:pop_div_methods}.  As before, we test whether observed patterns are statistically significant using a set of ANOVAs (with a Holm--Bonferroni correction to account for multiple comparisons).

Population diversity does not significantly affect agent rewards for HarvestPatch, $F(3,16) = 0.06$, $p = 1.0$ (Figure~\ref{fig:521/pop/cross_play}a), and Traffic Navigation, $F(3,16) = 0.22$, $p = 1.0$ (Figure~\ref{fig:521/pop/cross_play}b). Agents trained through self-play ($N = 1$) perform equivalently to agents from the largest, most diverse populations ($N = 8$). Training in a diverse population thus appears to neither advantage nor disadvantage agents in these environments.

In contrast, agents trained in diverse populations outperform those trained in lower-variation populations for Overcooked, $F(3,76) = 5.2$, $p = 7.7 \times 10^{-3}$ (Figure~\ref{fig:521/pop/cross_play}c) and Capture the Flag (Figure~\ref{fig:521/pop/cross_play}d). For both environments, we observe a substantial jump in performance from $N = 1$ to $N = 2$ and diminishing increases thereafter. The diminishing returns of diversity resemble the relationship between environment diversity and performance observed for Overcooked and Capture the Flag in Section \ref{sec:env/cross_play}.

\section{Discussion}
In summary, this paper makes several contributions to multi-agent reinforcement learning research.
Our experiments extend single-agent findings on environment diversity and policy generalization to the multi-agent domain. We find that applying a small amount of environment diversity can lead to a substantial improvement in the generality of agents. %
However, this generalization reduces performance on agents' training set for certain environments and co-players.

The \textit{expected action variation} metric demonstrates how population size and the diversification of agent hyperparameters can influence behavioral diversity. As with environmental diversity, we find that training with a diverse set of co-players strengthens agent performance in some (but not all) cases.

Expected action variation measures population diversity by estimating the heterogeneity in a population's policy distribution. As recognized by hierarchical and options-based frameworks \cite{sutton1999between}, the mapping of lower-level actions to higher-level strategic outcomes is imperfect; in some states, different actions may lead to identical outcomes. Higher levels of expected action variation may capture greater strategic diversity. Nonetheless, future work could aim to directly measure variation in a population's strategy set.

These findings may prove useful for the expanding field of human-agent cooperation research. Human behavior is notoriously variable \cite{cronbach1957two,eid1999intraindividual}. Interindividual differences in behavior can be a major difficulty for agents intended to interact with humans \cite{egan1988individual}. This variance thus presents a major challenge stymying the development of \textit{human-compatible} reinforcement learning agents. Improving the generalizability of our agents could advance progress toward human compatibility, especially for cooperative domains \cite{dafoe2020open}.

Future work could seek to develop more sophisticated approaches for quantifying diversity.
For example, here we use the ``number of unique levels'' metric as a proxy of environment diversity, and therefore increased $L$ leads to monotonically increasing environment diversity. However, these levels may be unique in ways which are irrelevant to the agents. Scaling existing approaches to these settings, such as those that study how the environment influences agent behaviour \cite{wang2020enhanced}, may help determine which features correspond to \textit{meaningful} diversity.

The experiments presented here employ a rigorous statistical approach to test the consistency and significance of the effects in question. Consequently, they help scope the benefits of environment and population diversity for multi-agent reinforcement learning. Overall, we hope that these findings can improve the design of future multi-agent studies, leading to more generalized agents.

\section*{Acknowledgements}

We thank Ian Gemp, Edgar Duéñez-Guzmán, and Thore Graepel for their support and feedback during the preparation of this manuscript. We are also indebted to Mary Cassin for designing and creating the sprite art for the DeepMind Lab2D implementation of Overcooked.

\bibliographystyle{spmpsci}      %
\bibliography{main}   %

\newpage
\appendix
\renewcommand{\thefigure}{A\arabic{figure}}
\renewcommand{\thetable}{A\arabic{table}}
\setcounter{figure}{0}
\setcounter{table}{0}
\onecolumn

\section{Environments}
\label{app:environments}

\begin{figure*}[b!]
    \centering
    \begin{subfigure}{0.32\textwidth}
        \centering
        \includegraphics[height=9em]{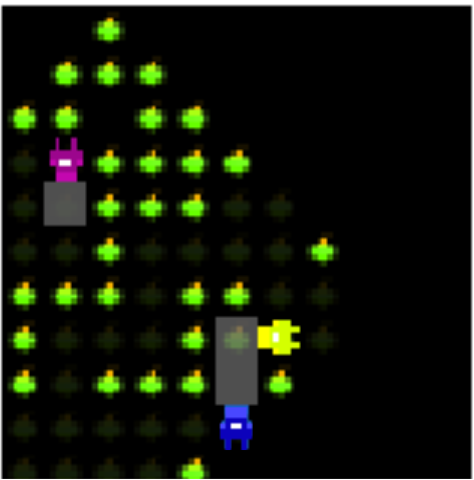}
        \caption{Example observation.}
        \label{fig:app/environments/hp/obs/example}
    \end{subfigure}
    \hfill
    \begin{subfigure}{0.32\textwidth}
        \centering
        \includegraphics[height=9em]{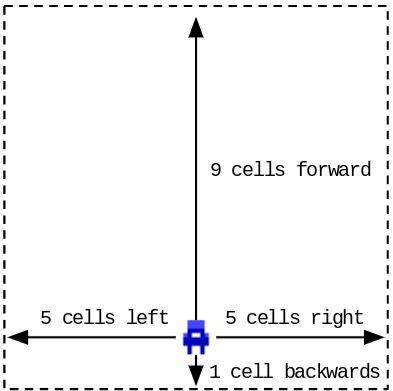}
        \caption{Observation dimensions.}
        \label{fig:app/environments/hp/obs/dimensions}
    \end{subfigure}
    \hfill
    \begin{subfigure}{0.32\textwidth}
        \centering
        \includegraphics[height=9em]{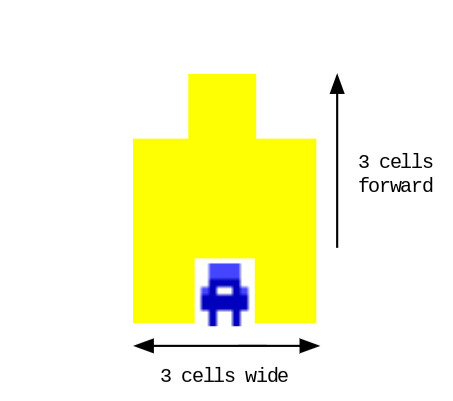}
        \caption{Beam structure.}
        \label{fig:app/environments/hp/obs/beam}
    \end{subfigure}    
    \caption{(a) Example observation for HarvestPatch. Players observe an $88\times88\times3$ egocentric view of the environment (i.e., $11\times11$ cells with $8\times8\times3$ sprites in each cell). (b) The observation window is offset from the player such that they can always see one cell behind them, five cells either side, and nine cells in front. (c) The beam is three cells wide and extends three cells forward from the player (until blocked by players or walls).}
    \label{fig:app/environments/hp/obs}
\end{figure*}

\subsection{HarvestPatch}
\label{app:environments/harvest_patch}

\subsubsection{Gameplay}
Players are placed in a $35 \times 35$ gridworld environment containing a number of apples. Players can harvest apples by moving over them, receiving $+1$ reward for each apple collected. Apples regrow after being harvested. The rate of apple regrowth is determined by the number of unharvested apples within the regrowth radius $r$. An apple cannot regrow if there are no apples within its radius $r$ or if a player is standing on its cell. This property induces a social dilemma for the players. The group as a whole will perform better if its members are abstentious in their apple consumption, but in the short term individuals can always do better by harvesting greedily.

In addition to basic movement actions, players can use a beam to tag out other players. Players are tagged out for 50 steps after being struck by another group member's tagging beam, and then are respawned at a random location within the environment. The ability to tag other players can be used to reduce the effective group size and mitigate the intensity of the social dilemma \cite{perolat2017multi}. There are no direct reward penalties for tagging or being tagged; any reward penalties experienced by the agents are indirectly imposed through opportunity costs.

\paragraph{Observations}
Players observe an egocentric view of the environment (Figure~\ref{fig:app/environments/hp/obs}).

\paragraph{Actions}

Players can take one of the following eight actions each step:
\begin{enumerate}
    \item \texttt{No-op}: The player stays in the same position.
    \item \texttt{Move forward}: Moves the player forwards one cell.
    \item \texttt{Move backward}: Moves the player backwards one cell.
    \item \texttt{Move left}: Moves the player left one cell.
    \item \texttt{Move right}: Moves the player right one cell.
    \item \texttt{Turn left}: Rotates the player 90 degrees anti-clockwise.
    \item \texttt{Turn right}: Rotates the player 90 degrees clockwise.
    \item \texttt{Use tag beam}: Fires a short yellow beam forwards from the player. The beam is three cells wide and is projected three cells forwards.
\end{enumerate}

The \texttt{use tag beam} action has a cooldown of four steps for each player. If the player tries to use the tag action during this cooldown period, the outcome is equivalent to the \texttt{no-op} action. Players who are hit by the beam are tagged out for 50 steps, after which they are respawned in a random location.

\paragraph{Rewards} Players receive $+1$ reward for each apple they harvest. (Players can harvest an edible apple by moving into its cell.) Players do not receive reward in any other way. Notably, neither using the \texttt{use tag beam} action nor being hit by the tagging beam yields any reward.

\paragraph{Apple regrowth probabilities}
In HarvestPatch, apples grow in ``patches''. Each apple in a patch is within $r$ distance of the other apples in the patch, and further than $r$ away from apples in all other patches. Based on the respawn rule described previously, on each step of an episode, harvested apples have a probability of regrowing based on the number of other non-harvested apples in their patch. These probabilities are as follows:

\begin{table}[h]
    \centering
    \begin{tabular}[h]{c c}
        \toprule
        \textbf{Number of} & \textbf{Regrowth} \\
        \textbf{apples in patch} & \textbf{probability} \\
        \midrule
        $0$ & $P = 0$ \\
        $1$ & $P = 0.001$ \\
        $2$ & $P = 0.005$ \\
        $3+$ & $P = 0.025$ \\
        \bottomrule
    \end{tabular}
    \caption{Regrowth rates for apples in HarvestPatch.}
\end{table}

Crucially, when the patch has been depleted (i.e., the number of apples in patch is zero), the apples in that patch cannot regrow for the rest of the episode.

\subsubsection{Additional details}

The HarvestPatch environment instantiates an intertemporal social dilemma \cite{hughes2018inequity}. A group can sustainably harvest from the environment by abstaining from fully depleting the apple patches. This strategy supplies the group with an indefinite stream of reward over time. However, for the group to reap the benefits of a sustainable harvesting strategy, \textit{every} group member must abstain from depleting apple patches. In contrast, an individual is immediately and unilaterally guaranteed reward for eating an ``endangered apple'' (the last unharvested apple remaining in a given patch) if it acts greedily. Overall, there is a strong tension between the short-term individual temptation to maximize reward through unsustainable behavior and the long-term strategy of maximizing group welfare by acting sustainably.

\subsubsection{Procedural generation}

\paragraph{Procedure}

To generate levels for HarvestPatch, we use the following procedure given a number of patches $p  \in [1, 14]$ with a patch radius $r_p \in [3, 7]$ and density $d \in [0.90, 1.00]$:

\begin{enumerate}
    \item Generate a $35 \times 35$ area.
    \item Place $p$ points randomly, ensuring a distance of at least $3 \times r_p$ between all points.
    \item Around each point, assign apples in radius $r_p$ with probability $d$ per apple.
    \item Place $10$ player spawn points in random non-assigned cells.
\end{enumerate}

\begin{figure}[ht]
    \centering
    \includegraphics[width=0.93\linewidth]{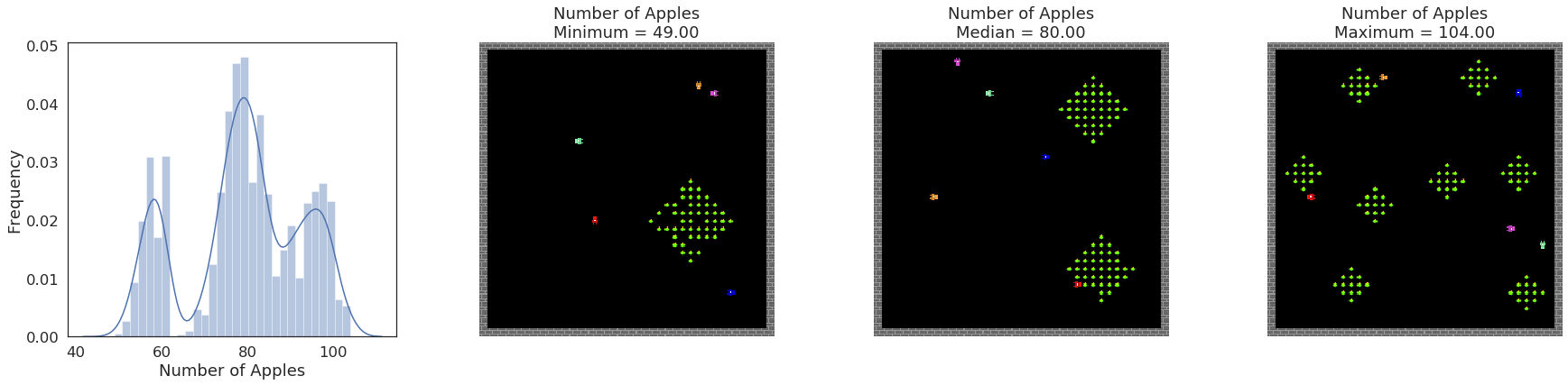}
    \caption{Distribution over environmental features, alongside an example level at the minimum, median, and maximum of these distributions.}
    \label{fig:app/environments/hp/levels_features}
\end{figure}

\begin{figure}[ht]
    \centering
    \includegraphics[width=0.9\linewidth]{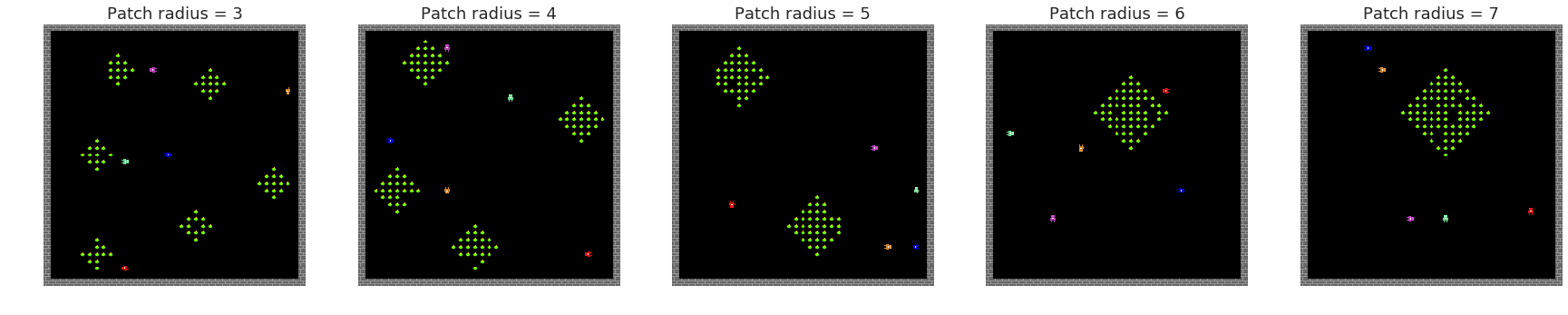}
    \caption{Example levels for each of the possible patch radius $r$ values.}
    \label{fig:app/environments/hp/levels_features_2}
\end{figure}

\paragraph{Distribution over environmental features}
Figure~\ref{fig:app/environments/hp/levels_features} presents the distribution of apple counts in HarvestPatch levels, alongside visualizations of the levels at the minimum, median, and maximum of this distribution.  Similarly, Figure~\ref{fig:app/environments/hp/levels_features_2} presents an example level for each of the various values of the patch radius parameter $r$.

\begin{figure}[ht]
    \centering
    \includegraphics[width=\linewidth]{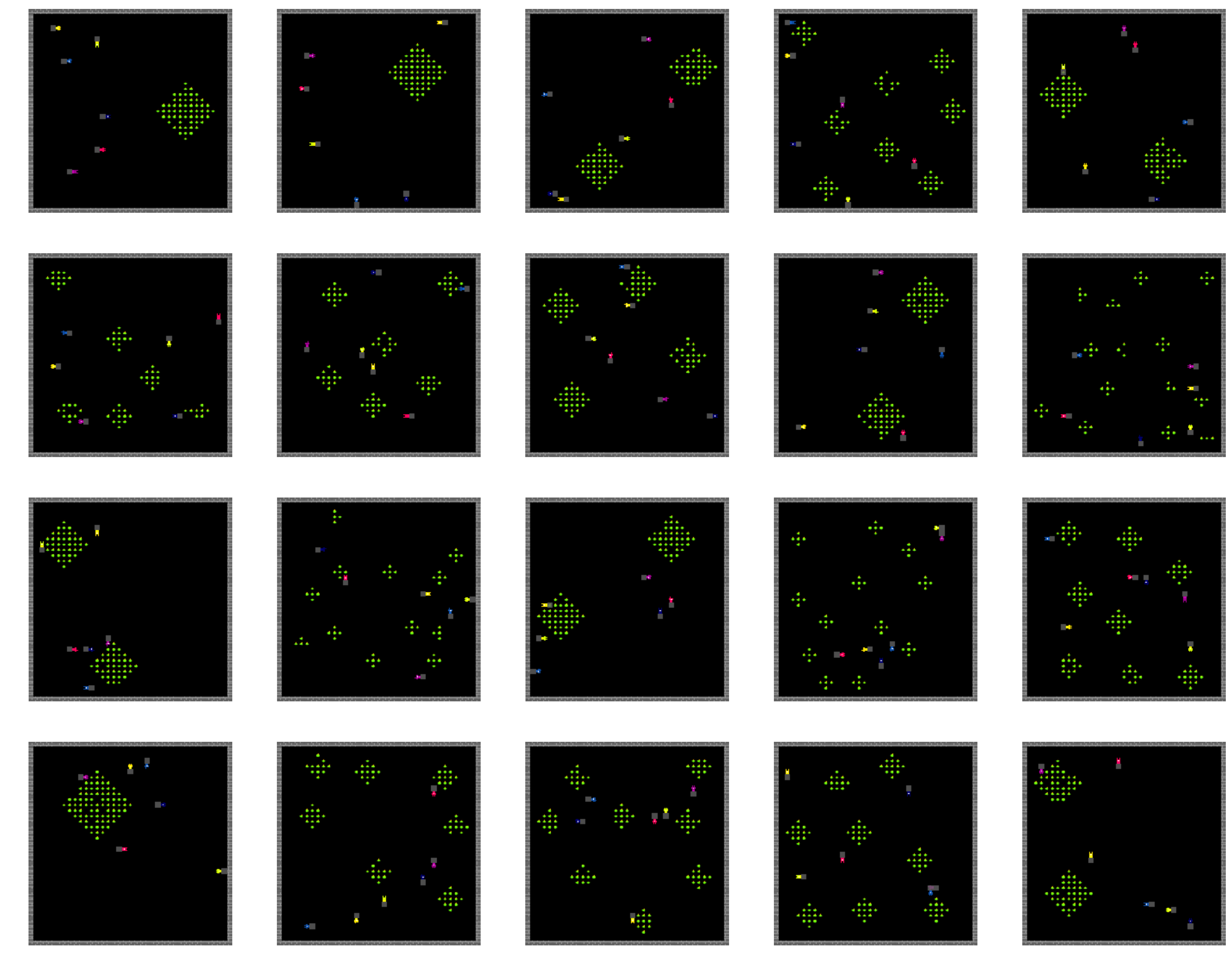}
    \caption{Example levels procedurally generated for the HarvestPatch environment.}
    \label{fig:app/environments/hp/levels}
\end{figure}

\newpage
\paragraph{Example levels}
Presented in Figure~\ref{fig:app/environments/hp/levels} are 20 randomly sampled levels created through the previously described generation procedure.

\clearpage
\subsection{Traffic Navigation}
\label{app:environments/traffic_navigation}

\subsubsection{Gameplay}
Eight players are placed along the edges of a gridworld environment. The goal for each player is to navigate through the level to their given ``goal location'' while avoiding collisions with other players. Goal locations are randomly sampled from the available edge cells in the game. More than one player can have the same goal at any one time. Upon a player reaching their given goal, they receive $+1$ reward and are assigned a new goal location.

Levels created by the procedural generator can contain large open-spaces (making navigation and coordination easy), as well as small corridors (requiring additional coordination to avoid collisions with other players). While the number of players remains fixed (at eight), the size of the levels ranges from $10 \times 10$ (100 cells in total) to $20 \times 20$ (400 cells in total). For the smaller levels, while a players may chart a short path to their goal, the tight proximity of other players increases the likelihood of colliding into another player. In large levels, the goals are further apart, on expectation necessitating further travel for players to receive their reward. However, due to the larger size of the environment, the likelihood of players encountering one another is reduced.

\begin{figure*}[h]
    \centering
    \begin{subfigure}{0.35\textwidth}
        \centering
        \includegraphics[height=9em]{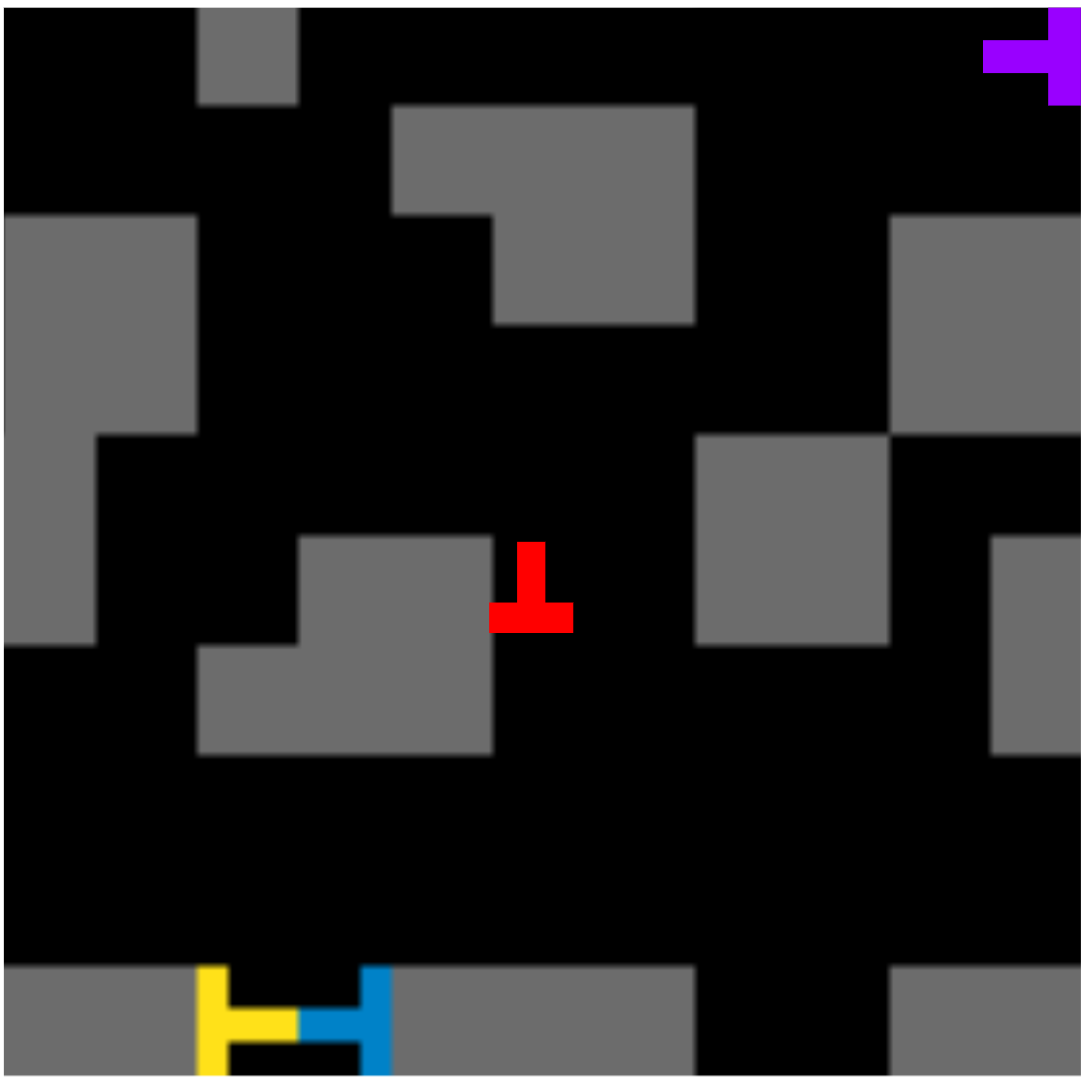}
        \caption{Example observation.}
        \label{fig:app/environments/tn/example}
    \end{subfigure}
    \begin{subfigure}{0.35\textwidth}
        \centering
        \includegraphics[height=9em]{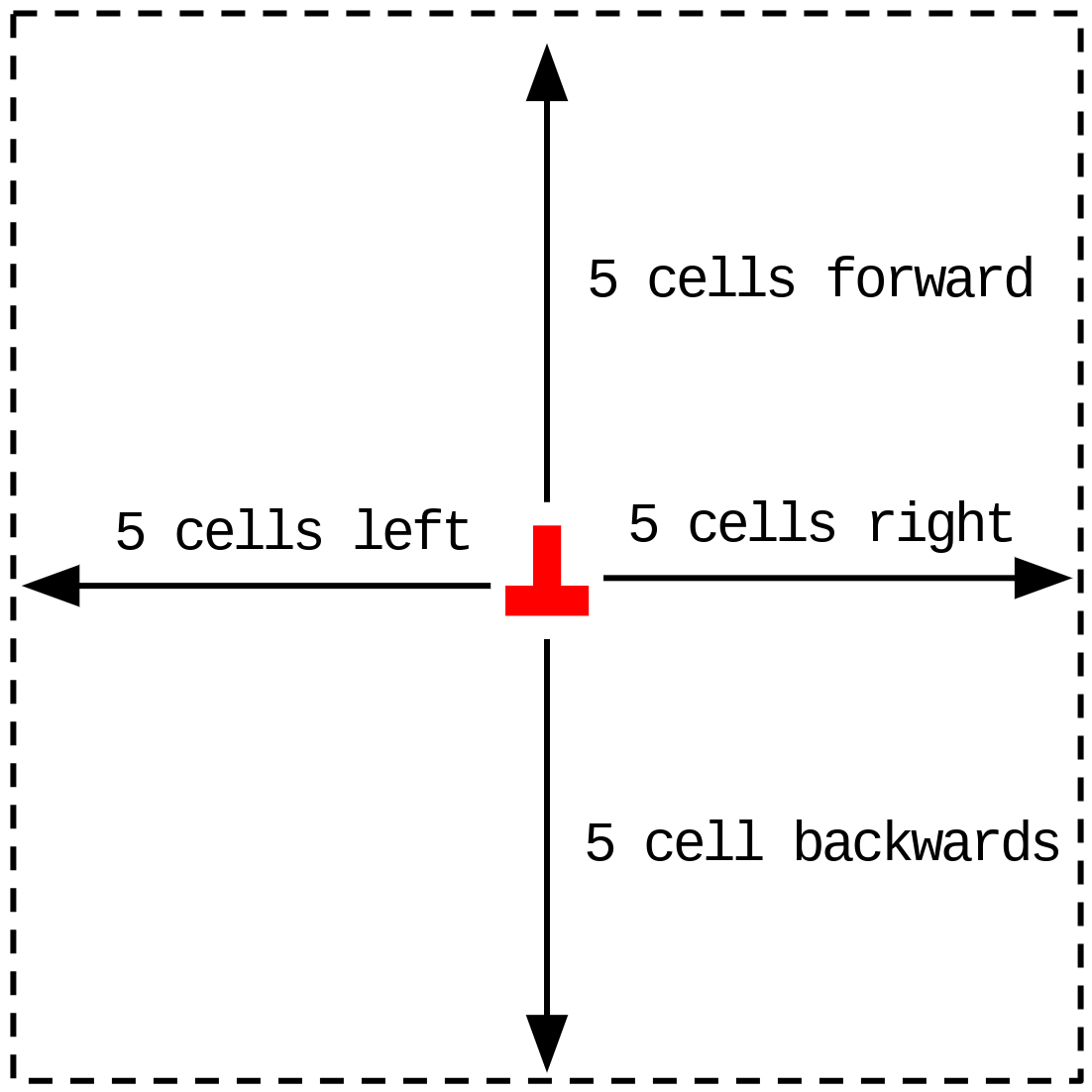}
        \caption{Observation dimensions.}
        \label{fig:app/environments/tn/dimensions}
    \end{subfigure}
    \caption{(a) Example observation for Traffic Navigation. Players observe a $33\times33\times3$ egocentric view of the environment (i.e., $11\times11$ cells with $3\times3\times3$ sprites in each cell). (b) The player is centered in the middle of their observation window and can see equally far in each direction.}
    \label{fig:app/environments/tn/obs}
\end{figure*}

\paragraph{Observations}
Players observe an egocentric view of the environment (Figure~\ref{fig:app/environments/tn/obs}) as well as their relative offset to the current goal location.

\paragraph{Actions}

Players can take one of the following five actions on each step:
\begin{enumerate}
    \item \texttt{No-op}: The player stays in the same position.
    \item \texttt{Move north}: Moves the player north one cell.
    \item \texttt{Move south}: Moves the player south one cell.
    \item \texttt{Move west}: Moves the player west one cell.
    \item \texttt{Move east}: Moves the player east one cell.
\end{enumerate}

\paragraph{Rewards}
Players receive $+1$ reward when they successfully occupy the same cell as their goal location. They receive $-1$ reward whenever they collide into another player (regardless of whether they or the other player caused the collision).

\subsubsection{Procedural generation}

\paragraph{Procedure}
To generate levels for Traffic Navigation, we use the following procedure:

\begin{enumerate}
    \item Generate an area with a width $\in [10, 20]$ and height $\in [10, 20]$. 
    \item Add walls along the outer edge of this area.
    \item Remove two neighbouring walls for every player, adding a player spawn point at each newly available cell and adding those cells as available goal locations.
    \item Create a random number of $3 \times 3$ wall blocks in the available area, ensuring that a majority of the blocks are more than two cells away from each other.
    \item Randomly scatter a number of individual walls within the available area.
    \item Check that the level is solvable:  all goals are reachable from all starting locations.
\end{enumerate}

\newpage
\paragraph{Distribution over environmental features}
To demonstrate the diversity of the procedurally generated levels for Traffic Navigation, this section introduces four descriptive features, presents their distributions from 100,000 samples, and visualizes the minimum, medium, and maximum level of each distribution. 

The two features are as follows:
\begin{enumerate}
    \item \textbf{Openness:} The proportion of non-wall cells in the level.
    \item \textbf{Number of Walls:} The number of walls in the level.    
\end{enumerate}

\begin{figure}[ht]
    \centering
    \includegraphics[width=\linewidth]{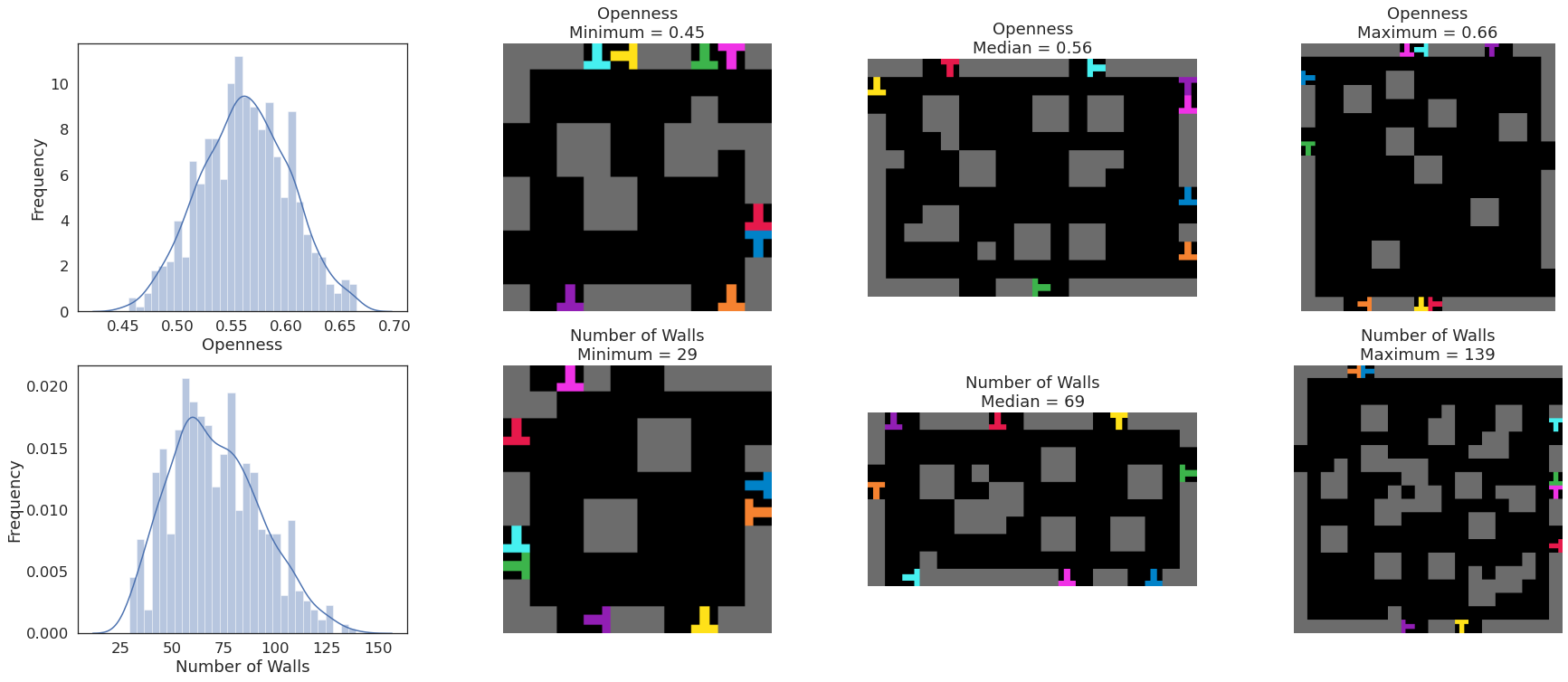}
    \caption{Distribution over environmental features, alongside an example level at the minimum, median, and maximum of these distributions.}
    \label{fig:app/environments/tn/levels_features}
\end{figure}

\begin{figure}[ht]
    \centering
    \includegraphics[width=\linewidth]{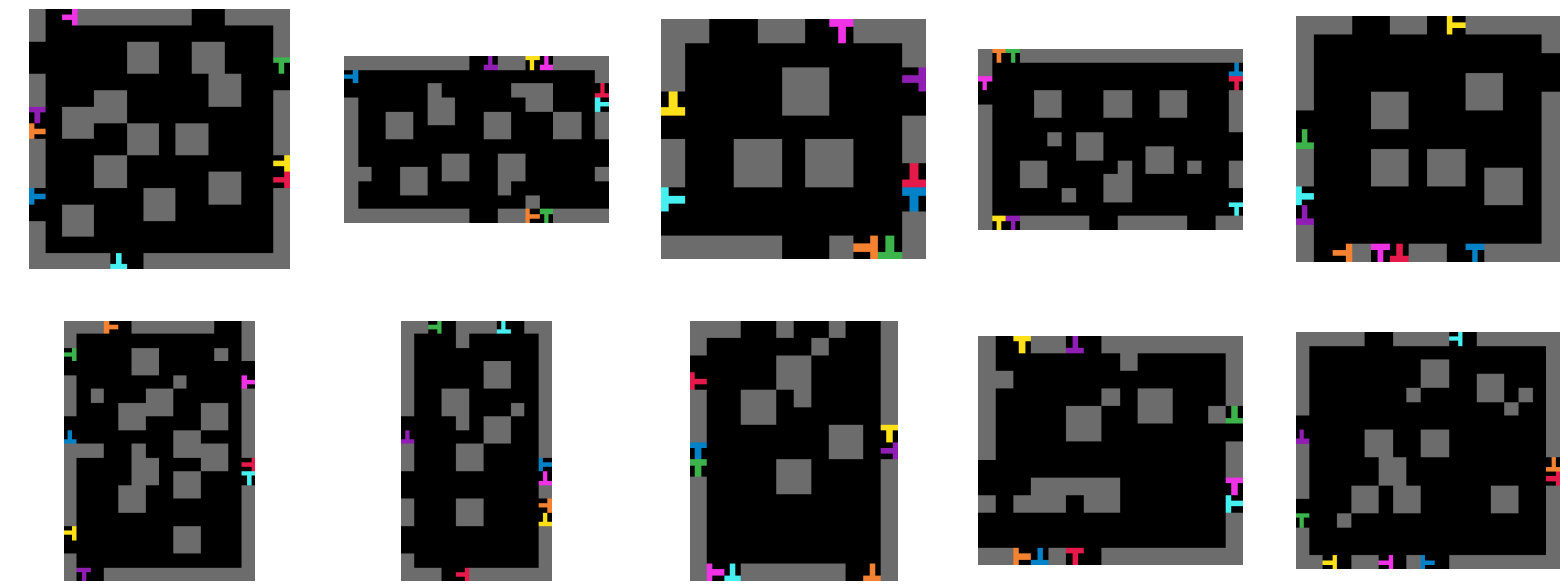}
    \caption{Example levels procedurally generated generated for the Traffic Navigation environment.}
    \label{fig:app/environments/tn/levels}
\end{figure}

\paragraph{Example levels}
Presented in Figure~\ref{fig:app/environments/tn/levels} are 10 randomly sampled levels created through the previously described generation procedure.

\clearpage
\subsection{Overcooked}
\label{app:environments/overcooked}

\subsubsection{Gameplay}
Players are placed in a gridworld environment containing cooking pots, dispensing stations (tomatoes or dishes), delivery stations, and empty counters. Players can move around and interact with these objects. By sequencing certain object interactions, players can pick up and deposit items. Each player (and each counter) can only hold one item at a time.

The objective of each episode is for the players to deliver as many tomato soup dishes as possible through delivery stations. In order to create a tomato soup dish, players must pickup tomatoes and deposit them into the cooking pot. Once there are three tomatoes in the cooking pot, it begins to cook for 20 steps. After 20 steps, the soup is fully cooked. Cooking progress for a dish is tracked by a loading bar overlaying the cooking pot. The loading bar increments over the cooking time and then turns green when the dish is ready for collection.

When the tomato soup is ready for collection, a player holding an empty dish can interact with the cooking pot to pick up the soup. The player can then deliver the soup by interacting with a delivery station while holding the completed dish. A successful delivery rewards both players and removes the dish from the game.

\begin{figure*}[h]
    \centering
    \begin{subfigure}{0.35\textwidth}
        \centering
        \includegraphics[height=9em]{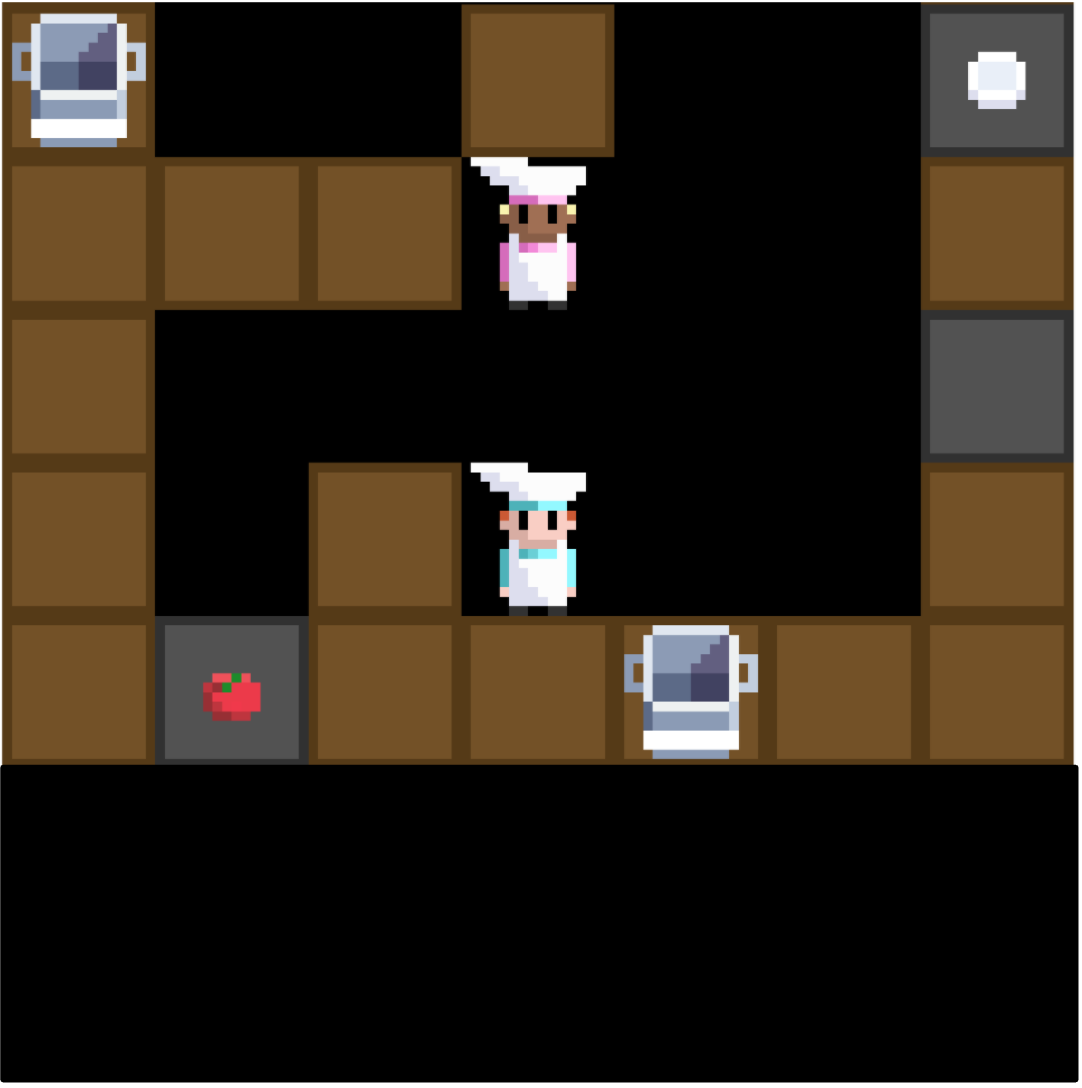}
        \caption{Example observation.}
        \label{fig:app/environments/oc/obs/example}
    \end{subfigure}
    \begin{subfigure}{0.35\textwidth}
        \centering
        \includegraphics[height=9em]{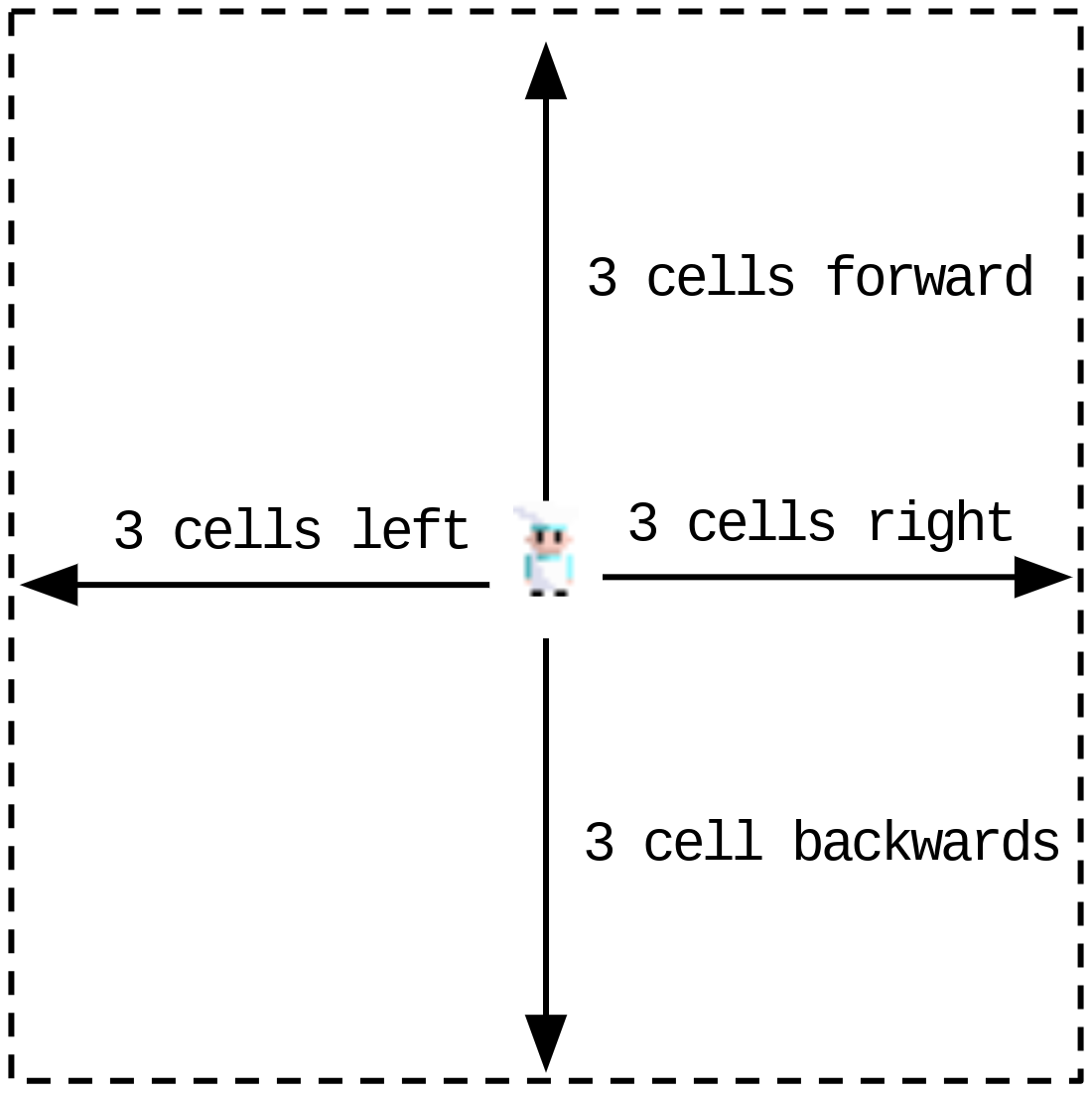}
        \caption{Observation dimensions.}
        \label{fig:app/environments/oc/dimensions}
    \end{subfigure}
    \caption{(a) Example observation for Overcooked. Players observe a $56\times56\times3$ egocentric view of the environment (i.e., $7\times7$ cells with $8\times8\times3$ sprites in each cell). (b) The player is centered in the middle of their observation window such that they can see equally far in all directions.}
    \label{fig:app/environments/oc/obs}
\end{figure*}

\paragraph{Observations}
Players observe an egocentric view of the environment (Figure~\ref{fig:app/environments/oc/obs}).

\paragraph{Actions}

Players can take one of the following six actions each step:
\begin{enumerate}
    \item \texttt{No-op}: The player stays in the same position.
    \item \texttt{Move north}: Orients and moves the player north one cell.
    \item \texttt{Move south}: Orients and moves the player south one cell.
    \item \texttt{Move west}: Orients and moves the player west one cell.
    \item \texttt{Move east}: Orients and moves the player east one cell.
    \item \texttt{Interact}: The player interacts with the cell that they are facing.
\end{enumerate}

The outcome of the \texttt{Interact} action depends on the current item held by the player (none, empty dish, tomato, or soup), as well as the type of object which they are facing (counter, cooking pot, tomato station, dish station, or delivery station). Depending on these two conditions, the player will either deposit the held item to the object or pick up an item from the object.

\paragraph{Rewards}
Players receive a shared $+20$ reward for every soup they deliver through the delivery station. As a result, they are incentivized to create and deliver as many soups as possible within an episode. To scaffold learning, players also receive $+1$ reward each time they deposit a tomato into the cooking pot.

\subsubsection{Procedural generation}

\paragraph{Procedure}
To generate levels for Overcooked, we use the following procedure:

\begin{enumerate}
    \item Generate an area with a width $\in [4, 9]$ and height $\in [4, 9]$.
    \item Add counters along the outer edge of this area.
    \item Create a random number of counters within the available area, either as a block of counters in the middle or randomly scattered throughout the area.
    \item For each object in \{cooking pot, tomato station, dish station, delivery station\}, convert one to three counters into that object.
    \item Place two player spawn points in random non-assigned cells.
    \item Check that the level is solvable: tomato soups can be created and delivered by both players.
\end{enumerate}

\paragraph{Distribution over environmental features}
To demonstrate the diversity of the procedurally generated levels for Overcooked, this section introduces two descriptive features, presents their distributions from 100,000 samples, and visualizes the minimum, medium, and maximum level of each distribution. 

The two features are as follows:
\begin{enumerate}
    \item \textbf{Solution's Estimated Path Length:} The estimated number of required movement actions between the kitchen objects to create and delivery one tomato soup.
    \item \textbf{Openness:} The proportion of non-counter cells in the level.
\end{enumerate}

\begin{figure}[ht]
    \centering
    \includegraphics[width=\linewidth]{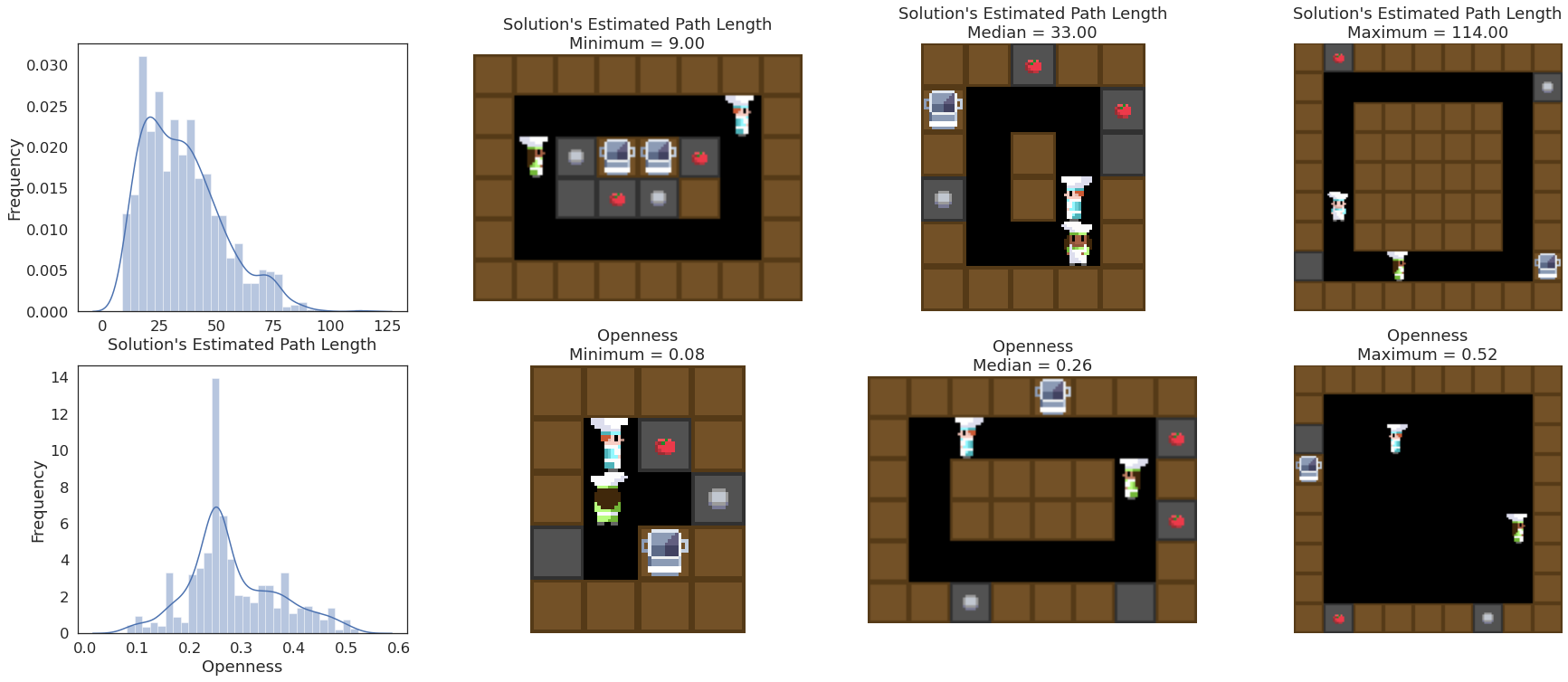}
    \caption{Distribution over environmental features, alongside an example level at the minimum, median, and maximum of these distributions.}
    \label{fig:app/environments/oc/levels_features}
\end{figure}

\begin{figure}[ht]
    \centering
    \includegraphics[width=\linewidth]{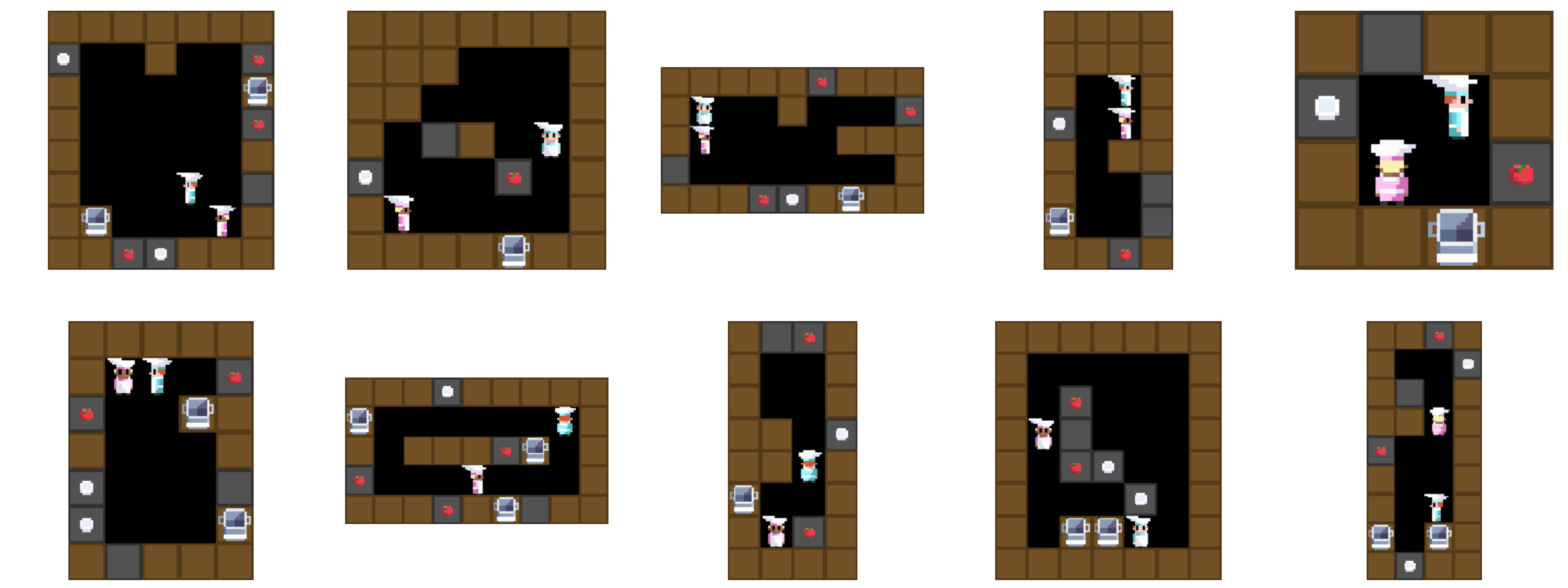}
    \caption{Example levels procedurally generated for the Overcooked environment.}
    \label{fig:app/environments/oc/levels}
\end{figure}

\paragraph{Example levels}
Presented in Figure~\ref{fig:app/environments/oc/levels} are 10 randomly sampled levels created through the previously described generation procedure.

\clearpage
\subsection{Capture the Flag}
\label{app:environments/ctf}

\subsubsection{Gameplay}
Capture the Flag is played in a gridworld environment segmented by impassable walls and containing two bases. Players are divided into two teams and tasked with capturing the opposing team's flag while simultaneously defending their own flag. Flags spawn within team bases. Players can capture the opposing team's flag by first moving onto it (picking it up) and then returning it to their own base, while their own flag is there. Players observe an egocentric window around themselves. On each step, a player can move around the environment and fire a tag-out beam in the direction they are facing.

At the beginning of an episode, players start with three units of health. A player's health is reduced by one each time it is tagged by a member of the opposing team. Upon reaching zero health, the player is tagged out for 20 steps. After 20 steps, the tagged-out player respawns at their base with three health. Players are rewarded for flag captures and returns, as well as for tagging opposing players.

\begin{figure*}[h]
    \centering
    \begin{subfigure}{0.32\textwidth}
        \centering
        \includegraphics[height=9em]{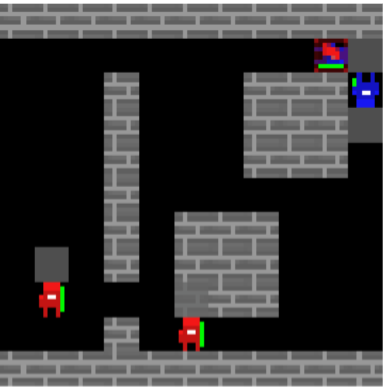}
        \caption{Example observation.}
        \label{fig:app/environments/ctf/obs/example}
    \end{subfigure}
    \hfill
    \begin{subfigure}{0.32\textwidth}
        \centering
        \includegraphics[height=9em]{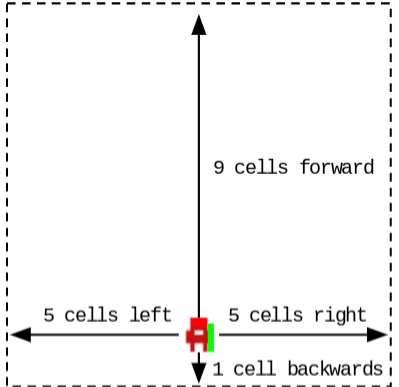}
        \caption{Observation dimensions.}
        \label{fig:app/environments/ctf/obs/dimensions}
    \end{subfigure}
    \hfill
    \begin{subfigure}{0.32\textwidth}
        \centering
        \includegraphics[height=9em]{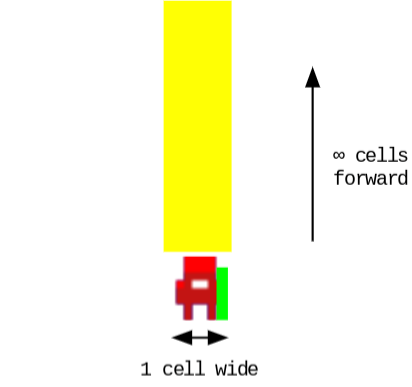}
        \caption{Beam structure.}
        \label{fig:app/environments/ctf/obs/beam}
    \end{subfigure}    
    \caption{(a) Example observation for Capture the Flag. Players observe an $88\times88\times3$ egocentric view of the environment (i.e., $11\times11$ cells with $8\times8\times3$ sprites in each cell). (b) The observation window is offset from the player such that they can always see one cell behind them, five cells either side, and nine cells in their facing direction. (c) The beam is one cell wide and extends infinitely forward from the player (until it hits either a wall or a player).}
    \label{fig:app/environments/ctf/obs}
\end{figure*}

\paragraph{Observations}
Players observe an egocentric view of the environment (see Figure~\ref{fig:app/environments/ctf/obs}), as well as a boolean value for each flag indicating whether it is being held by the opposing team.

\paragraph{Actions}

Players can take one of the following eight actions each step:
\begin{enumerate}
    \item \texttt{No-op}: The player stays in the same position.
    \item \texttt{Move forward}: Moves the player forwards one cell.
    \item \texttt{Move backward}: Moves the player backwards one cell.
    \item \texttt{Move left}: Moves the player left one cell.
    \item \texttt{Move right}: Moves the player right one cell.
    \item \texttt{Turn left}: Rotates the player 90 degrees anti-clockwise.
    \item \texttt{Turn right}: Rotates the player 90 degrees clockwise.
    \item \texttt{Use tag beam}: Fires a yellow beam forwards from the player until it hits either a wall or another player.
\end{enumerate}

The \texttt{use tag beam} action has a cooldown of three steps for each player. If the player tries to use the action during this cooldown period, the outcome is equivalent to the \texttt{no-op} action. Players on the opposing team who are hit by the beam have their health points reduced by one and are tagged out when their health points are reduced to zero. Tagged players are respawned back at their teams' base with three units of health after 20 steps.

\paragraph{Rewards}
Players are rewarded following the Quake III Arena points system presented in Jaderberg et al. \cite{jaderberg2019human}:

\begin{table}[h]
    \centering
    \begin{tabular}[h]{lc}
        \toprule
        \multicolumn{1}{c}{\textbf{Event}} & \textbf{Reward} \\
        \midrule
        I captured the flag & $+6$ \\
        I picked up the flag & $+1$ \\
        I returned the flag & $+1$ \\
        Teammate captured the flag & $+5$ \\
        I tagged opponent with the flag & $+2$ \\
        I tagged opponent without the flag & $+1$ \\
        \bottomrule
    \end{tabular}
    \caption{Event-based rewards given to players in Capture the Flag, following \cite{jaderberg2019human}.}
    \label{tab:app/environments/ctf/rewards}
\end{table}

While tagged out, a player receives all-black visual observations. Players can still receive reward from their teammate capturing the flag while tagged out.

\subsubsection{Procedural generation}

\paragraph{Procedure}
To generate levels for Capture the Flag, we adapt the procedural generation procedure introduced in \cite{jaderberg2019human}:

\begin{enumerate}
    \item Generate an area with an odd width $\in [15, 25]$ and height $\in [9, 15]$. 
    \item Create random sized rectangular rooms within this area.
    \item Fill the space between rooms using the backtracking-maze algorithm (to produce corridors).
    \item Remove deadends and horseshoes in the maze.
    \item Searching from the top left, declare the first room encountered to be the base room.
    \item Add a flag base and three player spawn points to the base room, corresponding to the red team.
    \item Make the level symmetric by taking the left-half of the level and concatenating it with its 180\degree-rotated self. The flags and base on the right side of the level correspond to the blue team.
    \item Check that the level is solvable: both flags can be captured and returned by each team, with the extra constraint that the flags must be at least six cells apart.
\end{enumerate}

\paragraph{Distribution over environmental features}
To demonstrate the diversity of the procedurally generated levels for Capture the Flag, this section introduces four descriptive features, presents their distributions from 100,000 samples, and visualizes the minimum, medium, and maximum level of each distribution. 

The four features are as follows:
\begin{enumerate}
    \item \textbf{Crow Distance:} The euclidean distance between the red flag and blue flag.
    \item \textbf{Path Distance:} The path distance between the red flag and blue flag.
    \item \textbf{Path Complexity:} The crow distance divided by the path distance. 
    \item \textbf{Openness:} The proportion of non-wall cells in the level.
\end{enumerate}

\begin{figure}[ht]
    \centering
    \includegraphics[width=0.94\linewidth]{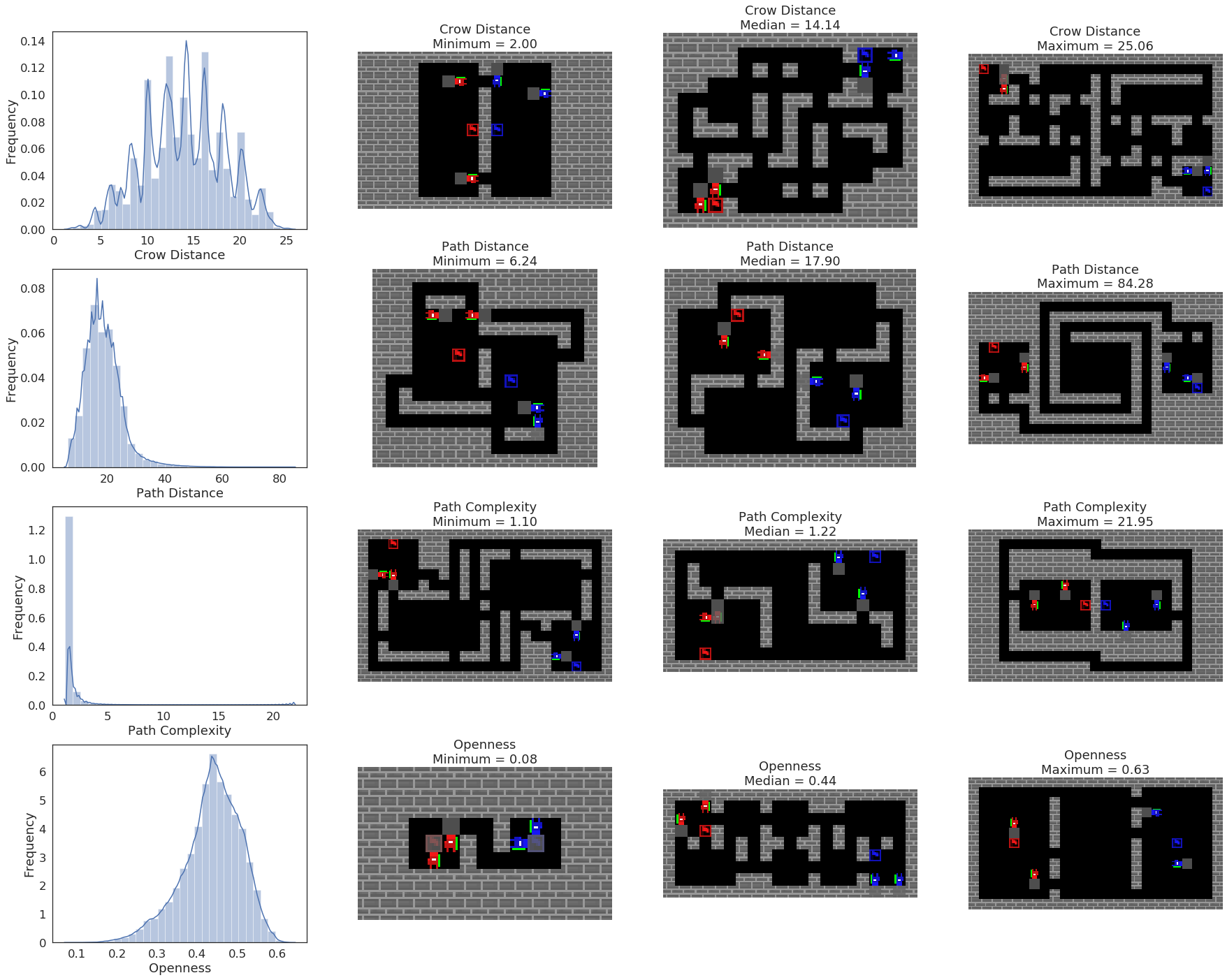}
    \caption{Distribution over environmental features, alongside an example level at the minimum, median, and maximum of these distributions.}
    \label{fig:app/environments/ctf/levels_features}
\end{figure}

\begin{figure}[ht]
    \centering
    \includegraphics[width=0.95\linewidth]{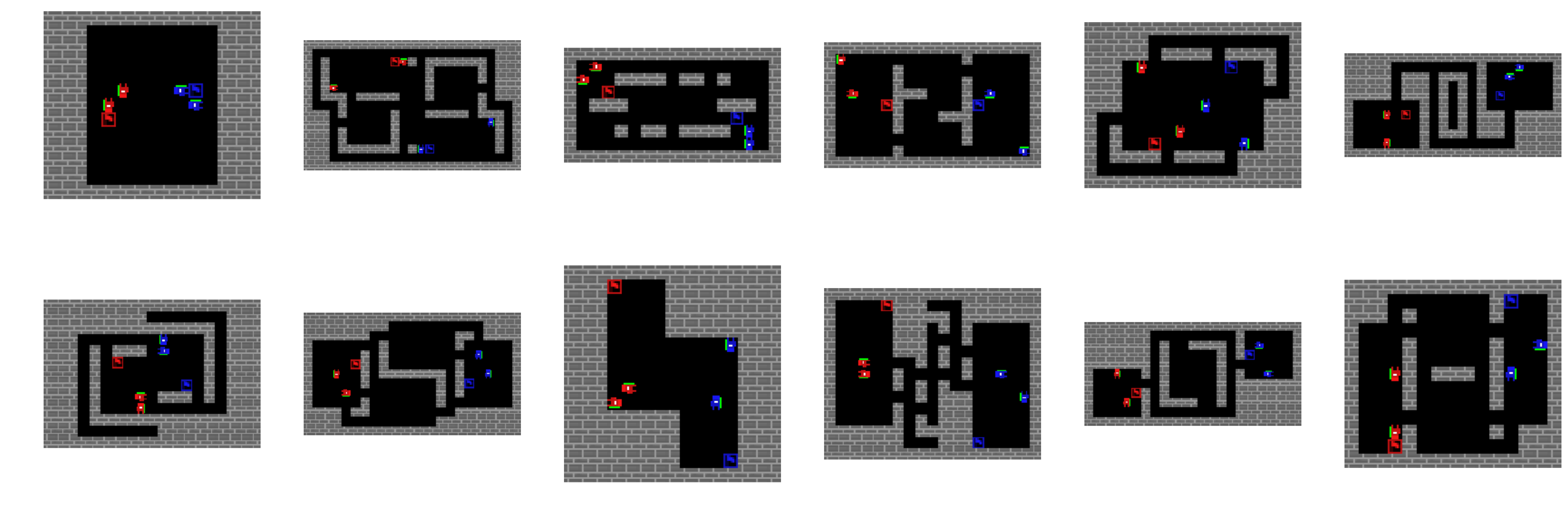}
    \caption{Example levels procedurally generated for the Capture the Flag environment.}
    \label{fig:app/environments/ctf/levels}
\end{figure}

\paragraph{Example levels}
Presented in Figure~\ref{fig:app/environments/ctf/levels} are 12 randomly sampled levels generated by the previously described procedure.

\clearpage
\section{Agents}
\label{app:agents}

\subsection{Training and evaluation}
We use a distributed asynchronous training framework consisting of ``arenas'' that run in parallel to train populations of reinforcement learning agents. In Tables \ref{tab:app/agents/training_details} and Table~\ref{tab:app/agents/training_details_shared}, we provide details of this experimental setup:

\begin{table*}[h]
    \centering
    \begin{tabular}{l r c c c c}
        \hline
         \multicolumn{2}{c}{\textbf{Setting}} & \textbf{HarvestPatch} & \textbf{Traffic Navigation} & \textbf{Overcooked} & \textbf{Capture the Flag} \\
         \hline  
         & Game type & mixed-motive & coordination & common-payoff & competitive \\         
         & \# players (players $n$) & 6 & 8 & 2 & 4 (2 versus 2) \\
         & \# training steps per agent & 1e9 & 5e8 & 3e9 & 3e9 \\
         & training time per experiment & 1 day & 12 hours & 3 days & 3 days \\
         & \# steps per episode & 1000 & 1000 & 540 & 2400 \\         
         \hline
    \end{tabular}
    \caption{Training details for each experiment by environment.}
    \label{tab:app/agents/training_details}
\end{table*}

\begin{table*}[h]
    \centering
    \begin{tabular}{l r c}
        \hline
         \multicolumn{2}{c}{\textbf{Setting}} & \textbf{Value} \\
         \hline  
         & \# agents (population size $N$) & \{1, 2, 4, 8\} for HarvestPatch, Traffic Navigation, Overcooked; \\
         & & \{1, 2, 4, 8, 16\} for Capture the Flag \\
         & \# levels (training levels $L$) & \{1, 1e1, 1e2, 1e3, 1e4\}\\
         & sample agents with replacement? & yes \\         
         & \# arenas & \{200, 400, 800, 1600, 1600\} for $N \in \{1, 2, 4, 8, 16\}$, respectively \\
         & \# GPUs per agent & 1 \\         
         \hline
    \end{tabular}
    \caption{Shared settings for all experiments.}
    \label{tab:app/agents/training_details_shared}
\end{table*}

\subsection{V-MPO}

\subsubsection{Architecture and hyperparameters}

We used the following architecture for V-MPO \cite{song2019v}. The agent's visual observations were first processed through a 3-section ResNet used in \cite{hessel2019multi}. Each section consisted of a convolution and $3 \times 3$ max-pooling operation (stride 2), followed by residual blocks of size 2 (i.e., a convolution followed by a ReLU nonlinearity, repeated twice,
and a skip connection from the input residual block input to the output). The entire stack was passed through one more ReLU nonlinearity. All convolutions had a kernel size of 3 and a stride of 1. The number of channels in each section was (16, 32, 32).

The resulting output was then concatenated with the previous action and reward of the agent (along with any extra observations in Capture the Flag and Traffic Navigation), and processed by a single-layer MLP with 256 hidden units. This was followed by a single-layer LSTM with 256 hidden units (unrolled for 100 steps), and then a separate single-layer MLP with 256 hidden units to produce the action distribution. For the critic, we used a single-layer MLP with 256 hidden units followed by PopArt normalization (ablated in Section~\ref{app:popart}).

To train the agent, we used a discount factor of 0.99, batch sizes of 16, and the Adam optimizer (\cite{kingma2014adam}; learning rate of 0.0001). We configured V-MPO with a target network update period of 100, $k = 0.5$, and an epsilon temperature of 0.1. For PopArt normalization, we used a scale lower bound of $\expnumber{1}{-2}$, an upper bound of $\expnumber{1}{6}$, and a learning rate of $\expnumber{1}{-3}$.

\subsubsection{PopArt normalization}
\label{app:popart}

\begin{figure}[ht]
    \centering
    \includegraphics[height=3.5cm]{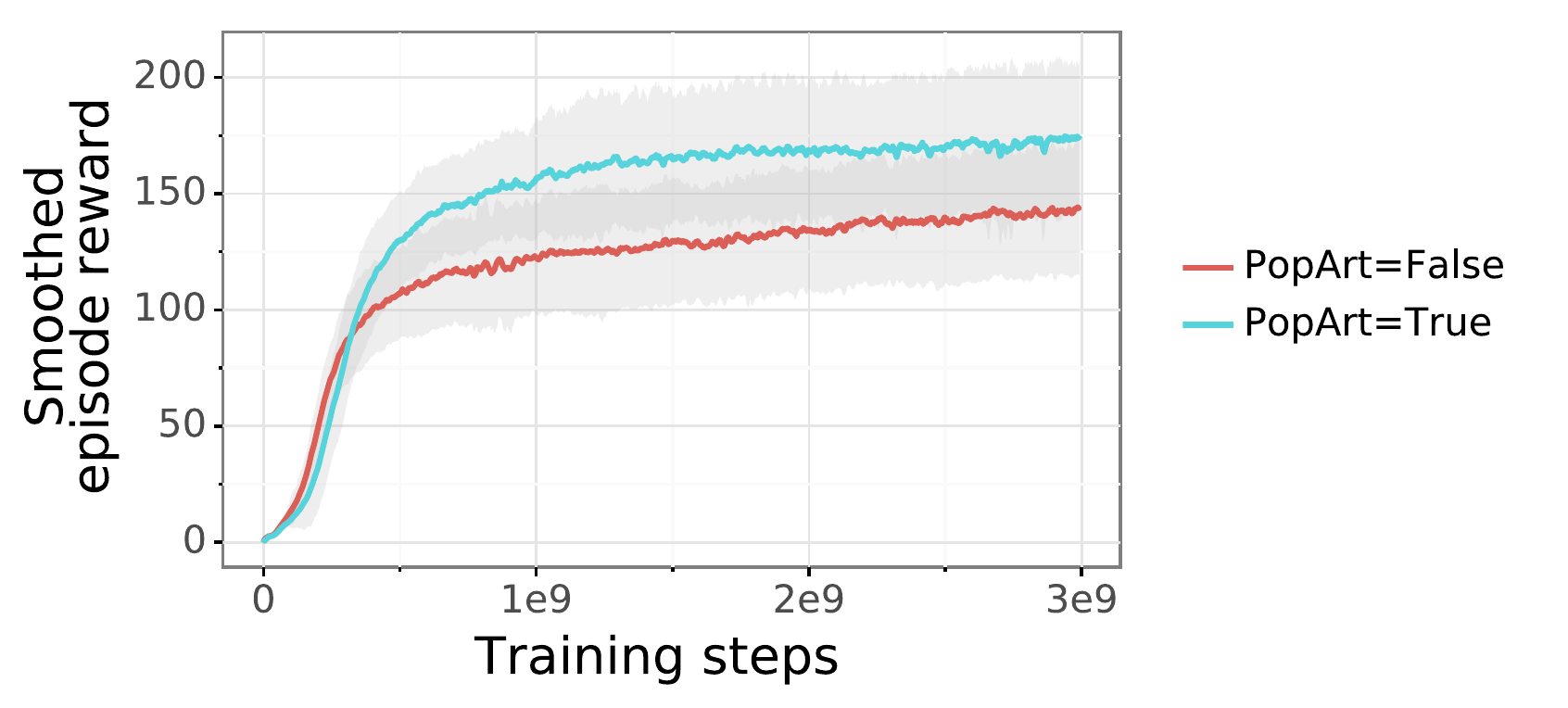}
    \caption{Training curves for V-MPO on Capture the Flag with PopArt normalization enabled (True) and disabled (False). Results averaged over 25 independent runs (five runs per $L \in \{1, \expnumber{1}{1}, \expnumber{1}{2}, \expnumber{1}{3}, \expnumber{1}{4}\}$). Error bands reflect $\pm 1$ standard deviation.}
    \label{fig:app/agents/vmpo/popart}
\end{figure}

PopArt normalization is typically used in multi-task settings \cite{hessel2019multi}. While we did not investigate multi-task settings in this work, we found that including PopArt (as in \cite{song2019v}) led to improved performance on the procedurally generated environments considered here. In Figure~\ref{fig:app/agents/vmpo/popart}, we show the reward curves across training on both the training and test levels with PopArt normalization enabled (PopArt = True) and disabled (PopArt = False). As can be seen, PopArt consistently leads to a higher reward and therefore more performant agents. Consequently, we used PopArt for all agents.

\subsubsection{Social Value Orientation}
\label{app:svo}
For the ``identical'' populations of Social Value Orentation agents, we set the \textit{target reward angle} $\theta$ hyperparameter with an identical distribution across agents: $\theta \in \{45\degree, 45\degree, 45\degree, 45\degree\}$. For the ``heterogeneous'' populations of Social Value Orientation agents, $\theta$ is set with a diverse distribution across agents: $\theta \in \{0\degree, 30\degree, 60\degree, 90\degree\}$.

\clearpage
\section{Methods}

\subsection{Expected action variation}
\label{app:eav}

For reproducibility, we include pseudocode to calculate expected action variation (Algorithm \ref{alg:behavioral_diversity}). Starting from a set of $k$ populations of interest trained on a shared set of levels $\mathcal{L}$, we simulate a number of episodes within each population to generate a representative pool of agent states (Algorithm \ref{alg:get_pool}). We then approximate action-policy distributions over these states for each agent in each population (Algorithm \ref{alg:get_hists}). Finally, for each population, we compare the action-policy distributions between each pair of agents within the population to calculate that population's expected action variation (Algorithm \ref{alg:tvd}).

Because the environments that we consider in this paper use discrete action spaces, Algorithm \ref{alg:tvd} uses total variation distance to estimate pairwise divergence between action-policy distributions. For environments with continuous action spaces, pairwise divergence could be estimated with earth mover's distance instead.

\begin{algorithm}
\caption{Calculate \textit{expected action variation} for multiple populations}
\label{alg:behavioral_diversity}
\begin{algorithmic}[1]
    \STATE Input: Set of populations of interest $\mathcal{P} = \{P_0, \dots, P_k\}$, shared set of training levels $\mathcal{L}$, number of episodes to simulate $E$, number of states to draw from each episode $J$, number of players for each episode $n$, number of action samples $R$
    \STATE $\mathcal{S} \leftarrow \texttt{get\_pool}(\mathcal{P}, \mathcal{L}, E, J, n)$
    \STATE $\texttt{policy\_dists}_{\mathcal{P}, \mathcal{S}} \leftarrow \texttt{approximate\_policy\_dists}(\mathcal{P}, R, \mathcal{S})$
    \STATE $\texttt{EAV}_{\mathcal{P}} \leftarrow 0 \,\, \forall P \in \mathcal{P}$
    \FORALL{$P \in P$}
        \STATE $\texttt{EAV}_{P} \leftarrow \texttt{intra\_population\_variation}(P, \mathcal{S}, \texttt{policy\_dists}_{\mathcal{P}, \mathcal{S}})$
    \ENDFOR
    \STATE Return: $\texttt{EAV}_{\mathcal{P}}$
\end{algorithmic}
\end{algorithm}

\begin{algorithm}
\caption{\texttt{get\_pool} Generate representative pool of states for policy comparison}
\label{alg:get_pool}
\begin{algorithmic}[1]
    \STATE Input: Set of populations of interest $\mathcal{P} = \{P_0, \dots, P_k\}$, shared set of training levels $\mathcal{L}$, number of episodes to simulate $E$, number of states to draw from each episode $J$, number of players for each episode $n$
    \STATE $\mathcal{S} \leftarrow \{ \}$
    \FORALL{$P \in \mathcal{P}$}
        \FOR{$e = 1:E$}
            \STATE $\vec{A} \leftarrow \{ \}$
                \FOR{$i = 1:n$}
                    \STATE $A_i \sim P$
                    \STATE $\vec{A} \leftarrow \vec{A} \cup A_i$
                \ENDFOR
            \STATE $l \sim \mathcal{L}$
            \STATE $\{(s,a,r,s')_t\} \leftarrow \texttt{sim}(\vec{A}, l)$
            \STATE $T = \vert \{(s,a,r,s')_t\} \vert$
            \FOR{$j = 1:J$}
                \STATE $t \sim \{1, \ldots , T\}$
                \STATE $\mathcal{S} \leftarrow \mathcal{S} \cup \{(s_t, l)\}$
            \ENDFOR
        \ENDFOR
    \ENDFOR
    \STATE Return: $\mathcal{S}$
\end{algorithmic}
\end{algorithm}

\begin{algorithm}
\caption{\texttt{approximate\_policy\_dists} Approximate action-policy distributions}
\label{alg:get_hists}
\begin{algorithmic}[1]
    \STATE Input: Set of populations of interest $\mathcal{P} = \{P_0, \dots, P_k\}$, number of action samples $R$, state pool $\mathcal{S}$
    \FORALL{$P \in \mathcal{P}$}
        \FORALL{$A \in P$}
            \FORALL{$(s, l) \in \mathcal{S}$}
                \STATE $\texttt{hist}_{A,(s,l)} \leftarrow 0$
                \FOR{$i = 1: R$}
                    \STATE $a \sim \pi_{A}(s,l)$
                    \STATE $\texttt{hist}_{A, (s,l)}(a) \leftarrow \texttt{hist}_{A, (s,l)}(a) + 1$
                \ENDFOR
            \STATE $\texttt{policy\_dists}_{A,(s,l)} \leftarrow \texttt{hist}_{A, (s,l)} / R$
            \ENDFOR
        \ENDFOR
    \ENDFOR
    \STATE Return: $\texttt{policy\_dists}$
\end{algorithmic}
\end{algorithm}

\begin{algorithm}
\caption{\texttt{intra\_population\_variation} Compute normalized total variation distance between all empirical action-probability distributions}
\label{alg:tvd}
\begin{algorithmic}[1]
    \STATE Input: Population of interest $P$, state pool $\mathcal{S}$, action-probability distributions for agents $\texttt{policy\_dists}$
    \STATE $\texttt{paired} \leftarrow \{ \}$
    \STATE $\texttt{TVD} \leftarrow 0$
    \FORALL{$A_1 \in P$}
        \FORALL{$A_2 \in P$}
            \IF{$A_1 \ne A_2$ and $\lnot (\{\{A_1, A_2\}\} \cap \texttt{paired})$}
                \FORALL{$(s,l) \in \mathcal{S}$}
                    \STATE $\texttt{tvd} \leftarrow \sum \vert \texttt{policy\_dists}_{A_1, (s,l)} - \texttt{policy\_dists}_{A_2, (s,l)} \vert$
                    \STATE $\texttt{norm\_tvd} \leftarrow \texttt{norm\_tvd} + \texttt{tvd} / \vert \mathcal{S} \vert$
                \ENDFOR
                \STATE $\texttt{paired} \leftarrow \texttt{paired} \cup \{\{A_1, A_2\}\}$
            \ENDIF
        \ENDFOR
    \ENDFOR
    \STATE Return: $\texttt{norm\_tvd} / \vert \texttt{paired} \vert$
\end{algorithmic}
\end{algorithm}

\newpage
\subsection{Calculating Elo rating}
\label{app:elo}
To calculate the Elo rating of each trained population, we evaluate every possible pairing of trained populations against each another with 100 matches per pairing. For each of these 100 matches, two agents are randomly sampled from the first population to form the red team, and two from the second population to form the blue team (sampling with replacement).

After these matches are completed, we iteratively calculate the Elo rating of each population using the following procedure on each match result (looping over all match results until convergence):

\begin{algorithm}
\caption{Update Elo rating}
\label{alg:elo}
\begin{algorithmic}[1]
    \STATE Input: Step size of Elo rating update $K$, population $i$ with Elo rating $r_i$ and match score $s_i$, population $j$ with Elo rating $r_j$ and match score $s_j$
    \STATE $s \leftarrow ($sign$(s_i, s_j) + 1) / 2$
    \STATE $s_{elo} \leftarrow 1/(1 + 10^{(r_j-r_i)/400})$ 
    \STATE $r_{i} \leftarrow r_{i} + K(s-s_{elo})$
    \STATE $r_{j} \leftarrow r_{j} - K(s-s_{elo})$
    \STATE return $r_{i}, r_{j}$
\end{algorithmic}
\end{algorithm}

\noindent where we initialised the rating of each population to 1000 and set $K=2$.

\clearpage
\section{Additional Results}
This section presents additional results and statistical analyses supporting the results in the main text.

For each ANOVA we conducted, we compare the categorical levels of the independent variable in pairs and apply Tukey's honestly significant differences (HSD) method to evaluate which pairs differ significantly \cite{tukey1949comparing}. Tukey's method adjusts the raw $p$-values to account for the increased probability of false positives when running multiple independent statistical tests.

\subsection{Environment Diversity}

\subsubsection{Generalization Gap Analysis}

\textbf{HarvestPatch}

\begin{table}[ht]
    \centering
    \begin{tabular}{ c c c c }
        \hline
         & \multicolumn{2}{c}{Individual reward on:} & Gen. gap \\
        $L$ & Train levels & Test levels & (train $-$ test) \\ 
        \hline
        $1$ & $152.4$ ($75.0$) & $96.9$ ($21.7$) & $55.5$ ($62.3$) \\
        $\expnumber{1}{1}$ & $115.1$ ($24.2$) & $101.0$ ($12.7$) & $14.2$ ($30.1$) \\
        $\expnumber{1}{2}$ & $113.5$ ($8.9$) & $110.0$ ($15.3$) & $3.5$ ($17.3$) \\
        $\expnumber{1}{3}$ & $114.1$ ($7.7$) & $109.8$ ($6.8$) & $4.3$ ($8.7$) \\
        $\expnumber{1}{4}$ & $114.6$ ($10.5$) & $112.5$ ($8.3$) & $2.2$ ($11.9$) \\
        \hline
    \end{tabular} \medskip
    \caption{Full HarvestPatch results from Figure~\ref{fig:511/env/generalization_gap}a: Performance metrics for environment diversity experiments. Mean values (and standard deviations, reported in parentheses) are calculated over 10 independent runs.}
    \label{tab:app/results/env/hp/performance}
\end{table}

\begin{table}[ht]
    \centering
    \begin{tabular}{ c c c c }
        \hline
        $L_a$ & $L_b$ & Mean difference & Adjusted $p$-value \\ 
        \hline
        $1$ & $\expnumber{1}{1}$ & $41.3$ & $5.0 \times 10^{-2}$ \\ 
        $1$ & $\expnumber{1}{2}$ & $52.1$ & $7.2 \times 10^{-3}$ \\ 
        $1$ & $\expnumber{1}{3}$ & $51.2$ & $8.6 \times 10^{-3}$ \\
        $1$ & $\expnumber{1}{4}$ & $53.4$ & $5.6 \times 10^{-3}$ \\
        $\expnumber{1}{1}$ & $\expnumber{1}{2}$ & $10.7$ & $0.95$ \\ 
        $\expnumber{1}{1}$ & $\expnumber{1}{3}$ & $9.8$ & $0.96$ \\ 
        $\expnumber{1}{1}$ & $\expnumber{1}{4}$ & $12.0$ & $0.92$ \\
        $\expnumber{1}{2}$ & $\expnumber{1}{3}$ & $-0.9$ & $1.00$ \\ 
        $\expnumber{1}{2}$ & $\expnumber{1}{4}$ & $1.3$ & $1.00$ \\
        $\expnumber{1}{3}$ & $\expnumber{1}{4}$ & $2.2$ & $1.00$ \\
        \hline
    \end{tabular} \medskip
    \caption{Pairwise comparisons of generalization gaps between level set sizes $L \in \{1, \expnumber{1}{2}, \expnumber{1}{3}, \expnumber{1}{4}\}$, calculated with Tukey's HSD method. Positive ``Mean difference'' values indicate that training with $L_a$ resulted in a higher generalization gap than training with $L_b$.}
    \label{tab:app/results/env/hp/perf_tukeys}
\end{table}

\newpage
\textbf{Traffic Navigation}

\begin{table}[ht]
    \centering
    \begin{tabular}{ c c c c }
        \hline
         & \multicolumn{2}{c}{Individual reward on:} & Gen. gap \\
        $L$ & Train levels & Test levels & (train $-$ test) \\ 
        \hline
        $1$ & $65.5$ ($7.9$) & $6.3$ ($7.1$) & $59.2$ ($8.3$) \\
        $\expnumber{1}{1}$ & $58.7$ ($2.5$) & $28.4$ ($3.6$) & $30.3$ ($3.9$) \\
        $\expnumber{1}{2}$ & $59.8$ ($1.2$) & $51.4$ ($2.5$) & 
        $8.4$ ($2.3$) \\
        $\expnumber{1}{3}$ & $58.7$ ($3.5$) & $58.5$ ($1.5$) & $0.2$ ($3.5$) \\
        $\expnumber{1}{4}$ & $59.0$ ($0.7$) & $58.9$ ($0.9$) & $0.1$ ($0.6$) \\
        \hline
    \end{tabular} \medskip
    \caption{Full Traffic Navigation results from Figure~\ref{fig:511/env/generalization_gap}b: Performance metrics for environment diversity experiments. Mean values (and standard deviations, reported in parentheses) are calculated over 10 independent runs.}
    \label{tab:app/results/env/tn/performance}
\end{table}

\begin{table}[ht]
    \centering
    \begin{tabular}{ c c c c }
        \hline
        $L_a$ & $L_b$ & Mean difference & Adjusted $p$-value \\ 
        \hline
        $1$ & $\expnumber{1}{1}$ & $28.9$ & $2.4 \times 10^{-13}$ \\ 
        $1$ & $\expnumber{1}{2}$ & $50.8$ & $2.4 \times 10^{-13}$ \\ 
        $1$ & $\expnumber{1}{3}$ & $59.0$ & $2.4 \times 10^{-13}$ \\
        $1$ & $\expnumber{1}{4}$ & $59.1$ & $2.4 \times 10^{-13}$ \\
        $\expnumber{1}{1}$ & $\expnumber{1}{2}$ & $21.9$ & $6.2 \times 10^{-13}$ \\ 
        $\expnumber{1}{1}$ & $\expnumber{1}{3}$ & $30.1$ & $2.4 \times 10^{-13}$ \\ 
        $\expnumber{1}{1}$ & $\expnumber{1}{4}$ & $30.2$ & $2.4 \times 10^{-13}$ \\
        $\expnumber{1}{2}$ & $\expnumber{1}{3}$ & $8.2$ & $1.8 \times 10^{-3}$ \\ 
        $\expnumber{1}{2}$ & $\expnumber{1}{4}$ & $8.3$ & $1.5 \times 10^{-3}$ \\
        $\expnumber{1}{3}$ & $\expnumber{1}{4}$ & $0.1$ & $1.00$ \\
        \hline
    \end{tabular} \medskip
    \caption{Pairwise comparisons of generalization gaps between level set sizes $L \in \{1, \expnumber{1}{2}, \expnumber{1}{3}, \expnumber{1}{4}\}$, calculated with Tukey's HSD method. Positive ``Mean difference'' values indicate that training with $L_a$ resulted in a higher generalization gap than training with $L_b$.}
    \label{tab:app/results/env/tn/perf_tukeys}
\end{table}

\textbf{Overcooked}

\begin{table}[h]
    \centering
    \begin{tabular}{ c c c c }
        \hline
         & \multicolumn{2}{c}{Individual reward on:} & Gen. gap \\
        $L$ & Train levels & Test levels & (train $-$ test) \\ 
        \hline
        $1$ & $359.7$ ($103.3$) & $0.4$ ($0.3$) & $359.2$ ($103.2$) \\
        $\expnumber{1}{1}$ & $345.5$ ($36.7$) & $2.8$ ($1.4$) & $342.8$ ($36.1$) \\
        $\expnumber{1}{2}$ & $351.8$ ($24.4$) & $103.0$ ($16.3$) & $248.8$ ($16.4$) \\
        $\expnumber{1}{3}$ & $296.6$ ($23.3$) & $244.6$ ($20.8$) & $52.0$ ($9.6$) \\
        $\expnumber{1}{4}$ & $284.4$ ($13.4$) & $274.0$ ($15.4$) & $10.4$ ($9.3$) \\        
        \hline
    \end{tabular} \medskip
    \caption{Full Overcooked results from Figure~\ref{fig:511/env/generalization_gap}c: Performance metrics for environment diversity experiments. Mean values (and standard deviations, reported in parentheses) are calculated over 10 independent runs.}
    \label{tab:app/results/env/oc/performance}
\end{table}

\begin{table}[h]
    \centering
    \begin{tabular}{ c c c c }
        \hline
        $L_a$ & $L_b$ & Mean difference & Adjusted $p$-value \\ 
        \hline
        $1$ & $\expnumber{1}{1}$ & $16.5$ & $0.95$ \\ 
        $1$ & $\expnumber{1}{2}$ & $110.4$ & $1.0 \times 10^{-4}$ \\ 
        $1$ & $\expnumber{1}{3}$ & $307.2$ & $2.4 \times 10^{-13}$ \\
        $1$ & $\expnumber{1}{4}$ & $348.8$ & $2.4 \times 10^{-13}$ \\
        $\expnumber{1}{1}$ & $\expnumber{1}{2}$ & $94.0$ & $1.1 \times 10^{-3}$ \\ 
        $\expnumber{1}{1}$ & $\expnumber{1}{3}$ & $290.8$ & $2.4 \times 10^{-13}$ \\ 
        $\expnumber{1}{1}$ & $\expnumber{1}{4}$ & $332.4$ & $2.4 \times 10^{-13}$ \\
        $\expnumber{1}{2}$ & $\expnumber{1}{3}$ & $196.8$ & $2.1 \times 10^{-10}$ \\ 
        $\expnumber{1}{2}$ & $\expnumber{1}{4}$ & $238.4$ & $8.6 \times 10^{-13}$ \\
        $\expnumber{1}{3}$ & $\expnumber{1}{4}$ & $41.6$ & $0.35$ \\
        \hline
    \end{tabular} \medskip
    \caption{Pairwise comparisons of generalization gaps between level set sizes $L \in \{1, \expnumber{1}{2}, \expnumber{1}{3}, \expnumber{1}{4}\}$, calculated with Tukey's HSD method. Positive ``Mean difference'' values indicate that training with $L_a$ resulted in a higher generalization gap than training with $L_b$.}
    \label{tab:app/results/env/oc/perf_tukeys}
\end{table}

\newpage
\textbf{Capture the Flag}

\begin{table}[ht]
    \centering
    \begin{tabular}{ c c c c }
        \hline
         & \multicolumn{2}{c}{Individual reward on:} & Gen. gap \\
        $L$ & Train levels & Test levels & (train $-$ test) \\ 
        \hline
        $1$ & $213.6$ ($165.4$) & $3.0$ ($2.1$) & $210.6$ ($164.0$) \\
        $\expnumber{1}{1}$ & $222.1$ ($65.2$) & $45.0$ ($20.3$) & $177.1$ ($59.2$) \\
        $\expnumber{1}{2}$ & $174.4$ ($19.4$) & $128.8$ ($20.3$) & $45.6$ ($10.3$) \\
        $\expnumber{1}{3}$ & $159.0$ ($47.6$) & $147.7$ ($23.3$) & $11.3$ ($38.1$) \\
        $\expnumber{1}{4}$ & $161.7$ ($20.9$) & $150.5$ ($21.5$) & $11.2$ ($5.5$) \\
        \hline
    \end{tabular} \medskip
    \caption{Full Capture the Flag results from Figure~\ref{fig:511/env/generalization_gap}d: Performance metrics for environment diversity experiments. Mean values (and standard deviations, reported in parentheses) are calculated over nine independent runs.}
    \label{tab:app/results/env/ctf/performance}
\end{table}

\begin{table}[h]
    \centering
    \begin{tabular}{ c c c c }
        \hline
        $L_a$ & $L_b$ & Mean difference & Adjusted $p$-value \\ 
        \hline
        $1$ & $\expnumber{1}{1}$ & $33.5$ & $0.90$ \\ 
        $1$ & $\expnumber{1}{2}$ & $165.0$ & $7.6 \times 10^{-4}$ \\ 
        $1$ & $\expnumber{1}{3}$ & $199.3$ & $4.5 \times 10^{-5}$ \\
        $1$ & $\expnumber{1}{4}$ & $199.4$ & $4.4 \times 10^{-5}$ \\
        $\expnumber{1}{1}$ & $\expnumber{1}{2}$ & $131.5$ & $1.0 \times 10^{-2}$ \\ 
        $\expnumber{1}{1}$ & $\expnumber{1}{3}$ & $165.7$ & $7.2 \times 10^{-4}$ \\ 
        $\expnumber{1}{1}$ & $\expnumber{1}{4}$ & $165.9$ & $7.1 \times 10^{-4}$ \\
        $\expnumber{1}{2}$ & $\expnumber{1}{3}$ & $34.3$ & $0.89$ \\ 
        $\expnumber{1}{2}$ & $\expnumber{1}{4}$ & $34.4$ & $0.89$ \\
        $\expnumber{1}{3}$ & $\expnumber{1}{4}$ & $0.2$ & $1.0$ \\
        \hline
    \end{tabular} \medskip
    \caption{Pairwise comparisons of generalization gaps between level set sizes $L \in \{1, \expnumber{1}{2}, \expnumber{1}{3}, \expnumber{1}{4}\}$, calculated with Tukey's HSD method. Positive ``Mean difference'' values indicate that training with $L_a$ resulted in a higher generalization gap than training with $L_b$.}
    \label{tab:app/results/env/ctf/perf_tukeys}
\end{table}

\newpage
\subsubsection{Cross-Play Evaluation}

\textbf{Capture the Flag}

\begin{figure*}[h]
    \centering
    \begin{subfigure}{0.35\textwidth}
        \centering
        \includegraphics[width=0.75\linewidth]{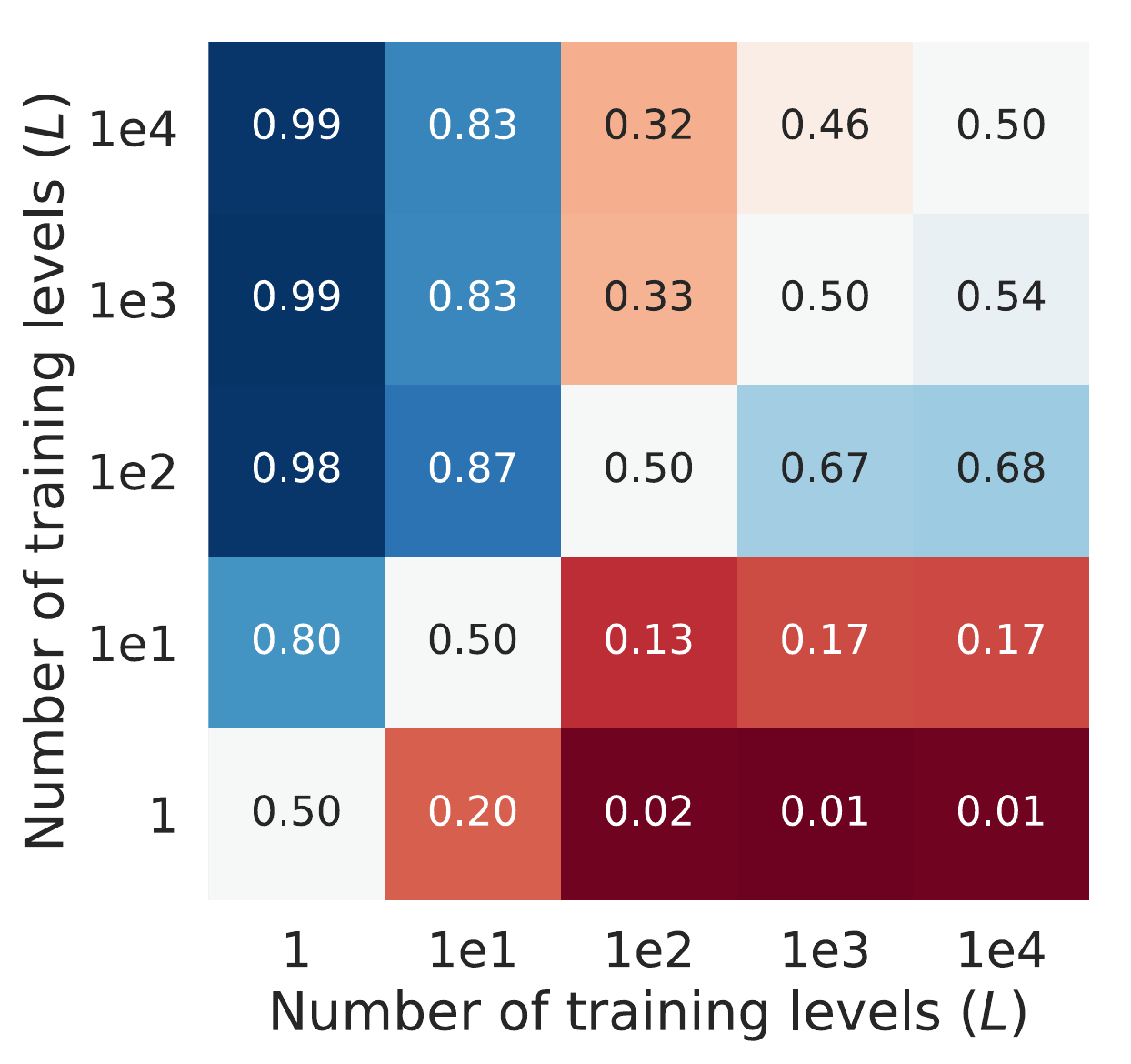}
        \caption{Win matrix on training level.}
    \end{subfigure}
    \begin{subfigure}{0.35\textwidth}
        \centering
        \includegraphics[width=0.75\linewidth]{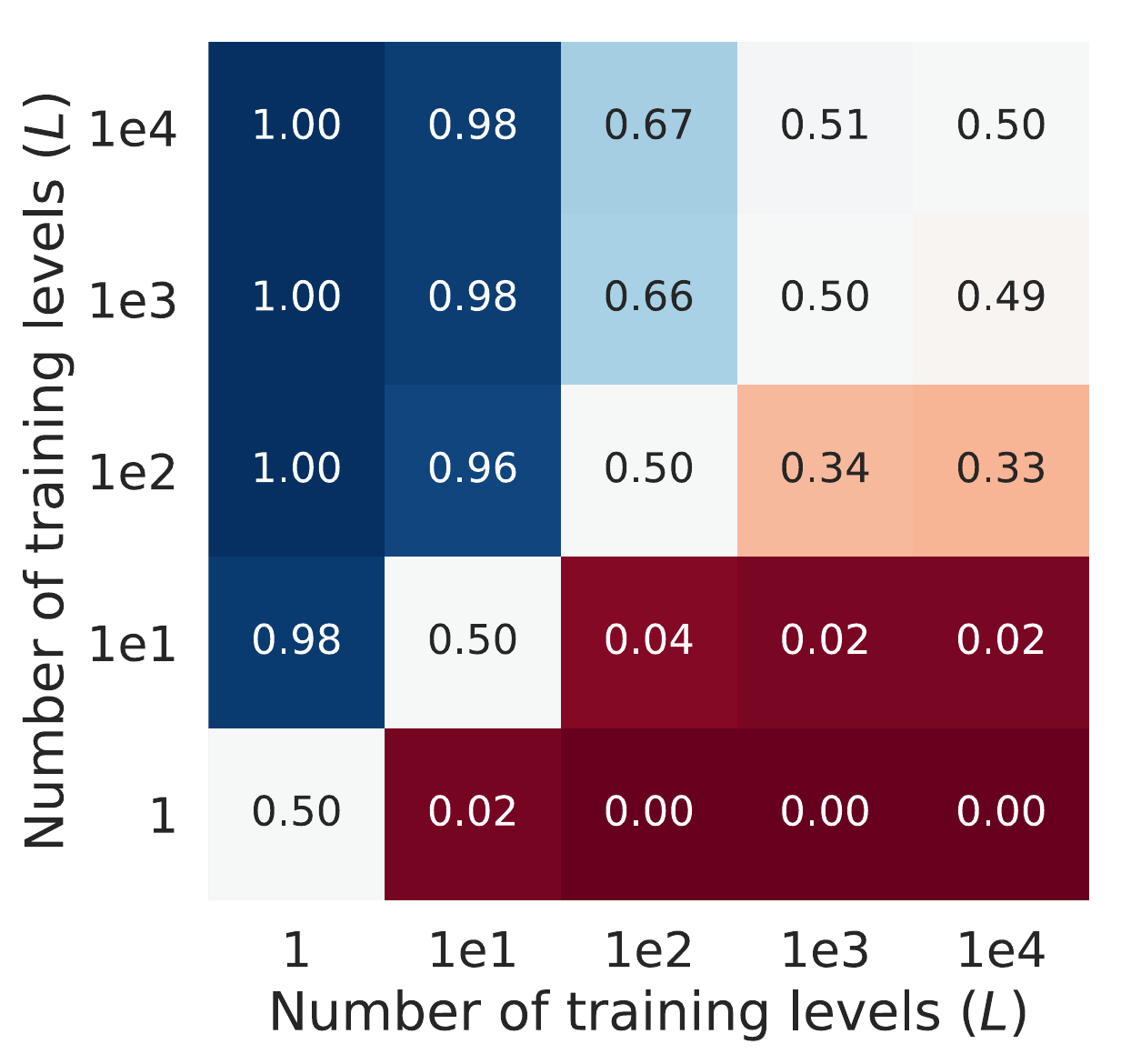}
        \caption{Win matrix on test set.}        
    \end{subfigure}
    \caption{Win matrices corresponding to the Elo ratings presented in Table~\ref{tab:512/env/cross_play/ctf/elos}:  Each win matrix contains the win rates of populations trained on the row value of $L$ over those trained on the column value of $L$.}
    \label{fig:app/results/env/ctf/win_matrix}
\end{figure*}

\subsection{Population Diversity}

\subsubsection{Expected Action Variation}

\textbf{All environments}

\begin{table*}[h]
    \centering
    \begin{tabular}{ c c c c c }
        \hline
         & \multicolumn{4}{c}{Expected Action Variation} \\
        $N$ & HarvestPatch & Traffic Navigation & Overcooked & Capture the Flag \\ 
        \hline
        $1$ & $0.00$ ($0.00$) & $0.00$ ($0.00$) & $0.00$ ($0.00$) & $0.00$ ($0.00$) \\
        $2$ & $0.20$ ($0.03$) & $0.08$ ($0.03$) & $0.15$ ($0.03$) & $0.19$ ($0.03$) \\
        $4$ & $0.36$ ($0.04$) & $0.13$ ($0.02$) & $0.36$ ($0.05$) & $0.34$ ($0.01$) \\
        $8$ & $0.49$ ($0.01$) & $0.18$ ($0.01$) & $0.54$ ($0.07$) & $0.44$ ($0.01$) \\
        $16$ & -- & -- & -- & $0.48$ ($0.01$) \\
        \hline
    \end{tabular} \medskip
    \caption{Expected action variation for population diversity experiments in each environment. Mean values (and standard deviations, reported in parentheses) are calculated over five independent runs.}
    \label{tab:app/results/pop/eav}
\end{table*}

\textbf{HarvestPatch}

\begin{table}[h]
    \centering
    \begin{tabular}{ c c c c }
        \hline
        $N_a$ & $N_b$ & Mean difference & Adjusted $p$-value \\ 
        \hline
        $1$ & $2$ & $-0.20$ & $5.4 \times 10^{-9}$ \\ 
        $1$ & $4$ & $-0.36$ & $6.8 \times 10^{-13}$ \\
        $1$ & $8$ & $-0.49$ & $2.4 \times 10^{-14}$ \\
        $2$ & $4$ & $-0.17$ & $6.7 \times 10^{-8}$ \\ 
        $2$ & $8$ & $-0.29$ & $1.5 \times 10^{-11}$ \\ 
        $4$ & $8$ & $-0.13$ & $2.8 \times 10^{-6}$ \\
        \hline
    \end{tabular} \medskip
    \caption{Pairwise comparisons of expected action variation between population sizes $N \in \{1, 2, 4, 8\}$, calculated with Tukey's HSD method. Positive ``Mean difference'' values indicate that training with $N_a$ resulted in higher behavioral diversity than training with $N_b$.}
    \label{tab:app/results/pop/hp/eav_tukeys}
\end{table}

\newpage
\textbf{Traffic Navigation}

\begin{table}[h]
    \centering
    \begin{tabular}{ c c c c }
        \hline
        $N_a$ & $N_b$ & Mean difference & Adjusted $p$-value \\ 
        \hline
        $1$ & $2$ & $-0.08$ & $1.1 \times 10^{-4}$ \\ 
        $1$ & $4$ & $-0.13$ & $1.8 \times 10^{-7}$ \\
        $1$ & $8$ & $-0.18$ & $1.3 \times 10^{-9}$ \\
        $2$ & $4$ & $-0.05$ & $6.1 \times 10^{-3}$ \\ 
        $2$ & $8$ & $-0.10$ & $3.1 \times 10^{-6}$ \\ 
        $4$ & $8$ & $-0.05$ & $4.4 \times 10^{-3}$ \\
        \hline
    \end{tabular} \medskip
    \caption{Pairwise comparisons of expected action variation between population sizes $N \in \{1, 2, 4, 8\}$, calculated with Tukey's HSD method. Positive ``Mean difference'' values indicate that training with $N_a$ resulted in higher behavioral diversity than training with $N_b$.}
    \label{tab:app/results/pop/tn/eav_tukeys}
\end{table}

\textbf{Overcooked}

\begin{table}[h]
    \centering
    \begin{tabular}{ c c c c }
        \hline
        $N_a$ & $N_b$ & Mean difference & Adjusted $p$-value \\ 
        \hline
        $1$ & $2$ & $-0.15$ & $5.1 \times 10^{-4}$ \\ 
        $1$ & $4$ & $-0.36$ & $1.1 \times 10^{-8}$ \\
        $1$ & $8$ & $-0.54$ & $2.1 \times 10^{-11}$ \\
        $2$ & $4$ & $-0.20$ & $1.9 \times 10^{-5}$ \\ 
        $2$ & $8$ & $-0.39$ & $3.3 \times 10^{-9}$ \\ 
        $4$ & $8$ & $-0.18$ & $7.3 \times 10^{-5}$ \\
        \hline
    \end{tabular} \medskip
    \caption{Pairwise comparisons of expected action variation between population sizes $N \in \{1, 2, 4, 8\}$, calculated with Tukey's HSD method. Positive ``Mean difference'' values indicate that training with $N_a$ resulted in higher behavioral diversity than training with $N_b$.}
    \label{tab:app/results/pop/oc/eav_tukeys}
\end{table}

\textbf{Capture the Flag}

\begin{table}[h]
    \centering
    \begin{tabular}{ c c c c }
        \hline
        $N_a$ & $N_b$ & Mean difference & Adjusted $p$-value \\ 
        \hline
        $1$ & $2$ & $-0.19$ & $2.7 \times 10^{-12}$ \\ 
        $1$ & $4$ & $-0.34$ & $1.9 \times 10^{-14}$ \\
        $1$ & $8$ & $-0.44$ & $1.9 \times 10^{-14}$ \\ 
        $1$ & $16$ & $-0.48$ & $1.9 \times 10^{-14}$ \\ 
        $2$ & $4$ & $-0.15$ & $1.4 \times 10^{-10}$ \\ 
        $2$ & $8$ & $-0.26$ & $2.8 \times 10^{-14}$ \\ 
        $2$ & $16$ & $-0.30$ & $2.0 \times 10^{-14}$ \\ 
        $4$ & $8$ & $-0.10$ & $9.1 \times 10^{-8}$ \\ 
        $4$ & $16$ & $-0.14$ & $3.6 \times 10^{-10}$ \\ 
        $8$ & $16$ & $-0.04$ & $1.5 \times 10^{-2}$ \\ 
        \hline
    \end{tabular} \medskip
    \caption{Pairwise comparisons of expected action variation between population sizes $N \in \{1, 2, 4, 8, 16\}$, calculated with Tukey's HSD method. Positive ``Mean difference'' values indicate that training with $N_a$ resulted in higher behavioral diversity than training with $N_b$.}
    \label{tab:app/results/pop/ctf/eav_tukeys}
\end{table}

\newpage
\textbf{Intrinsic Motivation and Behavioral Diversity}

\begin{table*}[h]
    \centering
    \begin{tabular}{ c c }
        \hline
        Social Value Orientation & Expected Action Variation \\ 
        \hline
        None & 0.30 (0.02) \\
        Identical & 0.30 (0.03) \\
        Diverse & 0.37 (0.01) \\
        \hline
    \end{tabular} \medskip
    \caption{Expected action variation for various distributions of SVO in HarvestPatch. Experiments are run with $N = 4$ and $L = \expnumber{1}{3}$. Mean values (and standard deviations, reported in parentheses) are calculated over five independent runs.}
    \label{tab:app/results/pop/eav/svo}
\end{table*}

\begin{table}[h]
    \centering
    \begin{tabular}{ c c c c }
        \hline
        $\mathrm{SVO}_a$ & $\mathrm{SVO}_b$ & Mean difference & Adjusted $p$-value \\ 
        \hline
        None & Identical & $0.00$ & $0.99$ \\ 
        None & Diverse & $-0.07$ & $6.2 \times 10^{-4}$ \\
        Identical & Diverse & $-0.07$ & $4.9 \times 10^{-4}$ \\
        \hline
    \end{tabular} \medskip
    \caption{Pairwise comparisons of expected action variation between different population distributions of SVO, calculated with Tukey's HSD method. Positive ``Mean difference'' values indicate that training with $\mathrm{SVO}_a$ resulted in higher behavioral diversity than training with $\mathrm{SVO}_b$.}
    \label{tab:app/results/pop/eav/svo_tukeys}
\end{table}

\subsubsection{Cross-Play Evaluation}

\textbf{HarvestPatch}

\begin{table}[h]
    \centering
    \begin{tabular}{ c c }
        \hline
         & \multicolumn{1}{c}{Individual reward on:} \\
        $N$ & Training level  \\ 
        \hline
        $1$ & $191.6$ ($32.8$) \\
        $2$ & $190.5$ ($32.1$) \\
        $4$ & $186.7$ ($28.4$) \\
        $8$ & $184.2$ ($36.4$) \\
        \hline
    \end{tabular} \medskip
    \caption{Full results from Figure~\ref{fig:521/pop/cross_play/hp}: Performance metrics for population diversity experiments in HarvestPatch. Mean values (and standard deviations, reported in parentheses) are calculated over five independent runs.}
    \label{tab:app/results/pop/hp/performance}
\end{table}

\begin{table}[h]
    \centering
    \begin{tabular}{ c c c c }
        \hline
        $N_a$ & $N_b$ & Mean difference & Adjusted $p$-value \\ 
        \hline
        $1$ & $2$ & $1.1$ & $1.00$ \\ 
        $1$ & $4$ & $4.9$ & $1.00$ \\
        $1$ & $8$ & $7.4$ & $0.98$ \\
        $2$ & $4$ & $3.8$ & $1.00$ \\ 
        $2$ & $8$ & $6.3$ & $0.99$ \\ 
        $4$ & $8$ & $2.5$ & $1.00$ \\
        \hline
    \end{tabular} \medskip
    \caption{Pairwise comparisons of agent performance between population sizes $N \in \{1, 2, 4, 8\}$, calculated with Tukey's HSD method. Positive ``Mean difference'' values indicate that training with $N_a$ resulted in higher performance than training with $N_b$.}
    \label{tab:app/results/pop/hp/perf_tukeys}
\end{table}

\newpage
\textbf{Traffic Navigation}

\begin{table}[h]
    \centering
    \begin{tabular}{ c c }
        \hline
         & \multicolumn{1}{c}{Individual reward on:} \\
        $N$ & Training level  \\ 
        \hline
        $1$ & $56.8$ ($6.7$) \\
        $2$ & $55.7$ ($7.1$) \\
        $4$ & $57.5$ ($4.7$) \\
        $8$ & $58.8$ ($6.1$) \\
        \hline
    \end{tabular} \medskip
    \caption{Full results from Figure~\ref{fig:521/pop/cross_play/tn}: Performance metrics for population diversity experiments in Traffic Navigation. Mean values (and standard deviations, reported in parentheses) are calculated over five independent runs.}
    \label{tab:app/results/pop/tn/performance}
\end{table}

\begin{table}[h]
    \centering
    \begin{tabular}{ c c c c }
        \hline
        $N_a$ & $N_b$ & Mean difference & Adjusted $p$-value \\ 
        \hline
        $1$ & $2$ & $1.0$ & $0.99$ \\ 
        $1$ & $4$ & $-0.8$ & $1.00$ \\
        $1$ & $8$ & $-2.1$ & $0.95$ \\
        $2$ & $4$ & $-1.8$ & $0.97$ \\ 
        $2$ & $8$ & $-3.1$ & $0.86$ \\ 
        $4$ & $8$ & $-1.3$ & $0.99$ \\
        \hline
    \end{tabular} \medskip
    \caption{Pairwise comparisons of agent performance between population sizes $N \in \{1, 2, 4, 8\}$, calculated with Tukey's HSD method. Positive ``Mean difference'' values indicate that training with $N_a$ resulted in higher performance than training with $N_b$.}
    \label{tab:app/results/pop/tn/perf_tukeys}
\end{table}

\textbf{Overcooked}

\begin{table}[h]
    \centering
    \begin{tabular}{ c c }
        \hline
         & \multicolumn{1}{c}{Individual reward on:} \\
        $N$ & Training level  \\ 
        \hline
        $1$ & $199.5$ ($78.5$) \\
        $2$ & $263.6$ ($85.5$) \\
        $4$ & $281.1$ ($86.4$) \\
        $8$ & $302.9$ ($97.8$) \\
        \hline
    \end{tabular} \medskip
    \caption{Full results from Figure~\ref{fig:521/pop/cross_play/oc}: Performance metrics for population diversity experiments in Overcooked. Mean values (and standard deviations, reported in parentheses) are calculated over 20 independent runs.}
    \label{tab:app/results/pop/oc/performance}
\end{table}

\begin{table}[h]
    \centering
    \begin{tabular}{ c c c c }
        \hline
        $N_a$ & $N_b$ & Mean difference & Adjusted $p$-value \\ 
        \hline
        $1$ & $2$ & $-64.1$ & $0.10$ \\ 
        $1$ & $4$ & $-81.6$ & $2.1 \times 10^{-2}$ \\
        $1$ & $8$ & $-103.4$ & $2.0 \times 10^{-3}$ \\
        $2$ & $4$ & $-17.5$ & $0.92$ \\ 
        $2$ & $8$ & $-39.3$ & $0.49$ \\ 
        $4$ & $8$ & $-21.8$ & $0.86$ \\
        \hline
    \end{tabular} \medskip
    \caption{Pairwise comparisons of agent performance between population sizes $N \in \{1, 2, 4, 8\}$, calculated with Tukey's HSD method. Positive ``Mean difference'' values indicate that training with $N_a$ resulted in higher performance than training with $N_b$.}
    \label{tab:app/results/pop/oc/perf_tukeys}
\end{table}

\newpage
\textbf{Capture the Flag}

\begin{figure}[h!]
    \centering
    \includegraphics[width=0.35\linewidth]{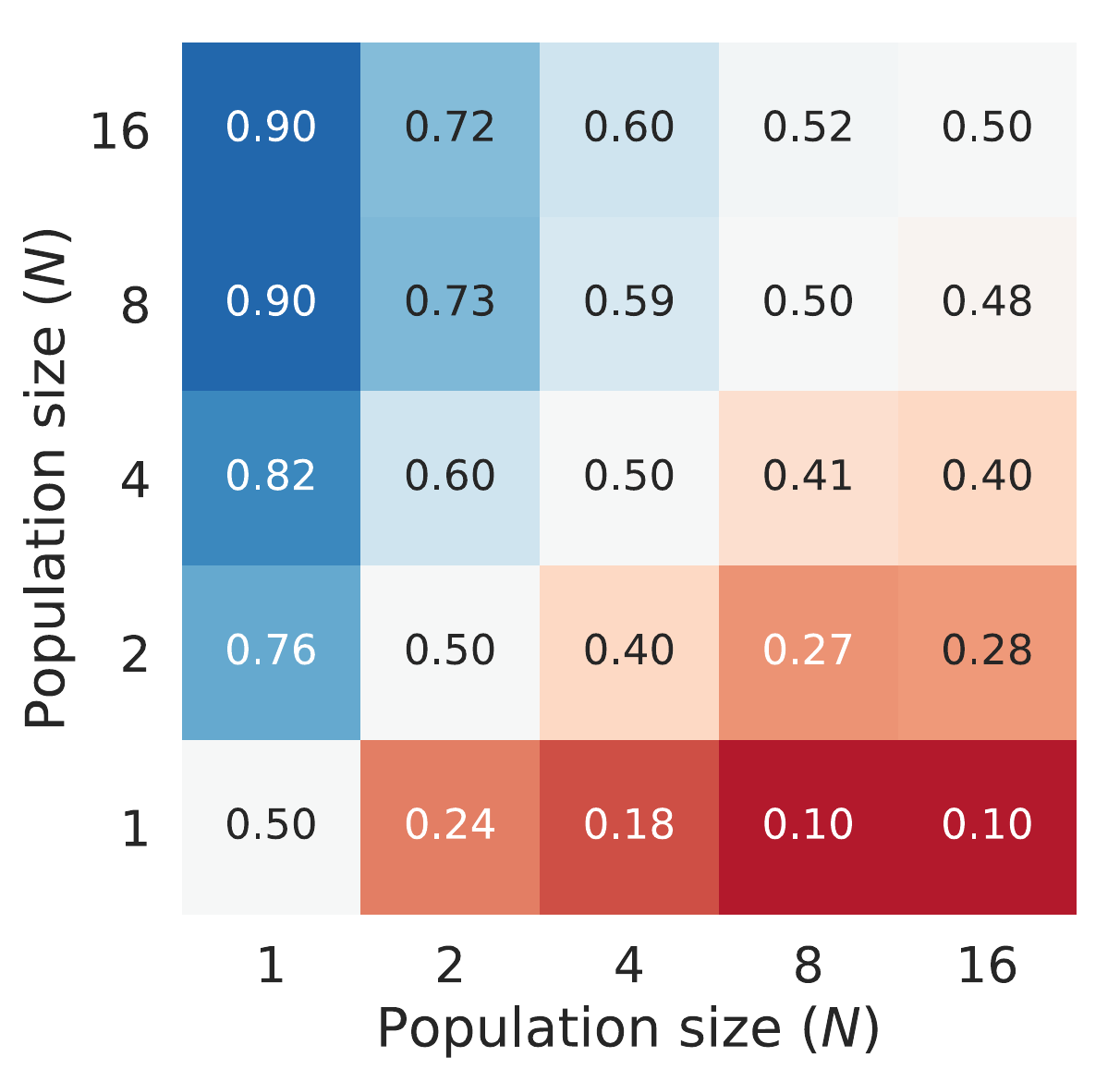}
    \caption{Win matrix corresponding to the Elo ratings for each population size presented in Figure~\ref{fig:521/pop/cross_play}d.}
    \label{fig:app/results/pop/ctf/win_matrix}
\end{figure}

\end{document}